\documentclass[prx, aps, twocolumn, amsmath, amssymb, superscriptaddress, nofootinbib, numbers, reprint]{revtex4-2} 

\newcommand\brpm{\mathbin{\vcenter{\hbox{\oalign{$\scriptstyle({+})$\cr
					\noalign{\kern-.3ex}
					\hfil$\scriptscriptstyle-$\hfil\cr}}}}}

\let\originalleft\left
\let\originalright\right
\renewcommand{\left}{\mathopen{}\mathclose\bgroup\originalleft}
\renewcommand{\right}{\aftergroup\egroup\originalright}

\newcommand{\pphy}{p_\text{phys}}
\newcommand{\pfail}{p_\text{fail}}

\newcommand{\Pfail}{P_\text{fail}}

\newcommand{\Var}[1]{\text{Var}\left({#1}\right)}
\newcommand{\Cov}[1]{\text{Cov}\left({#1}\right)}
\newcommand{\Prot}{\mathcal{P}}

\newcommand{\C}{\mathcal{C}}
\newcommand{\E}{\mathds{E}}

\usepackage[pdftex]{graphicx}
\usepackage{tikz}
\definecolor{myyellow}{rgb}{1, 0.9411764706, 0.337254902}
\definecolor{myblue}{rgb}{0.12156862745098039, 0.4666666666666667, 0.7058823529411765}
\usepackage{hyperref}
\hypersetup{
     colorlinks=true,
     linkcolor=myblue,
     filecolor=magenta,      
     urlcolor=myblue,
     pdftitle={Dynamical subset sampling of quantum error correcting protocols},
     citecolor = myblue
}

\usepackage{mathtools}
\usepackage{wasysym}
\usepackage{amssymb}

\usepackage{pifont}\newcommand{\cmark}{\color{green}\LARGE\ding{51}}\newcommand{\xmark}{\color{red}\LARGE\ding{55}}

\usepackage[utf8]{inputenc}
\usepackage{braket}
\usepackage[normalem]{ulem}
\usepackage{dsfont}
\usepackage{xcolor}
\usepackage{nicefrac}
\usepackage{enumerate}

\usepackage{float}
\makeatletter
\let\newfloat\newfloat@ltx
\makeatother
\usepackage{algorithm}
\usepackage{algpseudocode}

\usepackage{tikz-qtree}
\usepackage{forest}
\usetikzlibrary{quantikz}\usepackage{rotating}

\usepackage{wrapfig}
\usepackage[caption=false]{subfig}
\usepackage{ragged2e} \DeclareCaptionJustification{justified}{\justifying}

\makeatletter
\renewcommand*{\ALG@name}{Algorithm }
\makeatother

\begin{document}

\title{Dynamical subset sampling of quantum error correcting protocols}
\author{Sascha Heu\ss en}
\email{sascha.heussen@rwth-aachen.de}
\affiliation{\footnotesize Institute for Quantum Information, RWTH Aachen University, 52056 Aachen, Germany}
\affiliation{\footnotesize Institute for Theoretical Nanoelectronics (PGI-2), Forschungszentrum Jülich, 52425 Jülich, Germany}
\author{Don Winter}
\affiliation{\footnotesize Institute for Quantum Information, RWTH Aachen University, 52056 Aachen, Germany}
\affiliation{\footnotesize Institute for Theoretical Nanoelectronics (PGI-2), Forschungszentrum Jülich, 52425 Jülich, Germany}
\email{don.winter@rwth-aachen.de}
\author{Manuel Rispler}
\affiliation{\footnotesize Institute for Quantum Information, RWTH Aachen University, 52056 Aachen, Germany}
\affiliation{\footnotesize Institute for Theoretical Nanoelectronics (PGI-2), Forschungszentrum Jülich, 52425 Jülich, Germany}
\email{rispler@physik.rwth-aachen.de}
\author{Markus Müller}
\affiliation{\footnotesize Institute for Quantum Information, RWTH Aachen University, 52056 Aachen, Germany}
\affiliation{\footnotesize Institute for Theoretical Nanoelectronics (PGI-2), Forschungszentrum Jülich, 52425 Jülich, Germany}
\email{m.mueller@physik.rwth-aachen.de}

\begin{abstract}
Quantum error correcting (QEC) stabilizer codes enable protection of quantum information against errors during storage and processing. Simulation of noisy QEC codes is used to identify the noise parameters necessary for advantageous operation of logical qubits in realistic quantum computing architectures. Typical quantum error correction techniques contain intermediate measurements and classical feedback that determine the actual noisy circuit sequence in an instance of performing the protocol. Dynamical subset sampling enables efficient simulation of such non-deterministic quantum error correcting protocols for any type of quantum circuit and incoherent noise of low strength. As an importance sampling technique, dynamical subset sampling allows one to effectively make use of computational resources to only sample the most relevant sequences of quantum circuits in order to estimate a protocol's logical failure rate with well-defined error bars. We demonstrate the capabilities of dynamical subset sampling with examples from fault-tolerant (FT) QEC. We show that, in a typical stabilizer simulation with incoherent Pauli noise of strength $p = 10^{-3}$, our method can reach a required sampling accuracy on the logical failure rate with two orders of magnitude fewer samples than direct Monte Carlo simulation. Furthermore, dynamical subset sampling naturally allows for efficient simulation of realistic multi-parameter noise models describing faulty quantum processors. It can be applied not only for QEC in the circuit model but any noisy quantum computing framework with incoherent fault operators including measurement-based quantum computation and quantum networks. \end{abstract}

\maketitle

\section{Introduction}\label{sec:intro}

All quantum hardware inherently suffers from noise and thus will continue to be far from perfect for the upcoming decades~\citep{preskill2018quantum}. Today's physical architectures ranging from superconducting transmon qubits over photonic systems to ion trap and neutral atom platforms are suitable to implement noisy quantum algorithms for storage, communication or manipulation of quantum information and quantum simulation. Investigating the effect of noisy components via numerical simulation provides a path to practically realize useful quantum applications, equally important for circuit- and measurement-based quantum computation~\citep{briegel2009measurement,jozsa2006introduction} as well as quantum networks~\citep{duan2010colloquium,kozlowski2019towards}. 

\begin{figure*}[!htbp]
\centering
\includegraphics[width=0.99\textwidth]{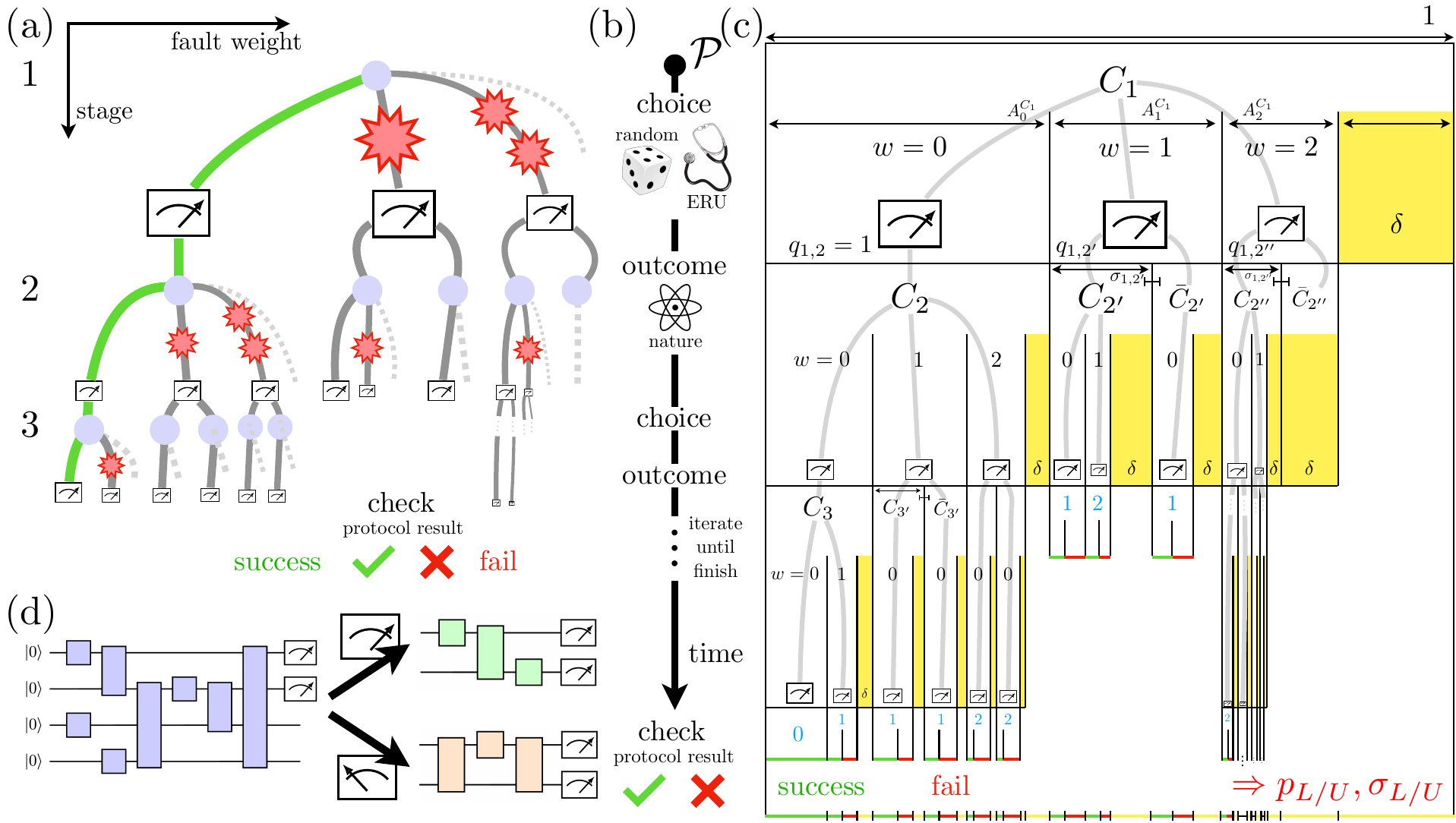}

\caption{\textbf{Protocol illustrations for dynamical subset sampling.} (a) Schematic event tree. Purple nodes symbolize individual quantum circuits. For each circuit one can choose to apply a fault with weight $w \in \mathds{N}_0$ (red stars), which increases from left to right. In the vertical direction, measurement outcomes determine which circuit is run as the next part of a protocol. We assume, for simplicity, that any measurement leads to a binary decision which circuit is run next. After termination, we check whether a failure has occurred. The fault-free path is highlighted in green. For applications outside of QEC, multiple fault-free paths might exist. (b) Protocol sampling consists of alternating steps of choosing a fault subset for the given circuit (``choice'') and evaluating the measurement at the end of that faulty circuit (``outcome'') to determine the next circuit until the protocol finishes. The choice can be taken at random or by using a diagnostic criterion such as the expected reduction of uncertainty (ERU) that we propose in Sec.~\ref{sec:dss}. Measurement outcomes are stochastically determined by ``nature''. (c) Box representation of the tree in (a) containing circuits $C$ and measurements that cause branchings. The horizontal width of any box, labelled with the fault weight $w$, symbolizes its binomial factor $A_w$. The failure rate lower bound $p_L$ is evaluated as the sum over all individual path failure rates (red portion of lower horizontal edge). The true failure rate is underestimated by at most all boxes that were not explored (yellow), i.e.~the cutoff error $\delta$, which allows calculation of an upper bound to the failure rate $p_U$ (red + yellow portion of lower horizontal edge). Both bounds have respective sampling uncertainties $\sigma_{L/U}$ that are rooted in the sampling uncertainties $\sigma_i$ of the branching ratios $q_i$. The fault-free path never leads to failure. All quantities are introduced in Sec.~\ref{sec:dss}. (d) Example of a generic dynamical sequence of two circuits. After measurement of the first two qubits in the first circuit (blue), either the upper (green) or the lower (orange) circuit is run on the last two qubits depending on the intermediate measurement result.} 
\label{fig:introDSS}

\end{figure*}

Quantum algorithms typically consist of sequences of faulty quantum circuits and measurements, visualized as a tree structure in Fig.~\ref{fig:introDSS}(a). Fig.~\ref{fig:introDSS}(b) shows that, crucially, the effect of random faults occurring in a given circuit is not known a priori but can subsequently lead to random measurement outcomes, which determine the next circuit of the sequence until the protocol terminates. 

Widely used numerical simulation methods that perform advantageously in certain regimes of physical fault rates are summarized in Ref.~\citep{bravyi2013simulation}: For direct Monte Carlo sampling, every ideal circuit operation is followed by stochastically drawn fault operators \citep{steane2003overhead, raussendorf2007topological}. It can be used at relatively large fault rates. Alternatively, one may exhaustively iterate all possible fault events and determine if each one leads to failure of the algorithm. This way, the protocol failure rate can be reconstructed, which is only feasible at very low physical fault rates~\citep{van2012quantum,fowler2013analytic}. In Ref.~\citep{bravyi2013simulation} the authors introduce a Metropolis-type technique specific to the surface code to estimate intermediate to small physical fault rates where convergence and thus a reliable confidence interval on the result is not guaranteed. 

Instead of iterating \emph{all} possible fault events at low physical fault rates, one may estimate the effect of noise by selectively sampling faults belonging to distinct subsets in a Monte Carlo-type procedure using a \emph{finite} number of samples and approximately reconstruct failure rates with a finite sampling uncertainty. This is the concept of subset sampling, which can be employed for any quantum circuit of a fixed size \citep{li2017fault, trout2018simulating, gutierrez2019transversality, li2020fault, paetznick2022performance, heussen2023strategies}. However, this technique cannot deal with adaptive execution of noisy quantum circuits, which are typically run sequentially as part of a quantum algorithm.

A general numerical technique to efficiently simulate protocols with small physical fault rates as in experiments today or with expected future improvements on experimental capabilities to get reliable estimates of protocol failure rates is missing.

\subsection{Results summary \& paper structure}

We introduce \emph{dynamical subset sampling} as an importance sampling method to numerically simulate noisy quantum protocols at low noise strength building on previous works on subset sampling \citep{li2017fault, trout2018simulating, gutierrez2019transversality, li2020fault, paetznick2022performance, heussen2023strategies}. Dynamical subset sampling allows one to estimate protocol failure rates from few, most relevant, fault processes. A well-defined confidence interval on the failure rate estimator is maintained at all times during the sampling process. When dynamical subset sampling is employed at a given maximal noise strength $p = p_\text{max}$ to obtain a protocol failure rate estimator $\hat{p}(p_\text{max})$, all values for $p \rightarrow 0$ can be extracted analytically while keeping the confidence interval on $\hat{p}$ tight. This is especially useful if the true failure rate $p^* \propto p^{t+1}$ scales to zero fast as $p \rightarrow 0$ with $t \geq 1$ as for example in the context of fault-tolerant (FT) quantum error correction (QEC) \citep{terhal2015quantum}. Any noise model consisting of incoherent faults can be treated via dynamical subset sampling even for circuits that contain non-Clifford gates. Multi-parameter noise models are included naturally in the subset sampling approach so it can accommodate the physical processes predominant in the given hardware architecture~\citep{gutierrez2019transversality}. With the python package \texttt{qsample}\footnote{\url{https://github.com/dpwinter/qsample}}, we provide an openly accessible numerical implementation of dynamical subset sampling \citep{qsample}.

This manuscript is structured as follows. In the remaining part of this introduction we review the required background of subset sampling. Next, we outline our dynamical generalization of the subset sampling technique in Sec.~\ref{sec:dss} in the context of FT QEC. In Sec.~\ref{sec:examples} we present illustrative examples on how dynamical subset sampling can be utilized to obtain failure rates for GHZ state preparation~\citep{greenberger1989going, shor1996fault}~and FT initialization of a logical qubit in the Steane code via flag circuits~\citep{steane1996multiple, steane1996error, goto2016minimizing, heussen2023strategies, chao2018quantum, chamberland2018flag}. Both examples are highly relevant protocols for FT QEC and have been used in several experiments recently demonstrating fault-tolerant qubit initialization and stabilizer readout~\citep{ryan2021realization,postler2021demonstration}. We conclude and provide an outlook on future work in Sec.~\ref{sec:conclusion}.

\subsection{Review: Subset sampling}\label{sec:subsetsampling}

Direct Monte Carlo (MC) simulations are implemented numerically by traversing a quantum circuit and deciding with probability $p$ whether or not to place a fault operator at a circuit location\footnote{We use the term ``location" to refer to an operation (a gate, a measurement or an initialization) at a specific position in the circuit.}. Fault operators are drawn randomly from a distribution representing the noise model. Repeated realizations of the noisy quantum circuit (``samples'' or ``shots'') yield different output states. The MC estimator for the failure rate is then given by the number of samples that result in a protocol failure divided by the total number of samples
\begin{align}
    \hat{p} &= \frac{\# \text{protocol failures}}{\# \text{MC samples}}.
\end{align}

This is the standard method to investigate QEC code performance where all components are considered noisy. The handiness of MC comes at the cost of lacking efficiency when failures happen less frequently, i.e.~at lower physical fault rates $p$. To illustrate this, observe that for instance at a physical fault rate of $p = 2 \times 10^{-3}$ in a circuit of $g = 50$ gates, the circuit will be sampled without any fault at all $(1-p)^g \approx 90\%$ of the time. Provided further knowledge about properties of the circuit to be simulated, the contrast can become even more stark. For example, when simulating a fault-tolerant circuit no single fault can cause a failure so MC sampling leads to a trivial result \mbox{$(1-p)^{50} + 50p(1-p)^{50-1} \approx 99.5\%$} of the time. When post-processing of an MC run is needed to determine if the result is a failure or not, those $99.5\%$ of all samples would be discarded -- a very inefficient strategy~\citep{li2017fault, li2020fault}. The lower the physical fault rate $p \rightarrow 0$ the larger this ratio of trivial results will become for a given circuit~\citep{bravyi2013simulation}. 

\begin{figure}\centering
    \includegraphics[width=0.99\linewidth]{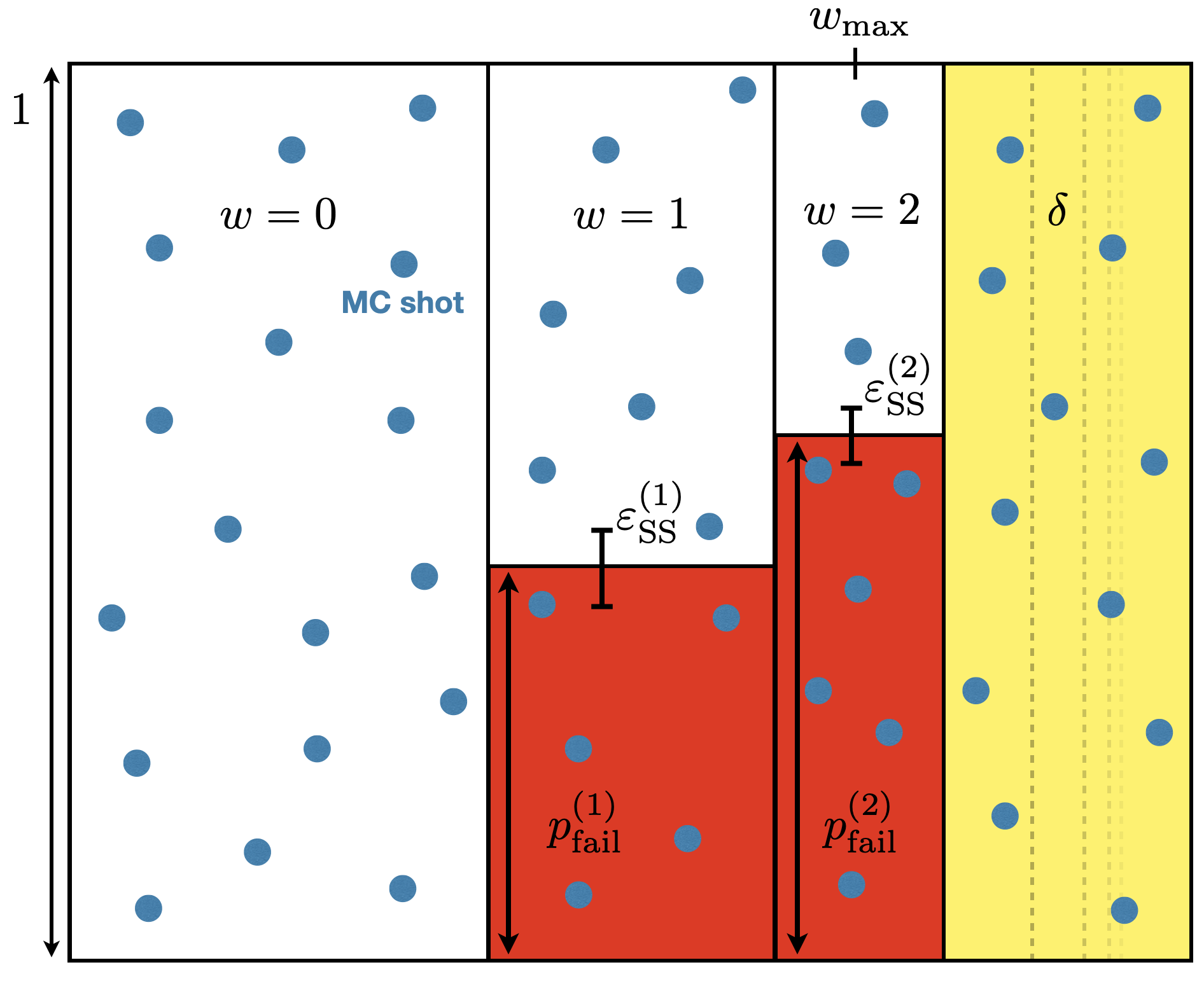}
    \caption{Representation of $w$-fault-subsets for a single circuit. Whereas direct MC would sample uniformly in the whole $w$-space (blue dots, most shots in $w=0$) subset sampling selectively determines the subset failure rates $\pfail^{(w)}$ (ratio of vertical rectangle colored red) up to a weight cutoff $w_\text{max}$. This is the key motivation for the subset sampling technique. Here, by splitting the sampling space we avoid the 0-fault-subset by design. The upper and lower bound on the true failure rate $p_c$ differ by the cutoff subsets illustrated by the yellow rectangle labelled $\delta$ and fainting dashed lines for $w > w_\text{max}$.}
    \label{fig:sampler_boxes}
\end{figure}

The number of faults $w$ happening during circuit execution can be used to uniquely label distinct \emph{subsets} in the space of sampled circuits. For sufficiently low physical fault rate $p = \pphy$ the ``largest" subset (in terms of sampling probability) is the 0-fault-subset as the above example illustrates. In stark contrast to direct MC, subset sampling never samples in the 0-fault-subset thus saving the more computational resources the larger the 0-fault-subset, i.e.~the lower $\pphy$. 
Within any $w$-fault-subset, the probability for a specific $w$-fault to occur is the probability to have exactly $w$ faulty locations $\pphy^w$ multiplied by the probability to have exactly $N_c-w$ non-faulty locations \mbox{$(1-\pphy)^{N_c-w}$} simultaneously. There are $\binom{N_c}{w}$ possibilities to choose a $w$-fault so the size of the $w$-fault-subset is given by the binomial factor 
\begin{align}
	A^c_{w}(\pphy) = \binom{N_c}{w}\pphy^w(1-\pphy)^{N_c-w} \label{eq:binom}
\end{align}
where $N_c$ is the number of circuit locations, which can cause a fault, and $w$ is the fault weight, which we discuss now. Faults are drawn randomly from a circuit-level Pauli noise model meaning that all single-qubit gates, two-qubit gates, initializations and measurements are performed ideally but followed by a fault operator according to the noise model under investigation. A fault configuration of weight $w$ belongs to the $w$-fault-subset where $w$ is the number of faulty locations in the circuit. The 1-fault-subset contains all faults of order $\pphy$, the 2-fault-subset contains all faults of order $\pphy^2$ and so on.\footnote{The fault weight $w$ is different from the Pauli weight of an operator: the number of qubits on which it has non-trivial support, i.e.~is not the identity $I$. For example, a $X \otimes Z$ fault on a CNOT gate (and all other circuit elements fault-free) is a 1-weight-fault of Pauli weight 2.} An illustration of the subsets is given in Fig.~\ref{fig:sampler_boxes}. The fault-free subset of a circuit $c$ is the largest when the physical fault rate is below the inverse number of potentially faulty circuit locations:
\begin{align}
    A_0^c(\pphy) &> A_1^c(\pphy) \notag \\
    \Leftrightarrow \pphy &< \frac{1}{N_c+1}. \label{eq:lowp}
\end{align}

The failure rate $p_c$ of the circuit is given by the subset failure rates $\pfail^{(w)}$ for each $w$-fault-subset independently as 
\begin{align}
	p_c &= \sum_{w=0}^{N_c} A^c_{w}(\pphy) \pfail^{(w)}\text{.} \label{eq:subset_pL}
\end{align}
Each subset failure rate $\pfail^{(w)}$ can be estimated numerically by randomly drawing $N_\text{SS}^{(w)}$ fault operators of weight $w$ according to the noise model under investigation and evaluating the ratio of $w$-weight-fault realizations that lead to a failure. 

Notably, the $\pphy$-dependence in Eq.~(\ref{eq:subset_pL}) is completely covered by the binomial factors $A$, which are known analytically. The subset failure rates $\pfail^{(w)}$ are independent of the physical fault rate $\pphy$ in this expression. As a consequence, the subset failure rates only need to be sampled once for a fixed numerical value $\pphy = p_\text{max}$, so the physical fault rate is naturally removed from the sampling in this formalism. The functional behavior $p_c(\pphy)$ can be reconstructed analytically after the sampling, which makes it easy to extract scalings. This is the major advantage of subset sampling, which becomes even clearer when we deal with fault-tolerant circuits. A QEC code with distance $d$ can correct $t = \lfloor \frac{d-1}{2} \rfloor$ errors and an FT circuit can tolerate $t$ faults without failing so that $p_\text{fail}^{(w)}=0~\forall w < t$. For the FT $d=3$ protocols considered in this work, we thus know that, by the definition of fault tolerance, $p_\text{fail}^{(w=0)}=0$ and $p_\text{fail}^{(w=1)}=0$.

We can bound the true failure rate $p_c$ by the sampled subset failure rates from above and below by assuming either $p_\text{fail}^{(\bar{w})}=0$ or $p_\text{fail}^{(\bar{w})}=1$ for all $\bar{w} > w_\text{max}$ in the circuit. These bounds on the true failure rate read
\begin{align}
    \sum_{w=0}^{w_\text{max}} A_w p_\text{fail}^{(w)} \leq p_c \leq \sum_{w=0}^{w_\text{max}} A_w p_\text{fail}^{(w)} + \overbrace{\sum_{w = w_\text{max}+1}^{N_c} A_w}^{\equiv\delta}, \label{eq:pL_ss_bounds}
\end{align}
where we have simplified the notation of $A^c_{w}(\pphy)$ by $A_w$ since the circuit $c$ and the physical fault rate are fixed quantities here. Using the normalization of the binomial distribution, the cutoff error $\delta$ for subset sampling, which is the rightmost sum in Eq.~(\ref{eq:pL_ss_bounds}), can be expressed as 
\begin{align}
    \delta = \delta(\pphy) = 1 - \sum_{w=0}^{w_\text{max}}\binom{N_c}{w}\pphy^w(1-\pphy)^{N_c-w}. \label{eq:ss_cutoff}
\end{align}
The bounds in Eq.~(\ref{eq:pL_ss_bounds}) tighten as $\pphy$ goes to zero, sharpening the advantage of subset sampling over MC at low $\pphy$. For larger $\pphy$ the cutoff $w_\text{max}$ needs to be larger in order to keep the cutoff error $\delta(p_\text{max})$ below a desired numerical value at fixed $\pphy = p_\text{max}$. Eventually, a large $w_\text{max}$ will require sampling in a large number of subsets such that employing direct MC will become the more efficient strategy to obtain the failure rate. 

We now compare the sampling errors in subset sampling and MC when estimating the failure rate.
The sampling error, i.e.~the standard deviation, for MC sampling 
\begin{align}
    \varepsilon_\text{MC} \sim \sqrt{\frac{\hat{p}\left(1-\hat{p}\right)}{N_\text{MC}}}~~~~~~(\text{as~}N_\text{MC} \rightarrow \infty) \label{eq:stddev_MC}
\end{align}
falls off slowly when $N_\text{MC}$ is the number of samples used to estimate $\hat{p}$. For subset sampling, the uncertainty stemming from sampling within a subset analogously reads
\begin{align}
    \varepsilon_\text{SS}^{(w)} \sim \sqrt{\frac{p_\text{fail}^{(w)}\left(1-p_\text{fail}^{(w)}\right)}{N_\text{SS}^{(w)}}}~~~~~~(\text{as~}N_\text{SS}^{(w)} \rightarrow \infty) \label{eq:subsetstddev}
\end{align}
where $N_\text{SS}^{(w)}$ is the number of samples used to estimate a single subset failure rate $p_\text{fail}^{(w)}$.
These subset sampling errors are suppressed by the binomial factors for the total sampling error on the failure rate
\begin{align}
	\varepsilon_\text{SS} &= \sqrt{\sum_{w=1}^{w_\text{max}} \left[ A^c_{w}(\pphy) \varepsilon^{(w)}_\text{SS} \right]^2} \label{eq:ss_samplingerror}
\end{align}
since the 0-fault-subset has the largest binomial factor but is never sampled ($\varepsilon^{(w=0)}_\text{SS} = 0$). 
The 68\%-confidence interval $\left[\hat{p} - \varepsilon, \hat{p} + \varepsilon\right]$ around the sampled quantity $\hat{p}$ given above is called the Wald interval. At extreme but finite subset failure rates $p_\text{fail}^{(w)} \approx 0$ or $p_\text{fail}^{(w)} \approx 1$ it is known that the Wald interval suffers from erratic behavior in the output variance $\varepsilon^2$ when the numerical input values only change slightly~\citep{brown2001interval, dauphin2014efficient}. In this case it is advantageous to use the more stable Wilson interval instead in order to calculate variances $\varepsilon^2$~\citep{wilson1927probable}. More details are given in App.~\ref{sec:wilson}. 

Provided that the binomial factors $A^c_{w}(\pphy)$ in Eq.~(\ref{eq:ss_samplingerror}) are small (which is always the case for low $\pphy$) we need much fewer samples $\sum_w N_\text{SS}^{(w)} \ll N_\text{MC}$ to estimate a failure rate with a given variance using subset sampling compared to direct MC. When considering circuits with small $N_c$ it might even be advantageous to exhaustively place all possible 1-faults and get the exact subset failure rate $p_\text{fail}^{(w=1)}$ instead of sampling in the 1-fault-subset. As a result, the subset sampling error $\varepsilon_\text{SS}^{(w=1)}$ vanishes. 

The approach can be extended to multiparameter noise models by distinguishing different physical fault rates in Eq.~(\ref{eq:binom}) and replace $A_w \rightarrow A_{\vec{w}}$ as described in~\citep{gutierrez2019transversality}. Furthermore, works, which have used subset sampling, include Refs.~\citep{trout2018simulating, paetznick2022performance} for surface code implementations in ion traps and Floquet code implementations in superconducting Majorana platforms respectively.

\subsection{Example: Single shot GHZ state preparation}\label{sec:examples_0}

\begin{figure}\centering
    \includegraphics[width=0.99\linewidth]{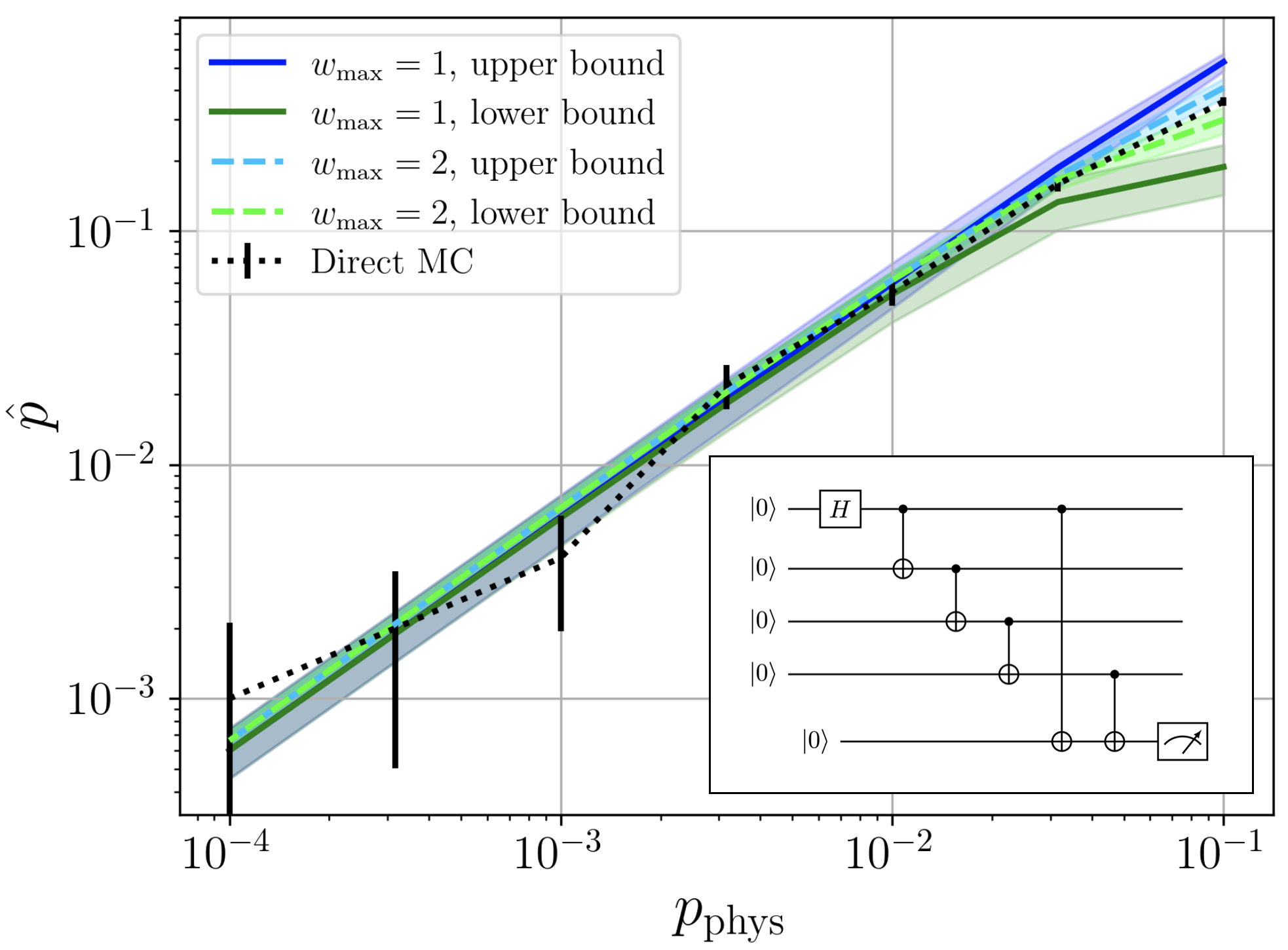}
    \caption{Behavior of the upper and lower bound on the true failure rate $p_c$ for two different weight cutoffs $w_\text{max} = 1$ (solid) and $w_\text{max} = 2$ (dashed) for subset sampling the FT four-qubit GHZ preparation circuit (inset, see Sec.~\ref{sec:examples_ghz}) where failure corresponds to measuring $-1$ on the auxiliary qubit; The cutoff error $\delta$ is determined at constant $p_\text{max}$. The sampling error $\varepsilon_\text{SS}$ (shaded areas around bounds) is independent of $\pphy$. Upper and lower bounds on the failure rate tighten when lowering $\pphy$. Uncertainty intervals for direct Monte Carlo (MC) sampling with at most 1000 shots grow larger when lowering $\pphy$.}
    \label{fig:ss_bounds}
\end{figure}

An example for bounds on the failure rate from Eq.~(\ref{eq:pL_ss_bounds}) and the combined sampling and cutoff error from Eqs.~(\ref{eq:ss_cutoff}) and (\ref{eq:ss_samplingerror}) is given in Fig.~\ref{fig:ss_bounds}. We show the bounds and errors for the inset quantum circuit that prepares the four qubit GHZ state under depolarizing circuit-level noise (see App.~\ref{sec:noise}) verified by an additional flag auxiliary qubit \citep{shor1996fault} (more detail on this procedure is provided in Sec.~\ref{sec:examples_ghz}). Measuring the flag qubit as $-1$ is interpreted as a protocol failure in this example. The tightening upper and lower bounds on the true failure rate $p_c(\pphy)$ for low $\pphy$ are clearly visible. For larger $w_\text{max} = 2$ the bounds are narrowing faster, equivalently the cutoff error $\delta$ at $\pphy = p_\text{max} = 10^{-2}$ is smaller than for $w_\text{max} = 1$. The sampling errors $\varepsilon_\text{SS}^{(w)}$ only depend on the weight $w$ and the number of samples $N_\text{SS}^{(w)}$; they are \emph{independent of the physical fault rate}, which is not the case for direct MC. In Fig.~\ref{fig:ss_bounds}, the size of the direct MC sampling error grows larger as we lower $\pphy$ while keeping $N_\text{MC}$ fixed because the relative error $\varepsilon_\text{MC}/\hat{p} \sim \hat{p}^{-1/2}$ as $\hat{p} \rightarrow 0$.

 \section{Dynamical Subset Sampling}\label{sec:dss}

With dynamical subset sampling (DSS) we present an approach that extends the capabilities of subset sampling by including only the most relevant fault-weight-subsets when estimating the failure rate of a \emph{non-deterministic} quantum protocol. DSS can be employed to maximize the gain of accuracy on the failure rate estimator with every shot while maintaining a well-defined uncertainty interval throughout the simulation. In the following discussion we do not explicitly distinguish between the true protocol failure rate $p^*$ from the numerical estimator of the failure rate $\hat{p}$ where there is no danger of confusion and only stress the difference explicitly when important.

Executing a quantum protocol $\Prot$ under the influence of noise can be viewed as running a non-deterministic sequence $\C$ of quantum circuits as an instance of the protocol $\Prot$, illustrated as an event tree in Fig.~\ref{fig:introDSS}. A single element of $\C$ is
\begin{align}
    \C_i &= (c_i,w_i), \label{eq:circutisequenceelement}
\end{align}
a tuple that specifies a quantum circuit $c_i$ and a $w_i$-fault-subset. A subset sampling approach samples a $w_i$-fault-subset of a circuit $c_i \in \C$ run as part of the protocol $N_{\C_i}$ times and determines the transition rate 
\begin{align}
    q_i \equiv \mathds{P}(\C_{i} \rightarrow c_{i+1} \,|\, \C_{<i})\label{eq:transprob}
\end{align}
from $\C_i$ to a next circuit $c_{i+1}$ of the sequence given all previous circuits $\C_{<i}$. Figure \ref{fig:introDSS}(c) depicts the sequential branching into subsets of different relevance in the vertical direction, quantified by the binomial factors $A(\C_i) = A_{w_i}^{c_i}$ and the transition rates $q_i$ in the horizontal direction. The latter have a sampling uncertainty of $\sigma_i = \sqrt{\Var{q_i}} = \sqrt{V_i}$ due to the finite number of samples\footnote{Since we assume there exists only one fault-free path (see Fig.~\ref{fig:introDSS}(a)), the uncertainties to branch off from this path are identical to zero.}, given by the Wilson score interval $V_i = \frac{N_{\C_i}q_i(1-q_i) + 1/4}{(1 + N_{\C_i})^2}$ (see Eq.~\eqref{eq:wilsonint}).

Locally, i.e.~at stage $i \in \mathds{N}$, it is impossible to determine by qubit measurements whether or not the execution of a circuit $c_i$ with fault-weight $w_i$ will lead to a failure after execution of the whole sequence $\C$. This is not only because the particular fault is drawn randomly but also because its effect on subsequent circuits cannot be determined a priori at stage $i$. We may only assume an expected value for a protocol failure from previous runs of the whole protocol and update that expectation value dynamically when acquiring additional information, i.e.~recording a measurement result of circuit $c_i$ with weight $w_i$. These transition rates are estimated by a finite number $N_{\C_i}$ of observed measurement outcomes of circuit subsets $\C_i$ within the protocol $\Prot$. 

After termination of the whole protocol, one can determine a posteriori whether a protocol failure has occurred as a result of the specific faulty circuit sequence that was realized in this particular shot, symbolized by a red outcome at the end of a tree path in Fig.~\ref{fig:introDSS}(c). The factors along this path determine the total contribution of this single failure event to the overall failure rate $p_L$. In total, the protocol failure rate is the weighted sum over all paths $\Pfail$ that lead to failures in this tree
\begin{align}
    p_L &= \sum_{\Pfail} \prod_{i \in \Pfail} A(\C_i) q_i\label{eq:dsspl}
\end{align}
where $q_l \equiv \mathds{P}(\C_{\leq l} \rightarrow \text{FAIL})$ when $l$ is the length of a path $\Pfail$.

The quantity $p_L$ is a lower bound to the true failure rate $p^*$ in the sense that we can estimate that
\begin{align}
    p^* \in [p_L - \sigma_{L},\,p_U + \sigma_{U}]
\end{align}
lies within a confidence interval of lower and upper bound estimators $p_L$ and $p_U$. Since they are always estimated from a finite number of shots, we can quantify their sampling uncertainty as
\begin{align}
    \sigma_{L/U} &= \sqrt{\Var{p_{L/U}}}. \label{eq:cisigs}
\end{align}
We determine our estimator $\hat{p}$ for the failure rate to lie in the center between upper and lower bound so that 
\begin{align}
    \hat{p} \equiv \frac{p_L + p_U}{2}. \label{eq:dssestim}
\end{align}
Note that $\hat{p} = p_L$ would also yield an unbiased estimator (see App.~\ref{sec:bias} for details).

The upper bound $p_U$ to the true failure rate $p^*$ can be set by considering all neglected subsets in a finite tree, illustrated in Fig.~\ref{fig:introDSS}(c) as yellow boxes. The worst case error made by neglecting these subsets is that they all exclusively produce failures all the time. At any circuit node $\C_i$ of the tree, its cutoff error $\delta_{\C_i}$ can be quantified, analogously to Eq.~\eqref{eq:ss_cutoff}, as the complement of the binomial factors of the non-neglected subsets like
\begin{align}
    \delta_{\C_i} &= 1 - \sum_{w_i = 0}^{w_{i,\text{max}}} A(\C_i) \label{eq:dssdelta}
\end{align}
(see, for example, the upper right yellow box width in Fig.~\ref{fig:introDSS}(c)). The upper bound $p_U$ is then given by adding the total value of all $\delta_{\C_i}$ to the lower bound
\begin{align}
    p_U &= p_L + \delta = p_L + \sum_{\C_i} \left(\prod_{k = 1}^{i} A(\C_k) q_k \right) \delta_{\C_i} \label{eq:dsspu}
\end{align}
where one must take into account the path product that leads to any $\delta_{\C_i}$ in the tree representation of Fig.~\ref{fig:introDSS}(c).

The variances of $p_L$ and $p_U$ can be calculated using standard rules for calculating sums of products of random variables
\begin{align}
    \Var{p_{L}} &= \Var{\sum_{\Pfail} \prod_{i \in \Pfail} A(\C_i) q_i} \\
    \Var{p_{U}} &= \Var{p_{L} + \delta},
\end{align}
which we lay out explicitly in App.~\ref{sec:errs} with examples. The central ingredient here is that two transition rates $q_i$ and $q_j$ where one is subsequent to the other in the same path are uncorrelated, i.e.~$\Cov{q_i, q_j} = 0$. At the branchings denoted with the measurement symbol in Fig.~\ref{fig:introDSS}(c), the branching ratios\footnote{We use the terms ``branching ratio'' and ``transition rate'' interchangeably.} $q_i$ are maximally anti-correlated since $\Cov{q_i, 1-q_i} = -\Var{q_i}$.

As mentioned above, it is not a priori clear which specific circuit sequence will be realized as an instance of the quantum protocol for a particular shot due to both the randomness in realizing a particular fault operator and measurement outcomes. In the following, we suggest two methods how to sensibly choose a fault-weight-subset in any stage $i$ as illustrated in Fig.~\ref{fig:introDSS}(b) (and provide more detail in App.~\ref{sec:choices}). 

A simple method is to draw a random fault weight $w_i$ according to the distribution of the binomial factors $A(\C_i)$ of the current circuit $c_i$. For a protocol that produces long circuit sequences, this method will produce small total fault weights $w_\text{tot} = \sum_i w_i$ for a single shot if $\pphy$ is not too large. If $\pphy$ is also not too small, the resulting $w_\text{tot}$ of any single shot will eventually be larger than zero and thus protocol failures can be expected. For a protocol that consists of just a single circuit or repeatedly runs the same circuit until the measurement result is accepted so that faults cannot propagate from one circuit run to the next, we can exclude drawing the fault-free subset completely. Note that in principle any distribution can be used to draw fault operators. In Ref.~\cite{paetznick2022performance}, a pre-sampling step is used to guess a distribution of most relevant subsets for a single circuit before actually performing the sampling procedure.

A more sophisticated method systematically decreases the total uncertainty of the failure rate estimation with each sample. It relies on the fact that at any point during the sampling procedure, we can determine expectation values of branching ratios $q_i$ and -- thus -- failure rates. Now, we may assume how these expectation values change given one additional sample in the current stage $i$ when the $w_i$-fault-subset is selected to sample in. Especially, the total estimation uncertainty \begin{align}
    \eta \equiv \sigma_L + \sigma_U + \delta \label{eq:eta}
\end{align}
changes when the subset $w_i$ is chosen if either a positive ($+$) or a negative ($-$) measurement outcome would be observed as a consequence in the next sample. We express the \emph{expected} uncertainty after the next sample as the average over the two possible measurement outcomes $q_i \eta|_{w_i \rightarrow +} + (1-q_i) \eta|_{w_i \rightarrow -}$. The two expected uncertainties $\eta|_{w_i \rightarrow \pm}$ can be calculated by replacing the respective affected transition rate $q_i$ in the tree like
\begin{align}
    q_i \rightarrow q_i^\pm = \begin{cases}
      \left(q_i + \frac{1}{N_{\C_i}}\right)\frac{N_{\C_i}}{N_{\C_i}+1}, & + \\
      q_i\frac{N_{\C_i}}{N_{\C_i}+1}. & -
      \end{cases} \label{eq:shiftq}
\end{align}
as we show in App.~\ref{sec:eruapp}. We can then choose to sample in the subset $w_i^*$ where the \emph{expected reduction of uncertainty} $\Delta(w_i)$ is maximized (``ERU criterion'') with the next sample compared to the current state, i.e.~the total uncertainty coming from the currently known event tree

\begin{align}
    \Delta(w_i) &= \eta - \left[ q_i \eta|_{w_i \rightarrow +} + (1-q_i) \eta|_{w_i \rightarrow -} \right] \label{eq:erv} \\
    w_i^* &= \text{argmax}_{w_i}(\Delta(w_i)).
\end{align}

It is sufficient to calculate $\Delta(w_i)$ for the $w_i$-fault-subsets that are already in the tree ($w_i = 0,\dots,w_{i, \text{max}}$) and the largest subset yet unsampled ($w_i = w_{i, \text{max}} + 1$) since sampling in this subset will decrease the uncertainty more than any other higher weight subset, assuming that $A_w > A_{w+1} \,\forall w \geq 0$, which is fulfilled for small enough $\pphy$. We consider the computational cost of this calculation further in Sec.~\ref{sec:analysis}.

\begin{figure}\centering
    \includegraphics[width=0.99\linewidth]{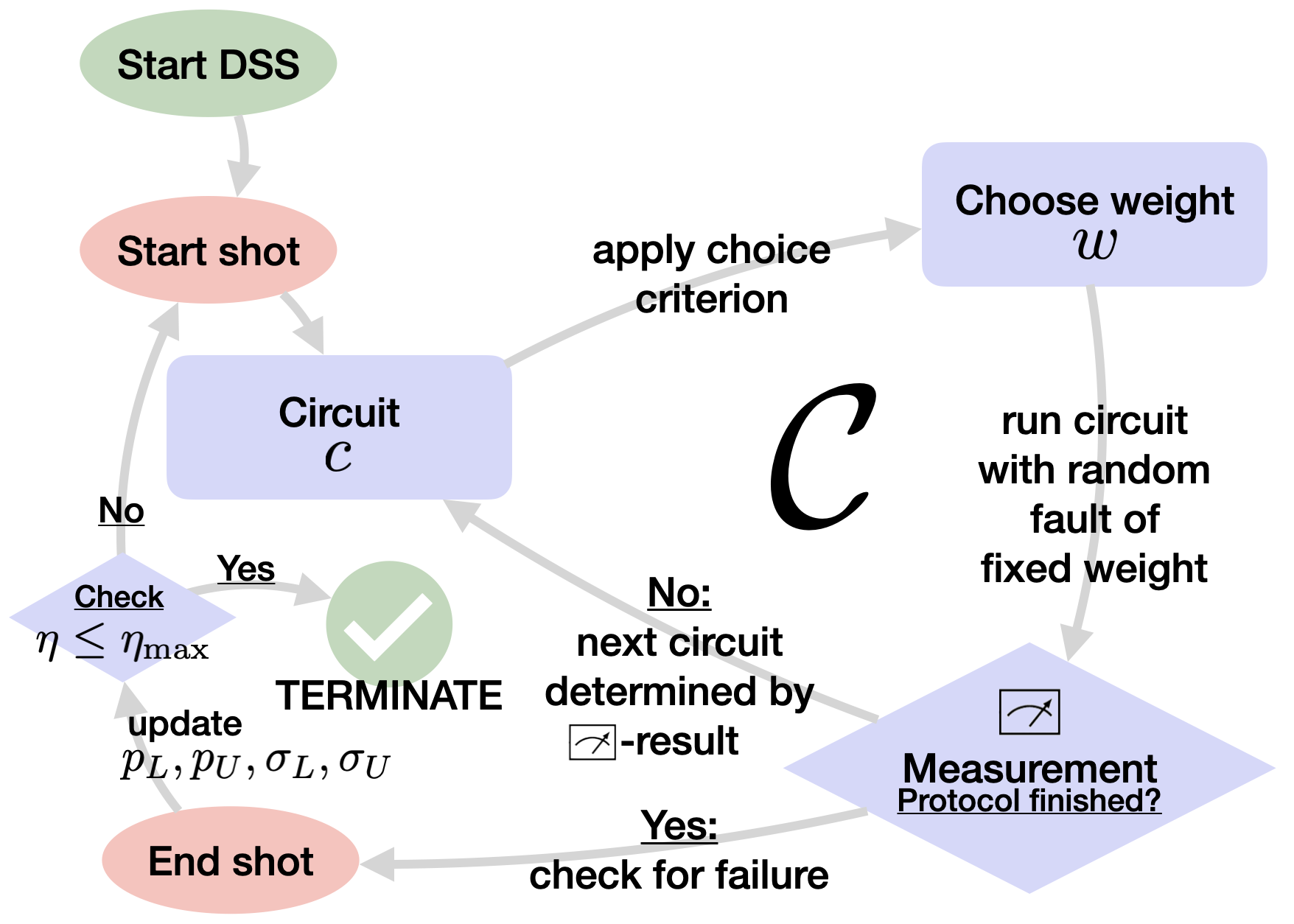}
    \caption{Circular visualization of the choice--measurement process from Fig.~\ref{fig:introDSS}(b). A single shot of dynamical subset sampling (DSS) consists of alternating between choosing the next circuit $c_i$ to sample and choosing the fault-weight-subset $w_i$ to sample this circuit in. The circuit sequence $\C$ is determined by the underlying protocol $\Prot$ and the stochastic intermediate measurement results. The fault-weight-subset can be chosen according to the binomial factor criterion or the ERU criterion (see main text). Then a random fault instance within that subset is drawn uniformly. Repeated execution of the inner cycle generates circuit sequences $\C$, see Eq.~\eqref{eq:circutisequenceelement}, and the event tree in Fig.~\ref{fig:introDSS} by sequentially running circuits that form an instance of a protocol. A new shot is run if the result of the previous shot does not lead to a resulting $\eta$ -- the total estimation uncertainty given in Eq.~\eqref{eq:eta} -- below the target uncertainty $\eta_\text{max}$ or the maximum number of shots $N_\text{max}$ is not reached yet.}
    \label{fig:dss_circ}
\end{figure}

The full dynamical subset sampling procedure is expressed in pseudocode in Alg.~\ref{alg:dss} and as a flow chart in Fig.~\ref{fig:dss_circ}. The numerical values of $p_L,\,p_U,\,\sigma_L$ and $\sigma_U$ are obtained at a fixed value of $p = p_\text{max}$ and the regime of low physical fault rates $\hat{p}(p < p_\text{max})$ is accessible analytically via Eq.~(\ref{eq:dsspl}) because the dependency on $p$ lies entirely in the binomial factors $A(\C)$ which themselves do not depend on the sampled transition rates $q_i$.

In the next section we discuss examples that illustrate the advantage of dynamical subset sampling over MC simulation in the context of fault-tolerant quantum error correction.

\begin{algorithm}[htp]
    \caption{Algorithm to dynamically estimate the failure rate $p^*$ of a non-deterministic QEC protocol $\Prot$ including feedback from classical measurement data using the ERU criterion}
    \begin{algorithmic}[1]
        \Statex \textbf{Input:} maximum number of samples $N_\text{max}$ or target uncertainty $\eta_\text{max}$, noise model $\mathcal{E}$, maximum noise strength $p_\text{max}$ \;
        \Statex \textbf{Output:} failure rate confidence interval $\left[p_L - \sigma_{L},\,p_U + \sigma_U \right]$ \; 
        \While{$N < N_\text{max}$ and $\eta > \eta_\text{max}$}
\State calculate $\Delta(w_i)$ for $w_{i} = 0,...,w_{i,\text{max}} + 1$ \;
		\State set $w_i \gets \text{argmax}_{w_i}( \Delta(w_i))$ \;
		\State run $\C_i$ with $w_i$-fault drawn from $\mathcal{E}$  \; \State determine $c_{i+1}$ from measurement outcome of $\C_i$ according to $\Prot$ \;
		\State update branching ratio $q_i$ \;
        \If{$c_{i+1}$ $\neq$ ``finish"}
            \State $i \gets i + 1$ go to next circuit
        \Else
            \State check for failure \;
			\State update $p_L,\,p_U,\,\sigma_L$ and $\sigma_U$ \;
			\State $N \gets N + 1$ go to next shot
        \EndIf
        \EndWhile
    \end{algorithmic}
    \label{alg:dss}
\end{algorithm} \section{Application to protocols for FT QEC}\label{sec:examples}

The numerical simulations in this section have been performed using our new python package \texttt{qsample}~\citep{qsample}. We first discuss the simple example of a flag circuit to fault-tolerantly prepare a GHZ state \citep{greenberger1989going, shor1996fault}. GHZ states are commonly used as resource states in QEC protocols, e.g.~to extract syndrome information for fault-tolerant stabilizer measurements in Shor-type error correction \citep{shor1996fault, divincenzo1996fault, plenio1997conditional}. 
As a second example, we investigate a protocol to fault-tolerantly encode the zero code word $\ket{0}_L$ in the Steane code via adaptive execution of either one or more quantum circuits \citep{steane1996multiple, goto2016minimizing, heussen2023strategies}. The third example is $\ket{0}_L$ state preparation by sequential stabilizer measurements with flag circuits where up to eight circuits can be run in total \citep{devitt2013quantum, chao2018quantum, chamberland2018flag}.

In our simulations we apply the standard circuit-level depolarizing noise model (see App.~\ref{sec:noise}). We, however, stress again that any noise model consisting of incoherent faults can be covered by dynamical subset sampling even for circuits that contain non-Clifford gates. 

\subsection{Background: Simulating noisy QEC}

Using QEC to reduce the occurrence of dangerous faults that corrupt the quantum data can allow one to push protocol failure rates below the threshold that enables advantageous operation of encoded blocks compared to physical qubits~\citep{Aharonov2008}. Estimating failure rates on the logical level through stabilizer simulation~\citep{aaronson2004improved, gidney2021stim}~is a way to evaluate the performance of stabilizer codes. These performance evaluations are needed for fault-tolerant quantum error correcting protocols that may be used in hardware implementations. The useful operating regime is below the break-even point where the failure rate becomes lower than the failure rate of the unencoded qubit~\cite{Gottesman2016}. If this break-even point converges to a finite value in the infinite system size limit, this is known as threshold. Depending on the code and noise model under investigation, finite thresholds for typical stabilizer codes can -- if existent -- range over several orders of magnitude ($10^{-9}$ to $10^{-2}$ faults per operation)~\citep{devitt2013quantum}. Thus different numerical techniques may be employed to find these thresholds, e.g.~for common stabilizer codes such as surface codes~\citep{kitaev2003fault, fowler2012surface}~or color codes~\citep{bombin2006topological,fowler2011two}.

Stabilizer simulations enable the simulation of quantum codes on classical computers in polynomial time for circuits that only contain Clifford operations, which is known as the Gottesman-Knill theorem~\citep{gottesman1998heisenberg, gidney2021stim}. When investigating noisy stabilizer codes, efficiently simulating faulty circuit operations poses numerical challenges beyond circuit depth or large numbers of qubits due to the large number of MC shots needed to get reliable failure rate estimators (see Sec.~\ref{sec:subsetsampling}). 

Here we consider uncorrelated circuit-level Pauli-noise in stabilizer simulations: Pauli fault\footnote{We distinguish between faults, errors and failures: A fault is an instance of noise on a circuit location, e.g.~a single Pauli operator applied after an ideal operation. An error is the result of a faulty location at the end of the circuit, e.g.~a Pauli fault propagated to a larger weight error. A failure is a wrongly decoded state as the result of an error of weight $w > \lfloor \frac{d-1}{2} \rfloor$ where $d$ is the code distance.} operators are stochastically placed on all individual circuit elements. The Pauli matrices form a generator basis of all $n$-qubit operators that may act as noise in a realistic quantum processor architecture. While other noise models such as phenomenological noise~\citep{dennis2002topological,ohno2004phase} or code capacity noise~\citep{katzgraber2009error} do not take the microscopic details of noisy circuitry for syndrome readout into account, circuit-level noise captures experimental conditions more closely because all circuit elements are modelled as prone to noise~\citep{raussendorf2007fault}. Taking into account the effect of noise to data qubits only thus cannot be expected to accurately model experimental conditions (for a summary of various noise models, see e.g.~\citep{landahl2011fault}). We restrict ourselves to uncorrelated circuit-level noise. All types of incoherent noise can be analyzed in this framework of circuit-level Pauli noise, e.g.~depolarizing noise~\citep{steane2003overhead,tomita2014low, bermudez2017assessing} or crosstalk~\citep{debroy2020logical,parrado2021crosstalk}. Furthermore, non-Clifford noise can be approximated in this way, e.g.~for the amplitude damping channel~\citep{gutierrez2013approximation}. To investigate coherent noise or circuits containing non-Clifford gates full wavefunction simulations can be employed at the price of exponentially growing simulation time and memory requirements~\citep{plenio1998quantum, jones2019quest}.

Placing Pauli faults in a quantum circuit stochastically in a direct Monte Carlo (MC) simulation and determining whether or not these faults cause logical failure has become the standard method to obtain failure rates, thresholds and compare the performance of QEC protocols~\citep{steane2003overhead,raussendorf2007fault}. Recall from Sec.~\ref{sec:subsetsampling} that at low physical fault rates however, direct MC is highly inefficient since most of the time the ideal fault-free circuit is sampled (no faults are placed at all). Subset sampling can obtain high accuracy on failure rates with well-defined error bars when the circuit is a \emph{deterministic} sequence of operations~\citep{gutierrez2019transversality,paetznick2022performance}. However, common QEC protocols such as Shor-type error correction or flag fault-tolerant stabilizer readout are \emph{non-deterministic} because they make use of intermediate measurement information (``feed-forward'')~\citep{shor1996fault, Chao2018}. 

In non-deterministic QEC protocols, these ``in-sequence'' measurements determine at runtime which circuit to execute next. Due to the stochastic nature of quantum measurements the resulting circuit sequence performed as an instance of the QEC protocol is not fixed a priori, e.g.~for adaptive execution of state preparation circuits as in \cite{foss2023experimental} or FT QEC cycles as demonstrated in~\citep{ryan2021realization}~where the circuits used for syndrome readout depend on in-sequence measurement information. Non-deterministic protocols are abundant in QEC. They include fault-tolerant initialization of logical qubits using flag schemes~\citep{goto2016minimizing,heussen2023strategies}~or teleportation elements where gates, which are applied conditioned on classical measurement data, play a role in the implementation of a fault-tolerant universal gate set~\citep{chamberland2019fault, postler2021demonstration}. 

In the following, we apply dynamical subset sampling to non-deterministic QEC with single- and multi-parameter noise models. Multi-parameter noise models can distinguish between different sources of errors and thus provide a more realistic description of the faulty hardware to simulate e.g.~different noise strengths on certain locations, crosstalk~\citep{debroy2020logical,parrado2021crosstalk} or biased Pauli noise~\citep{tuckett2018ultrahigh, tuckett2019tailoring, tuckett2020fault, bonilla2021xzzx,darmawan2021practical,xu2022tailored}.

\subsection{GHZ state preparation}\label{sec:examples_ghz}

GHZ states are commonly used resource states in quantum algorithms, for instance for FT syndrome readout in Shor-type QEC~\citep{shor1996fault, divincenzo1996fault, plenio1997conditional} or quantum teleportation schemes~\citep{gottesman1999demonstrating}. The preparation procedure of a four-qubit GHZ state 
\begin{align*}
    \ket{\text{GHZ}} &= \frac{\ket{0000} + \ket{1111}}{\sqrt{2}}
\end{align*}
as shown in the inset of Fig.~\ref{fig:ss_bounds} is considered successful when the flag measurement yields $+1$. Here the first four gates of the circuit prepare the state on the upper four qubits while the fifth qubit is the flag qubit that signals a fault in the preparation through a $-1$ measurement result, i.e.~when being measured in the computational $\ket{1}$ state. This additional flag qubit verification makes the preparation procedure fault-tolerant towards any 1-weight-fault in the circuit (``FT-1"), meaning that given a $+1$ measurement result, the prepared state is indeed $\ket{\text{GHZ}}$ up to a single correctable error unless at least two faults have happened. However, single faults can lead to a $-1$ measurement of the flag qubit.

\begin{figure}\centering\includegraphics[width=\linewidth]{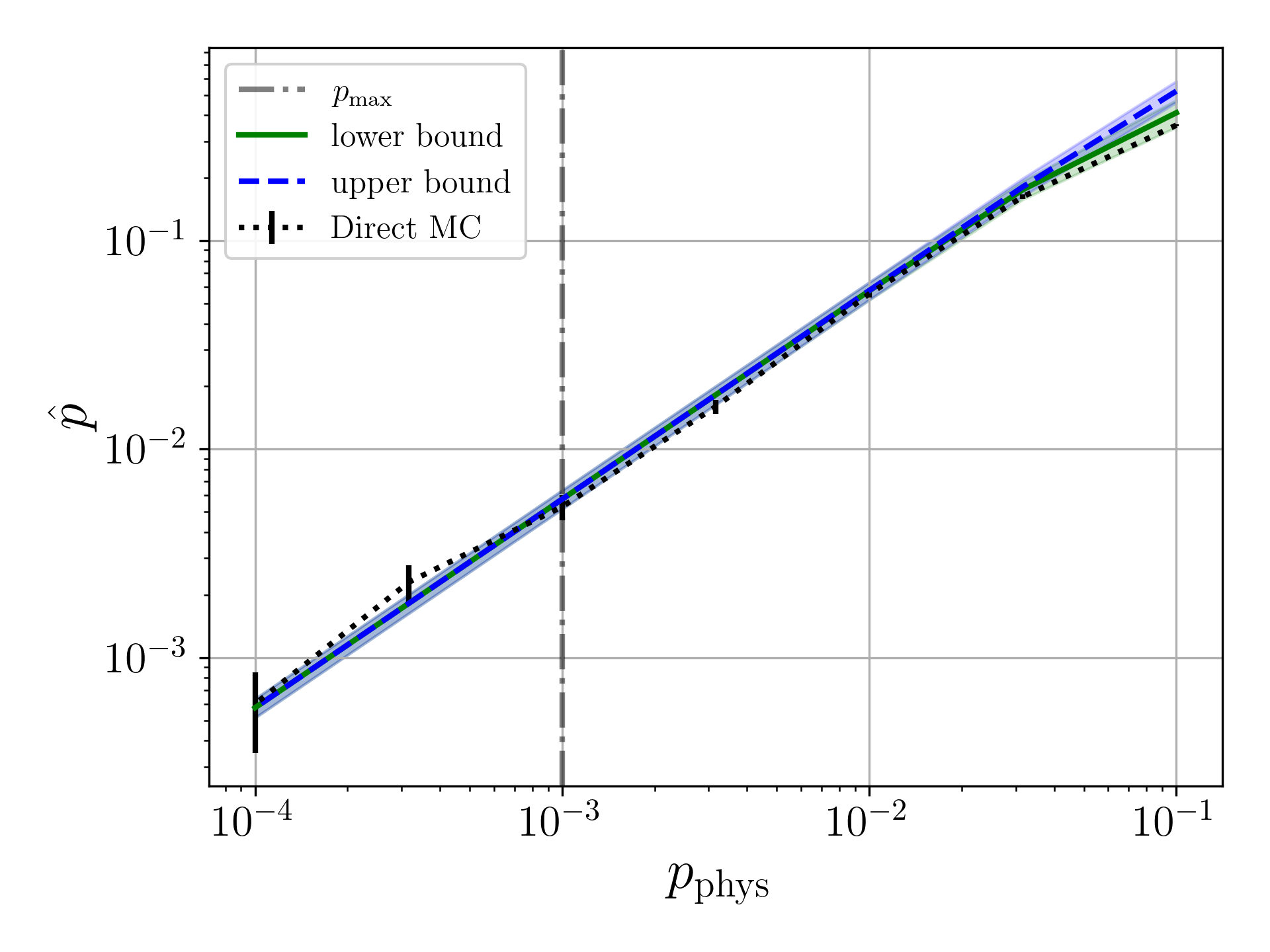}
	\caption{Upper and lower bound on the flag rate for GHZ state preparation are tight when using subset sampling at $p_\text{max} = 10^{-3}$. All rates for smaller $\pphy$ can be extracted analytically while the uncertainty intervals of the direct MC estimation grow larger for smaller $\pphy$.}
	\label{fig:ghz_scaling}
\end{figure}

Figure \ref{fig:ghz_scaling} shows the subset sampling estimation of the flag rate over the range of physical fault rates $\pphy \in [10^{-4}, 10^{-1}]$. Subset sampling is performed with 100 shots at $p_\text{max} = 10^{-3}$ and all other values of the failure rate are extracted analytically. We compare the subset sampling result to direct MC estimation, where $10^4$ shots are needed for comparable accuracy at $p_\text{max} = 10^{-3}$ and relative uncertainties are even larger for lower values of $\pphy$. 

\begin{figure}\centering\includegraphics[width=\linewidth]{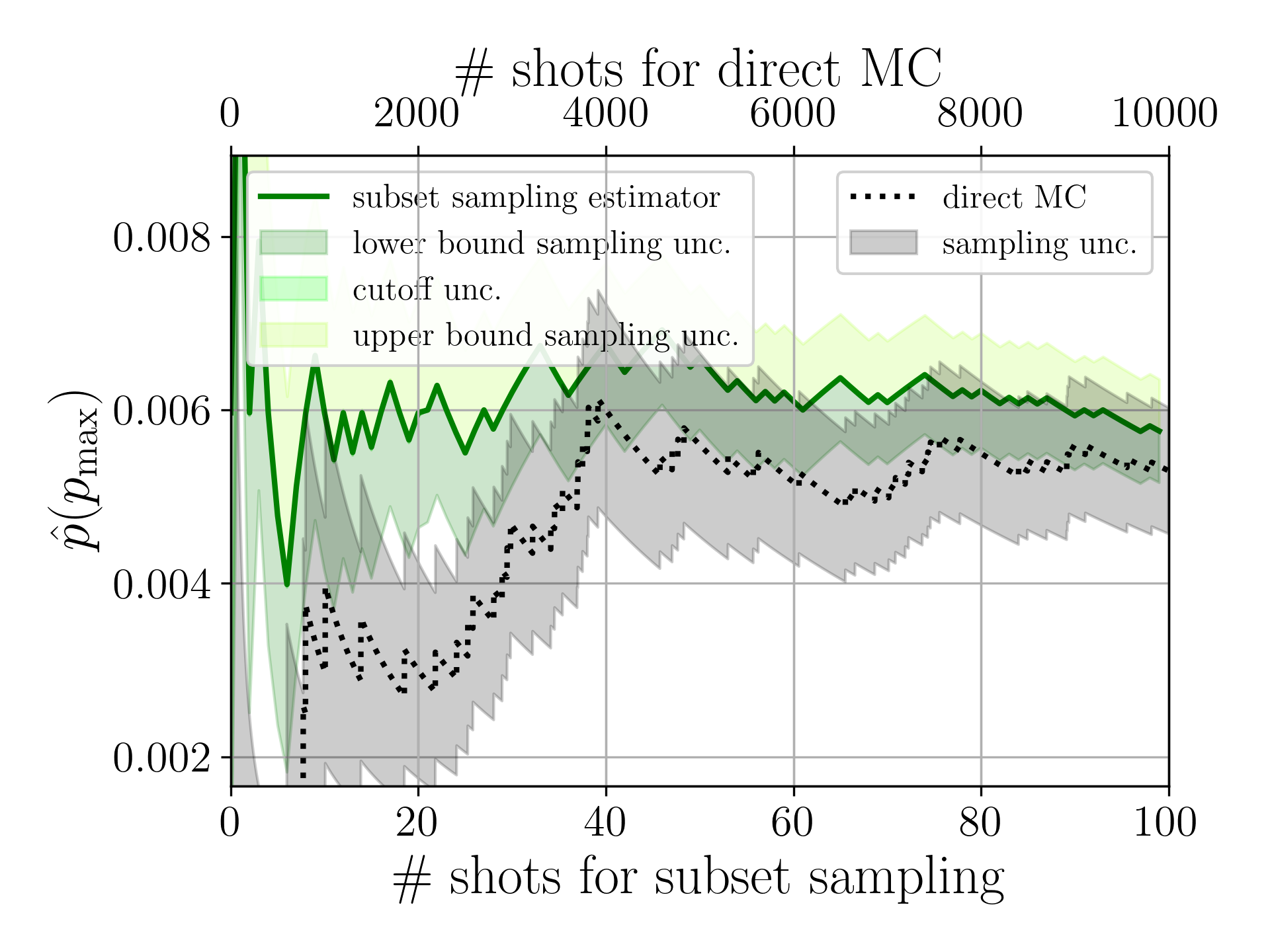}
	\caption{At $p_\text{max} = 10^{-3}$, direct MC (black dotted) needs $10^4$ shots to reach a comparable accuracy as subset sampling (green solid) after $10^2$ shots. Note that two different horizontal axes are used in the plot. The total estimation uncertainty (unc.) is similar after the sampling for both methods but the advantage of DSS over direct MC is two orders of magnitude in the number of samples.}
	\label{fig:ghz_failpershot}
\end{figure}

\begin{figure}\centering\includegraphics[width=\linewidth]{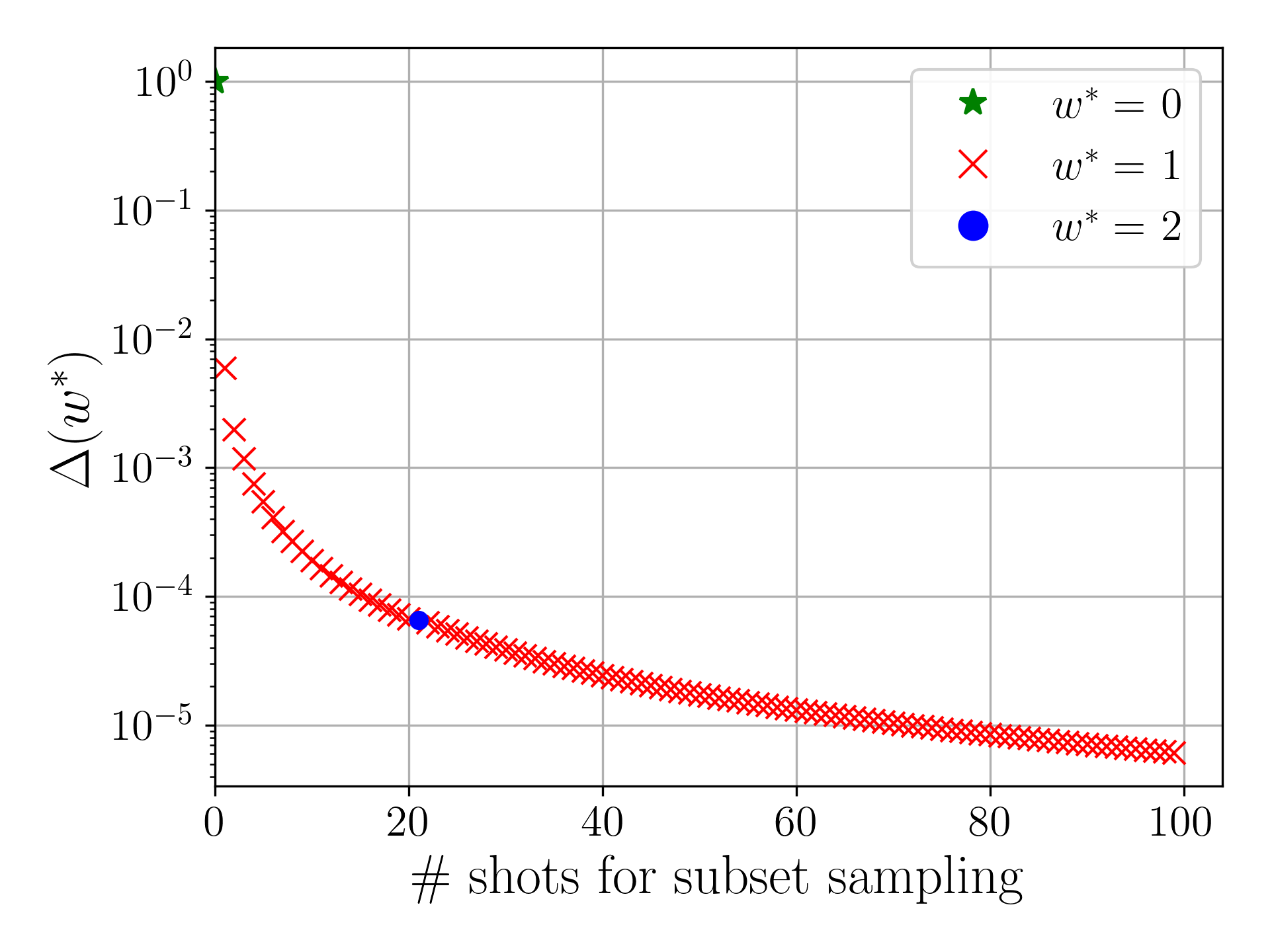}
	\caption{The ERU criterion, at $p_\text{max} = 10^{-3}$, chooses to open the next larger fault-weight-subset when decreasing the sampling uncertainty $\sigma$ does not improve the total uncertainty $\eta$ more than decreasing the cutoff error $\delta$.}
	\label{fig:ghz_ERUpershot}
\end{figure}

\begin{figure}\centering\includegraphics[width=\linewidth]{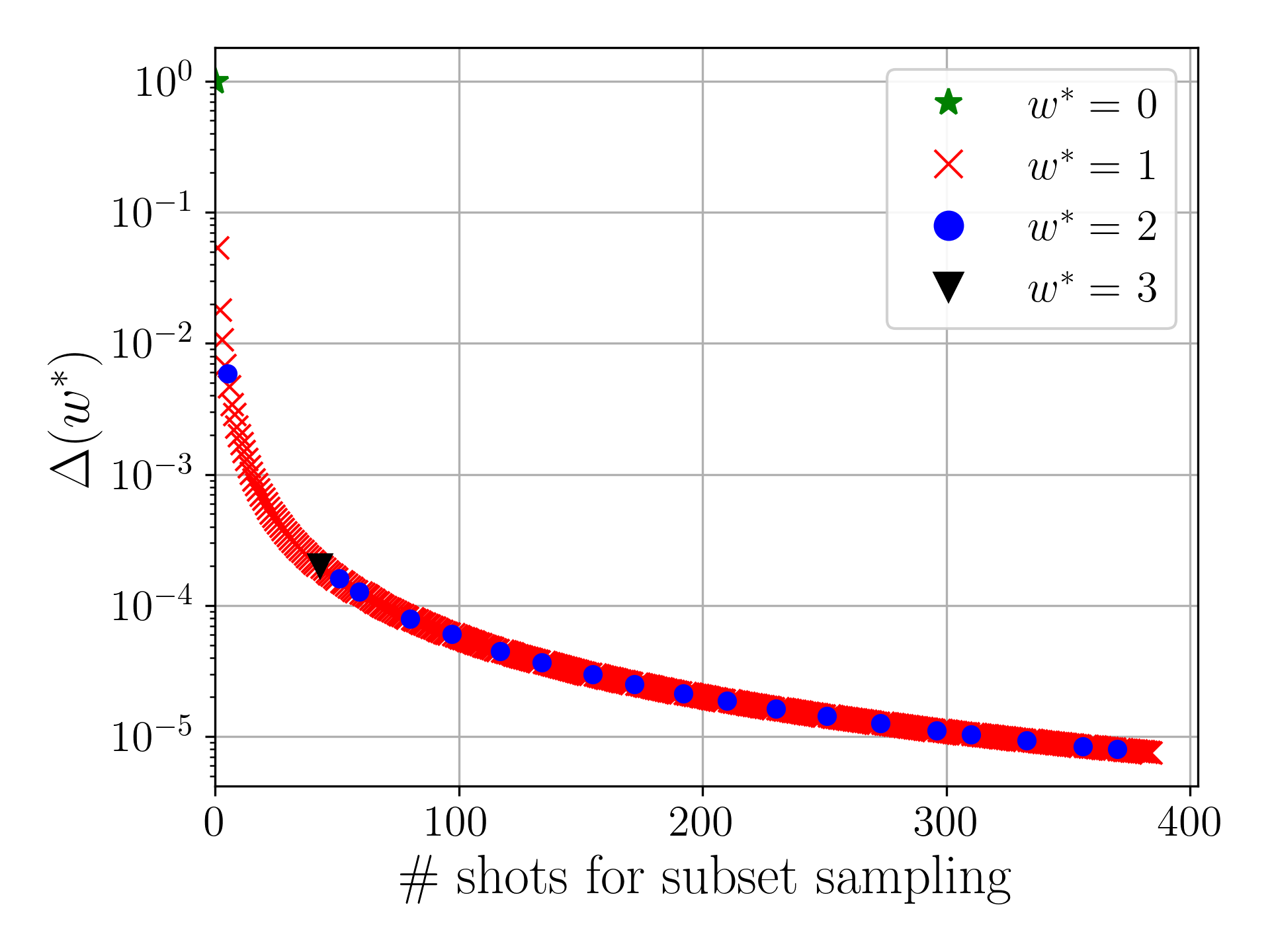}
	\caption{The ERU criterion, at $p_\text{max} = 10^{-2}$, mostly chooses the 1-fault-subset. Various times, the 2-fault subset is sampled. The 3-fault-subset is chosen once. The larger-weight subsets are more relevant as compared to Fig.~\ref{fig:ghz_ERUpershot} because their binomial factors $A_2$ and $A_3$ are larger at $p_\text{max} = 10^{-2}$ than at $p_\text{max} = 10^{-3}$.}
	\label{fig:ghz_ERUpershot_p1e-2}
\end{figure}

This is also reflected in Fig.~\ref{fig:ghz_failpershot}. This advantage of subset sampling over direct MC simulation becomes even larger the lower we choose $p_\text{max}$. For every shot, we choose the most ``relevant'' subset via the ERU criterion. The values of the ERU $\Delta$ for the selected subset $w^*$ for each shot are plotted in Fig.~\ref{fig:ghz_ERUpershot}. Here we observe that the fault-free circuit is chosen exactly once. After the very first shot there is no improvement expected from sampling the fault-free circuit. Then the 1-fault-subset is selected repeatedly. The 2-fault-subset is only sampled once, in the 21st shot, after the sampling uncertainty in the 1-fault-subset is so small that there is more gain to the overall accuracy to reduce the cutoff error by $A_2^\text{GHZ}$. Afterwards, it is again preferential to keep sampling in the 1-fault-subset. When we use the larger $p_\text{max} = 10^{-2}$, we observe in Fig.~\ref{fig:ghz_ERUpershot_p1e-2} that the 2-fault-subset is chosen more frequently since its binomial factor is now larger than at $p_\text{max} = 10^{-3}$ and thus the contribution of the 2-fault-subset to the logical failure rate and the total estimation uncertainty $\eta$ is deemed more relevant.

\subsection{Deterministic FT state preparation}\label{sec:examples_det0}

\begin{figure*}\centering\includegraphics[width=0.8\linewidth]{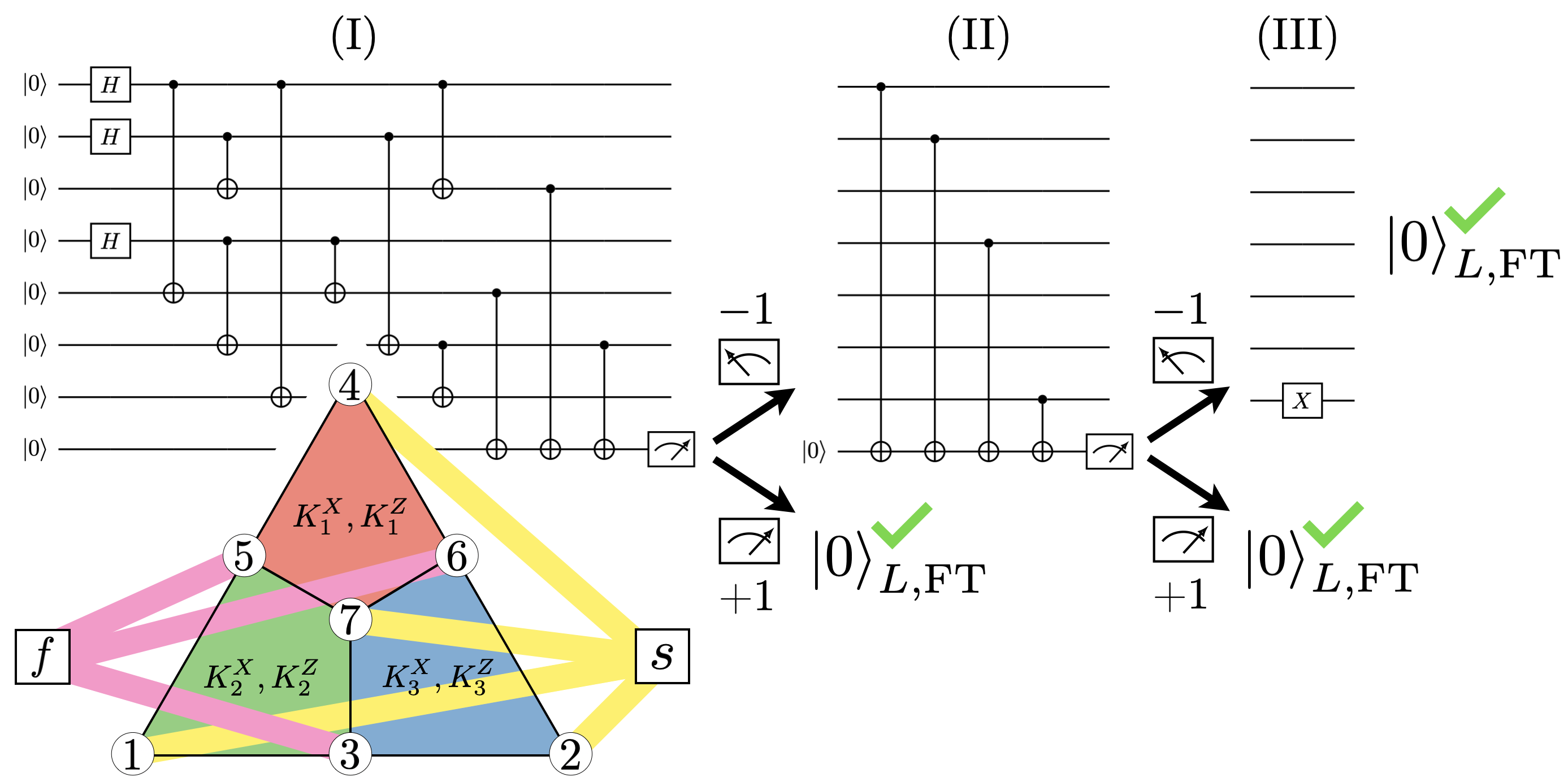}
	\caption{Possible circuit sequences for deterministic FT state preparation of $\ket{0}_L$ in the Steane code (triangle). If the flag (pink connections to $f$) is triggered in step (I), the stabilizer $K_1^ZK_2^ZK_3^Z = Z_1Z_2Z_4Z_7$ is measured (yellow connections to $s$) in step (II). Then, if the next measurement also yields $-1$, the correction $X_7$ is applied as step (III). Otherwise the protocol terminates. At any point of termination, $\ket{0}_L$ is prepared fault-tolerantly.}
	\label{fig:detprep}
\end{figure*}

We now present results on sampling the failure rate of a protocol \cite{heussen2023strategies} to fault-tolerantly and deterministically prepare the $\ket{0}_L$ state in the Steane code. For our stabilizer simulations of the Steane code we use the stabilizer generators
\begin{align}
    K_1^X &= X_4X_5X_6X_7~~~~~K_1^Z = Z_4Z_5Z_6Z_7 \notag\\
    K_2^X &= X_1X_3X_5X_7~~~~~K_2^Z = Z_1Z_3Z_5Z_7 \label{eq:stabs_devitt} \\
    K_3^X &= X_2X_3X_6X_7~~~~~K_3^Z = Z_2Z_3Z_6Z_7, \notag
\end{align}
which are illustrated in Fig.~\ref{fig:detprep}, and the according static look up table decoder. The $\ket{0}_L$ state is prepared fault-tolerantly in one, two or three steps: first, prepare the state non-fault-tolerantly and measure the operator $Z_3Z_5Z_6$ with a single flag qubit. If the flag qubit is measured as $+1$, we measure all data qubits in the $Z$-basis and the protocol terminates. Otherwise, the stabilizer $K_1^ZK_2^ZK_3^Z = Z_1Z_2Z_4Z_7$ is measured with a second auxiliary qubit. The correction $X_7$ is applied if both auxiliary qubits are measured as $-1$. Then, the state preparation protocol terminates. We then measure all data qubits in the $Z$-basis to determine whether a logical failure has incurred. 

Provided that at most a weight-1 fault happens during the overall protocol, the $\ket{0}_L$ state is prepared correctly up to a correctable weight-1 error. The result is a logical failure if the minimum Hamming distance of the measured bitstring $m$ to the bitstrings that label all eight basis states that make up the state
\begin{align}
     \ket{0}_L = \frac{1}{\sqrt{8}}(\ket{0000000} &+ \ket{1010101} + \ket{0110011} \notag \\ 
     + \ket{1100110} &+ \ket{0001111} + \ket{1011010} \notag \\ 
     + \ket{0111100} &+ \ket{1101001})
\end{align}
is larger than one. This is possible because all weight-2 $Z$-errors are of distance 1 to a logical $Z$-operator in the Steane code. 

For an FT protocol, at least two faults are necessary to cause failure and these faults can happen in different circuits. We must ensure that our knowledge about subsets that can never lead to failure is accounted for correctly in the calculation of branching uncertainties $\sigma_i$ and the cutoff error $\delta$. In particular, we must remove all paths of total weight $w_\text{tot} \leq 1$ from the cutoff error. This can be done by assuming a maximum sequence length $L$ for such paths (see App.~\ref{sec:cutoff_app} for details). Then, any FT path has a total path product smaller then $LA_1(1-M_0)$ where $M_0 = \text{min}_c~A^c_0$ is the minimal binomial factor at $\pphy = p_\text{max}$ out of all circuit's binomial factors and $A_1$ is the binomial weight of the circuit where a single fault was realized.

To illustrate dynamical subset sampling for a multi-parameter noise model, in the following we estimate failure rates for a two-parameter noise model with $\vec{p} = (p_1,\,p_2)$ where $p_1$ is the physical single-qubit gate fault rate and $p_2$ is the physical two-qubit gate fault rate. The binomial factor of Eq.~(\ref{eq:binom}) is extended by the now two-dimensional vector $\vec{w} = (w_1,\,w_2)$ labelling the fault-weight-subsets for single-qubit and two-qubit faults to read
\begin{align}
	A^{c}_{\vec{w}}(p_1,p_2) = &\binom{N_c^{(1)}}{w_1}\binom{N_c^{(2)}}{w_2}p_1^{w_1}p_2^{w_2} \times \notag \\
	&(1-p_1)^{N_c^{(1)}-w_1}(1-p_2)^{N_c^{(2)}-w_2} \label{eq:binom2}
\end{align}
where $N_c^{(1)}$ and $N_c^{(2)}$ are the number of single- and two-qubit gates in the circuit $c$ respectively. We employ dynamical subset sampling at $\vec{p}_\text{max} = (10^{-3},\,10^{-2})$ and choose fault-weight-subsets randomly according to the distribution of binomial factors as described in Sec.~\ref{sec:dss}.

\begin{figure}\centering\includegraphics[width=\linewidth]{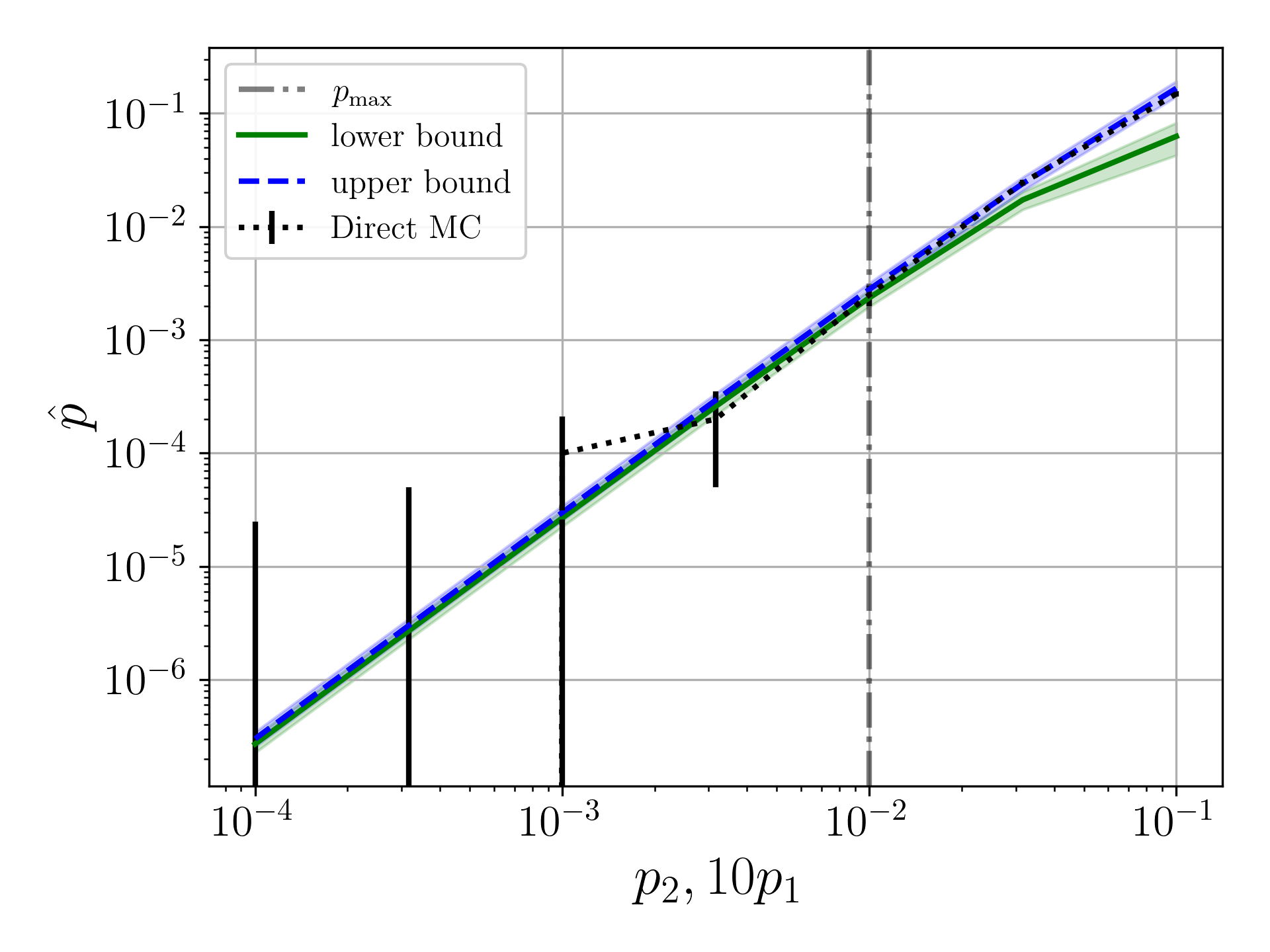}
	\caption{For the three-step $\ket{0}_L$ state preparation protocol, direct MC achieves a reasonable estimation for $p_2 > 10^{-2}$ (and $p_1 > 10^{-3}$). The direct MC estimator lies in between the upper and lower bounds estimated via subset sampling, which diverge for growing $\vec{p}$. At low $p_2$ (and $p_1$) however, direct MC records few or no failures after $N_\text{MC} = 10^4$ shots but the analytical scaling of subset sampling yields tight uncertainty intervals around the estimated failure rate bounds.}
	\label{fig:det0_scaling}
\end{figure}

In Fig.~\ref{fig:det0_scaling} we observe that the bounds estimated via dynamical subset sampling are tight in the relevant range for the two parameters $p_1 < p_{1,\text{max}}$ and $p_2 < p_{2,\text{max}}$ that are scaled together uniformly. Due to the quadratic scaling of the failure rate estimator $\hat{p} \sim p_i^2$ of the FT protocol, the uncertainty intervals on the direct MC estimators after a fixed number of shots at each $\vec{p}$-value are even larger than for the GHZ state preparation example. At low $p_1, p_2$, very few or even no logical failures are registered anymore by direct MC.

\begin{figure}\centering\includegraphics[width=\linewidth]{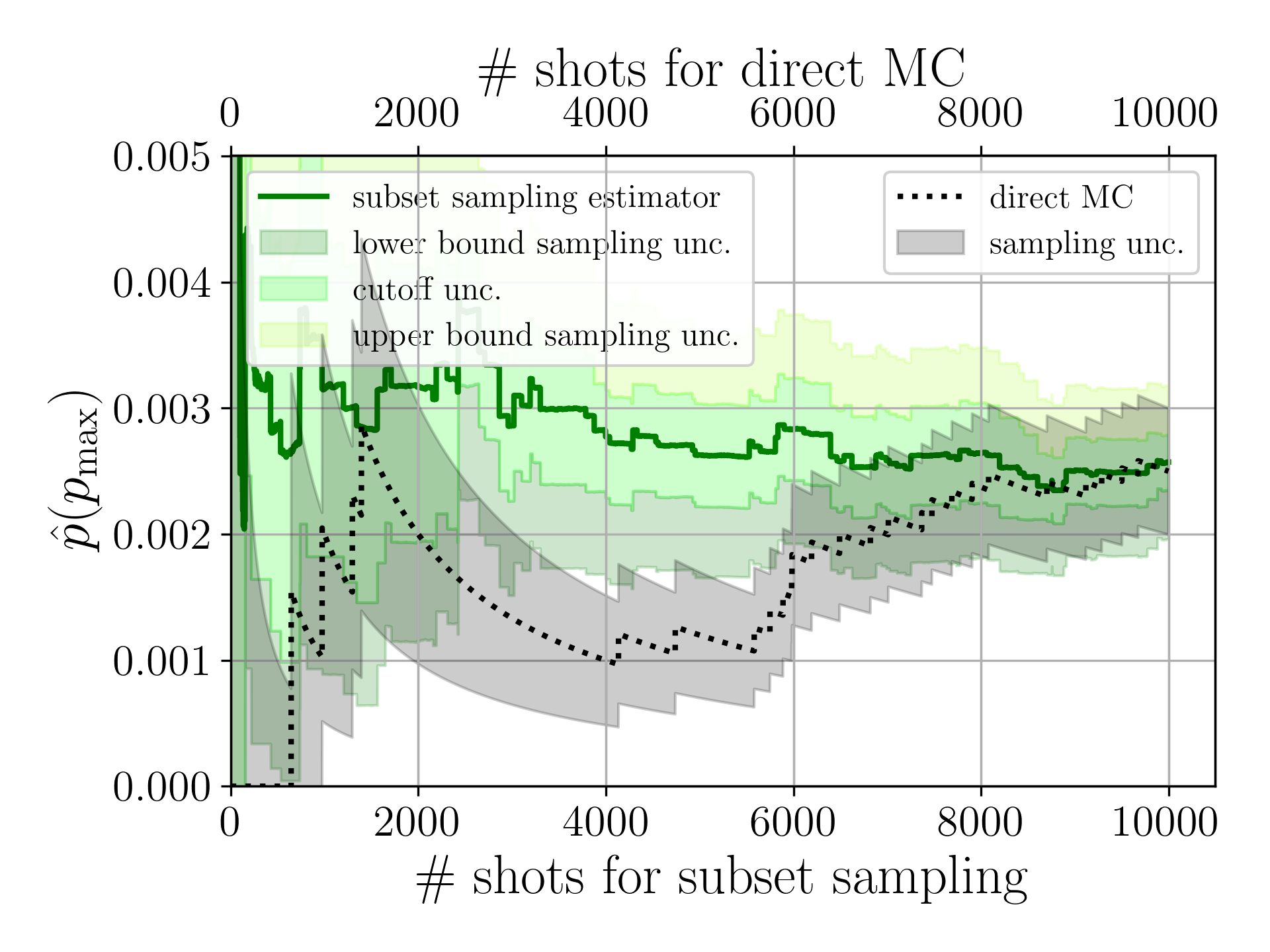}
	\caption{The subset sampling estimator to the logical failure rate $\hat{p}$ has an uncertainty interval $\eta$ (shades of green, solid) composed of the upper and lower bound sampling uncertainties $\sigma$ and the cutoff uncertainty $\delta$. After 10000 shots at $\vec{p}_\text{max} = (10^{-3},\,10^{-2})$, uncertainties are similar for both the direct MC estimation (black dotted) and dynamical subset sampling. For this example, we drew the fault weights randomly according to the circuits' binomial factors.}
	\label{fig:det0_failpershot}
\end{figure}

Although the total uncertainty interval $\eta$ of dynamical subset sampling is comparable to the direct MC interval after $10^4$ shots, as we can see in Fig.~\ref{fig:det0_failpershot}, we stress that many more MC shots would be required to use direct MC at lower values than $\vec{p}_\text{max}$ while the DSS estimator can be extracted analytically without taking new shots.

\subsection{Flag-FT stabilizer measurements}\label{sec:examples_flag}

\begin{figure}\centering\includegraphics[width=0.75\linewidth]{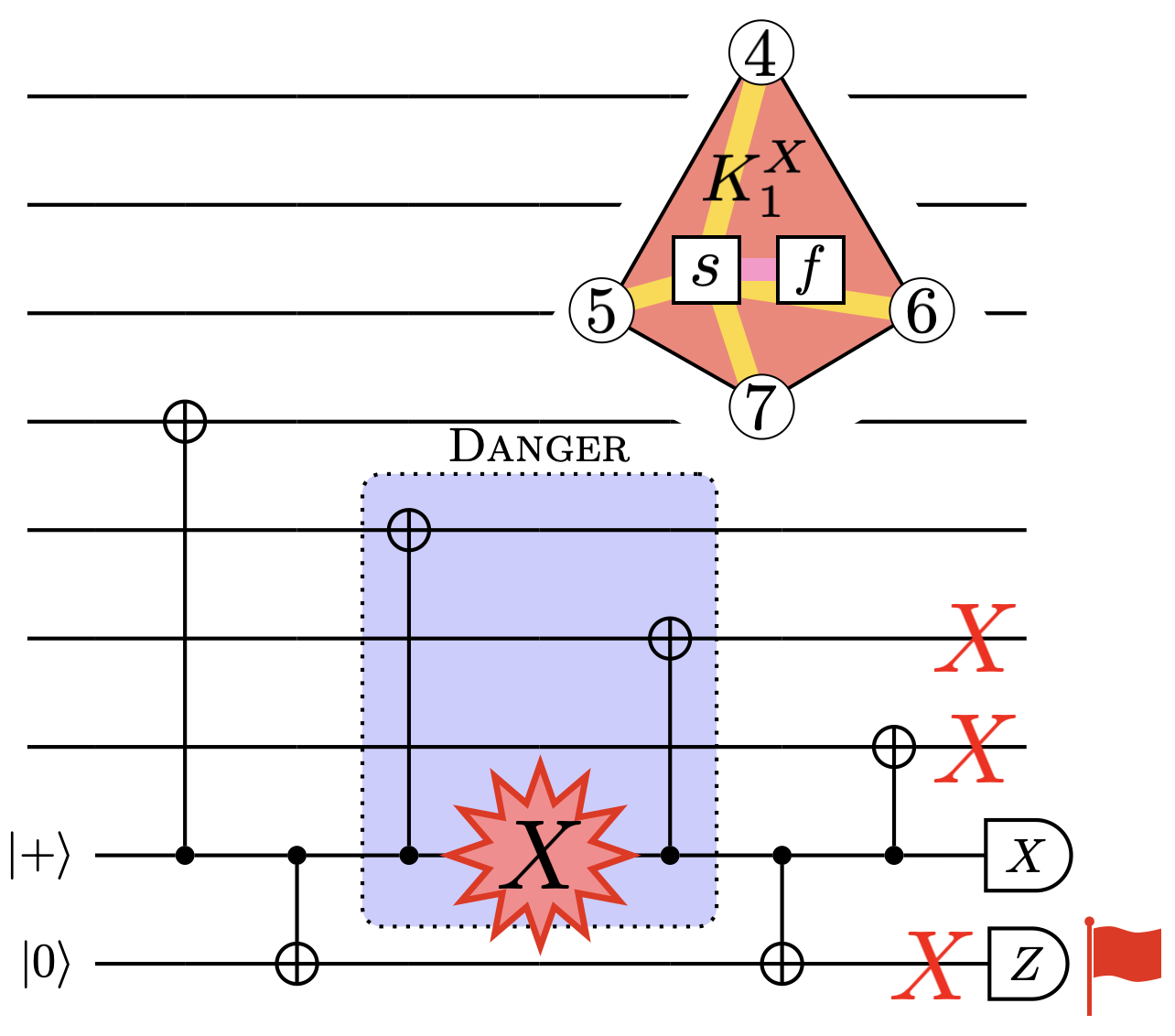}
	\caption{The stabilizer plaquette $K_1^X = X_4X_5X_6X_7$ of the Steane code can be readout fault-tolerantly using a measurement auxiliary qubit and a flag qubit. Dangerous faults may happen on the two highlighted middle CNOT gates; an example $X$-fault is marked, which propagates to trigger the flag for causing a weight-2 Pauli error on the data qubits. In the subsequent non-fault-tolerant stabilizer readout this weight-2 error results in the syndrome $\{+1,-1,+1\}$ and according to case 3(a) of the protocol the correction $X_6X_7$ is (correctly) applied. \textit{Inset: Visualization of the qubits that belong to the red plaquette $K_1$ of the Steane code (according to Eqs.~(\ref{eq:stabs_devitt})) and the syndrome readout qubit $s$, which is connected to the four data qubits (yellow) and the flag qubit $f$, which is only connected to $s$ (pink).}}
	\label{fig:dss_ft0_kx1}
\end{figure}

Another way to initialize a logical qubit into the Steane code's $\ket{0}_L$ state is performing sequential stabilizer measurements utilizing flag fault-tolerant circuits as introduced in Ref.~\citep{chamberland2018flag} and recently demonstrated experimentally \citep{hilder2022fault, ryan2021realization}. Here the data qubit state is projected from $\ket{0}^{\otimes 7}$ onto the simultaneous +1-eigenstate of the $X$-stabilizers. The flag protocol achieves fault-tolerance towards 1-faults (FT-1) through extending the Steane EC look up table decoder by additional correction operators that may be applied depending on the measurement result of flag qubits. An example circuit to fault-tolerantly read out the $K_1^X$ stabilizer is shown in Fig.~\ref{fig:dss_ft0_kx1}. A non-trivial $-1$ measurement outcome for this stabilizer only (while $K_2^X$ and $K_3^X$ yield trivial $+1$ measurements) would lead to a $Z_4$ Pauli correction (compare inset) that fixes the $K_1^X$-eigenspace. Note that the $Z$-stabilizers are already fulfilled by initializing all data qubits in the $\ket{0}$ state. The $X$-stabilizer measurements cannot cause uncorrectable weight-2 $X$-errors without triggering a flag as illustrated in Fig.~\ref{fig:dss_ft0_kx1}. 

The only possible dangerous errors in this scheme follow from faults on the bare readout auxiliary qubits. There, only Pauli-$X$ faults can spread to the data qubits through the subsequent CNOT gates used to measure the $X$-stabilizers. Dangerous faults on the auxiliary qubits that may propagate to weight-2 errors on the data qubits (and thus break FT-1) are accounted for by repeating syndrome measurements on the seven data qubits sequentially using the flag circuits for $K_1^X,\,K_2^X$ and $K_3^X$ until one of the following terminating cases are fulfilled:

\begin{enumerate}
    \item In two consecutive rounds $a$ and $b$, the same syndrome is observed and no flag was triggered. In this case we apply the $Z$-Pauli correction according to the Steane look up table.
    \item In two consecutive rounds $a$ and $b$, two different syndromes are observed but no flag was triggered. In this case we run a third round of non-fault-tolerant syndrome readout and apply the $Z$-Pauli correction according to the Steane look up table using the syndrome that is obtained in the third round of measurement.
    \item At any point a non-trivial flag measurement is observed. In this case we abort the flag circuit readout procedure and continue with a full non-fault-tolerant syndrome readout:
    \begin{enumerate}
        \item If there exists an $X$-error that is caused by one fault that triggers a flag and agrees with the observed syndrome we apply the $X$-error a second time to correct it. (The set of all errors caused by one fault that triggers the flag is called the flag error set, see example in Fig.~\ref{fig:dss_ft0_kx1}.)
        \item If no such $X$-error exists we apply the $Z$-Pauli correction according to the Steane look up table.
    \end{enumerate}
\end{enumerate}

\begin{figure*}\centering\includegraphics[width=0.99\linewidth]{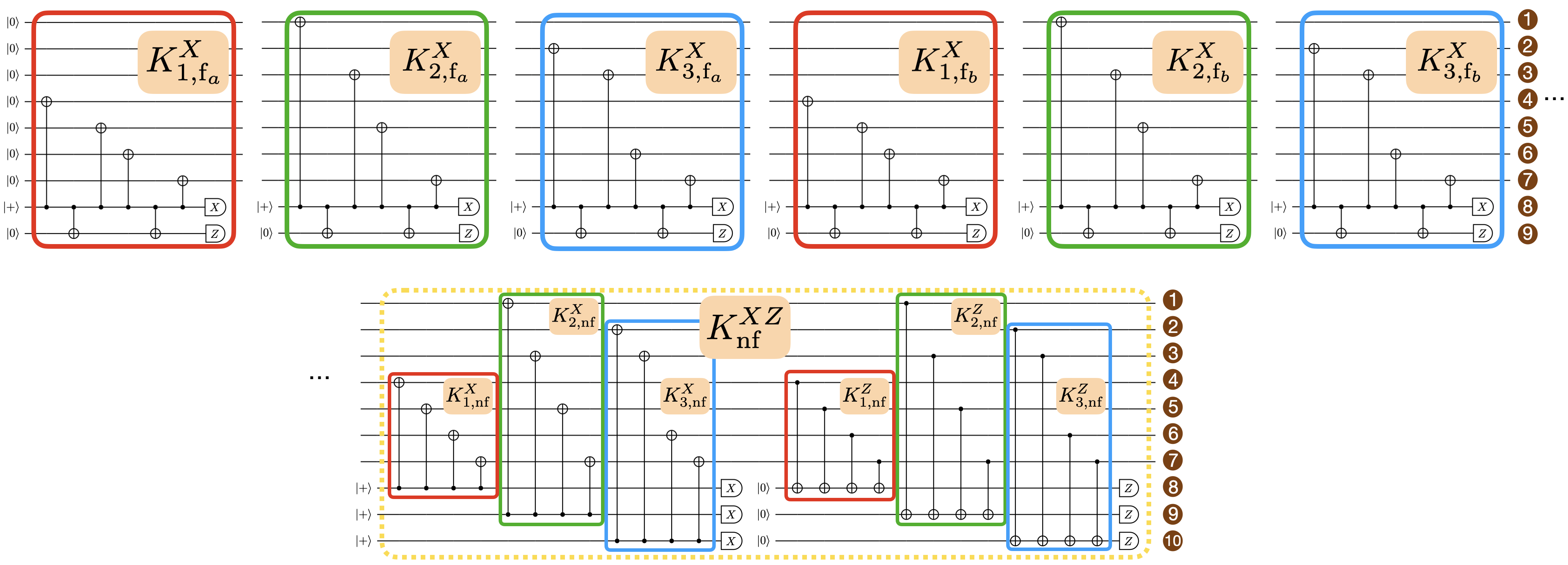}
	\caption{Longest possible circuit sequence that may occur when using the flag fault-tolerant protocol to encode the $\ket{0}_L$ state. RGB-colored boxes mark the respective Steane code plaquette $K_i^\sigma$ readout with a flag circuit ($._\text{f}$) or a non-flag circuit ($._\text{nf}$). Whenever the flag qubit (qubit number 9, protocol case 3) is measured as $-1$ or whenever two consecutive flagged measurements of the same stabilizers $K^\sigma_{i,\text{f}_a}$ and $K^\sigma_{i,\text{f}_b}$ disagree (qubit number 8, protocol case 2), the circuit run is interrupted and continues at the yellow boxed circuit. The non-fault-tolerant syndrome measurement (dashed yellow box ``$K_\text{nf}^{XZ}$'') in the end is only omitted if all six previous stabilizer readouts agree and no flag has been triggered (protocol case 1).}
	\label{fig:dss_ft0_longest}
\end{figure*}

\begin{figure}\centering\includegraphics[width=\linewidth]{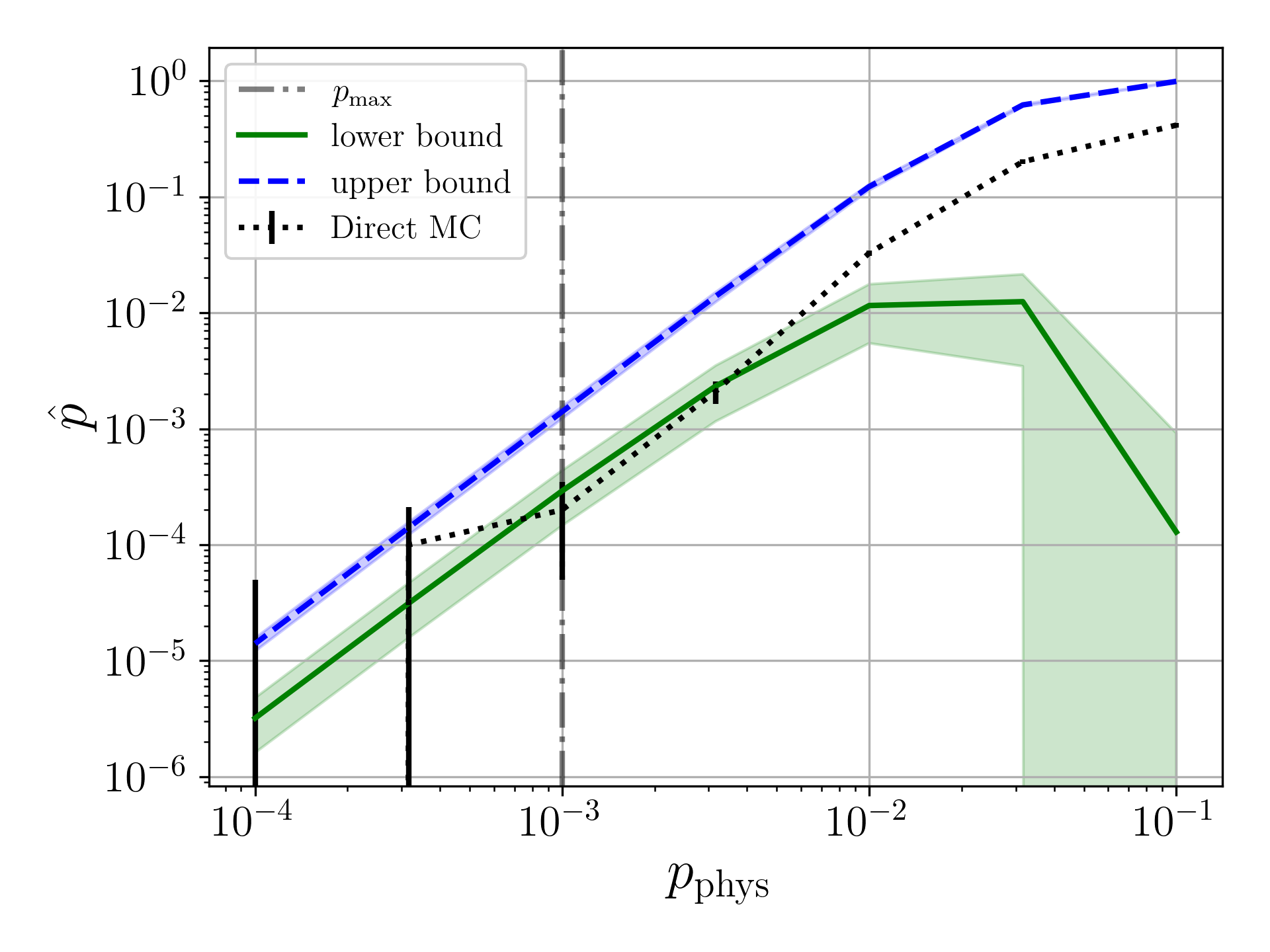}
	\caption{Logical failure rate estimation for the state preparation protocol via flag-FT stabilizer measurements. Although a relatively large cutoff error remains at $p_\text{max} = 10^{-3}$, the total uncertainty on $\hat{p}$ stays bound as $\pphy \rightarrow 0$. The direct MC estimators are practically useless in this regime due to too few shots or -- equivalently -- logical failure rates being too low for the limited number of $N_\text{MC} = 10^4$ shots. The cutoff error can be reduced by increasing $p_\text{max}$ as in Fig.~\ref{fig:flag_scaling_2}.}
	\label{fig:flag_scaling}
\end{figure}

The longest possible circuit sequence that may result from applying this protocol is drawn in Fig.~\ref{fig:dss_ft0_longest}. Actual instances of the noisy protocol can be much shorter: for instance, when the first circuit flags (``$K^X_{1,\text{f}_\text{a}}$''), we will directly continue to run the non-FT syndrome extraction circuit (``$K^{XZ}_\text{nf}$'') and the protocol terminates. The branching events in this protocol are determined by the measurement outcomes of the current flag circuit: Either the next flag circuit is run or one jumps ahead to the non-FT syndrome readout (``$K^{XZ}_\text{nf}$'') The variety of branching options makes dynamical subset sampling suitable to effectively explore subsets that contribute most significantly to the overall failure rate. Low-weight subsets are sampled preferentially by choosing the fault weights -- again -- randomly according to the distribution of their binomial factors.

Figure \ref{fig:flag_scaling} shows that the direct MC estimator lies in between the DSS bounds, which diverge for large $\pphy$. Since the failure rate scales quadratically $\hat{p} \sim \pphy^2$ as $\pphy \rightarrow 0$, direct MC cannot produce reliable estimations at low $\pphy$ with $N_\text{MC} = 10^4$ shots. Dynamical subset sampling is used at $p_\text{max} = 10^{-3}$ and yields upper and lower bounds on the true failure rate of the flag-FT protocol that do not diverge as $\pphy \rightarrow 0$ with the same number of shots. The bound is not as tight as for the previous protocols. This indicates that paths, which are not realized by employing the binomial factor criterion in the sampling procedure and are thus contained in the cutoff error $\delta$, have a relatively large product of binomial factors, given the chosen $\pphy$. Additionally, we assume a maximum FT path length $L = 8$ (6 flag circuits + 1 syndrome readout circuit + 1 circuit that contains only noisy measurements of all data qubits) in our simulations, which can lead to an overestimation of the upper bound (see App.~\ref{sec:cutoff_app} for details). As a result of fixing $L = 8$, we might overestimate the relevance of long circuit sequences to the cutoff error $\delta$ if, at a given $p_\text{max}$, the typical sequence length is much shorter. Both aspects are not fundamental problems of dynamical subset sampling but can be improved in a future version of \texttt{qsample} (and we show a tighter bound obtained from sampling at larger $\pphy$ in App.~\ref{sec:app_flagstab2}). \section{Remarks on efficiency and run time}\label{sec:analysis}

We consider dynamical subset sampling efficient and thus preferential over direct Monte Carlo simulation when, analogous to Eq.~\eqref{eq:lowp}, the paths that never lead to failure have the largest weight at a given $p_\text{max}$ and/or these paths can be excluded by a suitable choice criterion. 

Calculation of the total estimation uncertainty $\eta$ from a large event tree might take a relatively long time compared to running a single shot of a protocol. Parallelization of the sampling procedure for dynamical subset sampling can be employed two-fold: Firstly, during the numerical evaluation of the expected reduction of uncertainty $\Delta(w_i)$ at any stage $i$ within the protocol, one may continue to run further DSS shots, using the subset choices from a previous ERU evaluation or the binomial factor criterion, in order to keep lowering the sampling uncertainty, while waiting for the ERU criterion to testify. Secondly, the evaluations of $\Delta(w_i)$ at stage $i$ can themselves be parallelized for each of the $w_{i,\text{max}}+1$ subsets under consideration and also for both the assumed positive or negative measurement outcome to be probed. Anyways, for practical use of DSS, any $w_{i,\text{max}}$ should be relatively small.

\section{Conclusions \& outlook}\label{sec:conclusion}

In this work we have presented a novel importance sampling technique to efficiently estimate failure rates with well-defined error bars for QEC protocols with in-sequence measurements that make the realized circuit sequence non-deterministic while suffering from incoherent noise at low physical fault rates. Our dynamical subset sampling technique allows for a significant reduction in the number of samples needed to achieve a given accuracy on the estimated failure rate compared to direct Monte Carlo sampling. It can take dependencies between circuits that typically occur in QEC protocols, i.e.~circuit execution conditioned on a measurement result, into account. The examples provided allow insight into useful modes of operation for dynamical subset sampling. The first is to avoid the need for post-selecting on a specific measurement result. All circuit sequences resulting from any measurement result can be accounted for. The second is to avoid sampling fault-free circuits which does not provide useful information but takes up most of the computational resources in direct Monte Carlo simulation at low physical fault rates. In order to choose which fault-weight-subset we should sample, we introduce two criteria. 

While they proved effective for the examples provided in this work, it is worth exploring whether sampling efficiency could be increased even further, e.g.~by introducing a more elaborate criterion. For instance, it would be interesting to explore the possibility to let an artificial neural network choose fault-weight-subsets with a cost function that punishes sampling fault-free or fault-tolerant paths but rewards sampling low-weight paths that can in fact cause failure of a protocol. On a more speculative note, one may be able to employ non-Markovian sampling algorithms that are specifically designed to sample in a non-uniform energy landscape. In a manner similar to simulated annealing \cite{kirkpatrick1983optimization}, fault-weight-subsets could be chosen such that deviating just little from the fault-free or fault-tolerant paths is preferential over large fluctuations.

The need for efficient simulation techniques of protocols with intermediate branching options becomes evident in the light of progressing fault-tolerant quantum hardware realizations~\citep{postler2021demonstration, ryan2021realization, krinner2021realizing}. Further study on dynamical subset sampling should include protocols that can have multiple fault-free paths instead of just one. Furthermore, the bound used for the FT cutoff error can be improved by, instead of using the maximum FT sequence length $L$, inferring the longest possible ``rest of sequence'' $l \leq L$ that can still be run at any given point of an individual sample. Also, we conjecture that a generalization of the FT cutoff error to QEC codes of arbitrary distance is possible by straightforward combinatorics.

\section*{Code availability}
All codes used for data analysis are available from the corresponding author upon reasonable request.

The python package \texttt{qsample}, which was used to generate all data in this manuscript, is publicly available at \url{https://github.com/dpwinter/qsample}. Within the documentation, also accessible via \url{https://dpwinter.github.io/qsample}, we provide a tutorial on how to use \texttt{qsample} for modelling quantum protocols and determine failure rates by means of dynamical subset sampling and direct Monte Carlo simulation. The protocols for the examples of Sec.~\ref{sec:examples} are also given therein.
\\

\section*{Author contributions}
S.H. developed the DSS scheme, performed the numerical simulations and analyzed the data. D.W. authored the code for \texttt{qsample} under supervision of S.H. and all authors contributed to theory modelling and writing of the manuscript. M.M. supervised the project.

\section*{Acknowledgements}
S.H. would like to thank Friederike Butt, Lorenzo Cardarelli, Eliana Fiorelli, David Locher, Josias Old, Ciarán Ryan-Anderson and Seyong Kim for fruitful discussions that led to the development of DSS and Thomas Botzung for also providing useful feedback on the manuscript. This work was funded by the Deutsche Forschungsgemeinschaft (DFG, German Research Foundation) under Germany's Excellence Strategy – Cluster of Excellence Matter and Light for Quantum Computing (ML4Q) EXC 2004/1 – 390534769. We gratefully acknowledge support by the EU Quantum Technology Flagship grant under Grant Agreement No.~820495 (AQTION), the U.S. Army Research Office through Grant No.~W911NF-21-1-0007 and the ERC Starting Grant QNets through Grant No.~804247. The U.S. Government is authorized to reproduce and distribute reprints for governmental purposes notwithstanding any copyright annotation thereon. Any opinions, findings, and conclusions or recommendations expressed in this material are those of the author(s) and do not necessarily reflect the view of the U.S. Army Research Office. This research is also part of the Munich Quantum Valley (K-8), which is supported by the Bavarian state government with funds from the Hightech Agenda Bayern Plus. Furthermore, we receive funding from the European Union’s Horizon Europe research and innovation programme under grant agreement No.~101114305 (“MILLENION-SGA1” EU Project). We gratefully acknowledge computing time provided to us at the NHR Center NHR4CES at RWTH Aachen University (Project No.~p0020074). This is funded by the Federal Ministry of Education and Research and the state governments participating on the basis of the resolutions of the GWK for national high performance computing at universities.

\appendix
\onecolumngrid

\section{Bias of the protocol estimator}\label{sec:bias}

Subset sampling yields unbiased estimators of protocol failure rates, as we state for Eq.~\eqref{eq:dssestim}, in the following sense. Consider as the bias the difference between the expectation value obtained by an estimator $\hat{p}$ and the true value $p^*$ of the failure rate. The bias of $\hat{p}$ to $p^*$ \cite{statproofbook} is
\begin{align}
    \text{Bias}(\hat{p}, p^*) = \E_\text{Tree}(\hat{p}) - p^* = \E_\text{Tree}\left(\frac{p_L + p_U}{2}\right)- p^* = \E_\text{Tree}\left(p_L + \frac{\delta}{2} \right)- p^*
\end{align}
where $\E_\text{Tree}(.)$ denotes the expectation value deduced from an event tree created by running a finite number of samples. For a single circuit $c$ and a single noise parameter $\pphy$, we can explicitly plug in Eqs.~\eqref{eq:subset_pL} and \eqref{eq:ss_cutoff} to express the bias as
\begin{align}
    \text{Bias}(\hat{p}, p^*) = \sum_{w=1}^{w_\text{max}} A^c_{w}(\pphy) \hat{p}_\text{fail}^{(w)} + \frac{1}{2} \left( 1 - \sum_{w=0}^{w_\text{max}} A^c_{w}(\pphy) \right) - p^* \rightarrow 0.
\end{align}
In the limit of large numbers of shots in every subset, convergence of the subset failure rate estimators $\hat{p}_\text{fail}^{(w)}$ to their true values $p_\text{fail}^{*\,(w)}$ is assured by the central limit theorem and the law of large numbers. The cutoff error contribution $\delta \rightarrow 0$ in the limit of taking all subsets into account $w_\text{max} \rightarrow N_c$, which is equivalent to the limit of $\pphy \rightarrow 0$. Thus, it is clear that our choice $\hat{p} = \frac{p_L + p_U}{2}$ is an unbiased estimator of the true failure rate $p^*$.

\section{Discussion of the failure rate uncertainty interval}\label{sec:errs}

In this appendix, we refer to uncertainty intervals as ``errors'' since there is no danger of confusion with Pauli operators corrupting a QEC code in the following. The lower and upper bound to the logical failure rate estimator and the cutoff error, as defined in Eqs.~\eqref{eq:dsspl}, \eqref{eq:dssdelta} and \eqref{eq:dsspu} in Sec.~\ref{sec:dss}, are given by 
\begin{align}
    p_L &= \sum_{P_\text{fail}} \prod_i A(w_i) q_i \label{eq:pL} \\
    \delta &= 1 - \sum_P \prod_i A(w_i) q_i = \sum_{s=1}^L \sum_{C_s} \left[ \prod_{i}^s A(w_i)q_i \left( 1 - \sum_{w_{C_s} = 0}^{w_\text{max}^{(C_s)}} A(w_{C_s}) \right) \right] \label{eq:delta_both} \\
    p_U &= p_L + \delta = 1 - \sum_{P_\text{no-fail}} \prod_i A(w_i) q_i.
\end{align}

The quantities $p_L$ and $p_U$ can be used to bound the true failure rate $p^*$ of a protocol in the sense that
\begin{align}
    p^* \in \left[ p_L - \sigma_L,\, p_U + \sigma_U \right]
\end{align}
where $\sigma_{L/U}$ is the respective sampling error for the estimators $p_L$ and $p_U$ that results from estimating $p_{L/U}$ from a finite number of shots and the true failure rate $p^*$ likely lies between the upper and lower bound. In this section we derive these confidence intervals $[p_L - \sigma_L, p_U + \sigma_U]$ with 
\begin{align}
    \sigma_{L/U} &= \sqrt{\Var{p_{L/U}}} \label{eq:cidef}
\end{align}
from the individual branching uncertainties $\sigma_i$ of an event tree.

The two equalities in Eq.~\eqref{eq:delta_both} are two equivalent ways to calculate the cutoff error $\delta$ in a subtractive or additive way respectively, which we elaborate further in the subsequent sections. The additive cutoff error (r.h.s.) contains the sum over all stages $s$ of an event tree of length $L$ and all circuits $C_s$ within a stage $s$ with their respective binomial factors $A(w_{C_s})$. The subtractive cutoff error (l.h.s.) is just calculated via the sum over all paths $P$ in the tree. The summation index $i$ refers to the respective path $P$, which the binomial factor $A(w_i)$ is part of.

\subsection{Calculating event tree variances}

In order to calculate the variances in Eq.~\eqref{eq:cidef}, we first show general properties of variances that are useful to calculate expressions of the form of $\Var{p_L}$, which contain both sums and products of random variables. Then, we explicitly calculate illustrative examples of generic event trees.

\subsubsection{General formulas}

\textbf{Sums of random variables.} For a sum of two random variables $X$ and $Y$, each multiplied by constants $a$ and $b$ respectively, we calculate its variance as
\begin{align}
	\Var{aX+bY} &=a^2\,\Var{X}+b^2\,\Var{Y}+2ab\,\Cov{X,Y}\label{eq:varianceof2sum}
\end{align}
where
\begin{align}
    \Cov{X,Y} \equiv \E((X-\E(X))(Y-\E(Y))) \label{eq:covdef}
\end{align}
is the covariance between $X$ and $Y$. Generalizing to a sum of $n$ variables, the total variance is
\begin{align}
    \Var{\sum_i^n a_iX_i} &= \sum_i^n a_i^2\,\Var{X_i} + 2 \sum_{j=1}^n \sum_{i=1}^{j-1} a_i a_j\,\Cov{X_i,X_j}. \label{eq:varlincomb}
\end{align}
A useful observation is that
\begin{align}
    \Cov{X, 1-X} &= \Cov{X, -X} = - \Var{X}. \label{eq:cov_pm}
\end{align}
More generally, a relation that is useful at tree branchings is
\begin{align}
    \Var{aX + b(1-X)} = a^2 \,\Var{X} + b^2 \,\Var{1-X} + 2ab \,\Cov{X, 1-X} = (a-b)^2 \,\Var{X} \leq (a^2+b^2) \,\Var{X}.
\end{align}
From this we can see directly that the variance of two subtrees that branch off a node with transition rates $q = X$ and $1-q = 1-X$ is always overestimated if one were to ignore the covariance term.

\textbf{Products of random variables.} We can also calculate the variance of a product of two independent variables
\begin{align}
    \Var{XY} &= \Var{X}\Var{Y} + \Var{X}\E(Y)^2 + \Var{Y} \E(X)^2,
\end{align}
which generalizes to
\begin{align}
    \Var{\prod_i a_iX_i} &= \prod_i a_i^2 (\Var{X_i} + \E(X_i)^2) - \prod_i a_i^2 \E(X_i)^2 \label{eq:goodman}
\end{align}
as long as the $X_i$ and $X_i^2$ are independent \cite{goodman1960exact, goodman1962variance}. We also refer to Eq.~\eqref{eq:goodman} as Goodman's formula. Calculating a path variance with Goodman's formula requires an exponential number of additions. For a path with $L$ branching ratios, the first product creates $L$ factors of two-termed sums and only the term consisting of the product of all $q_i^2$ is subtracted again,
\begin{align}
    \Var{\prod_{i=1}^L q_i} &= \prod_{i=1}^L \left( \Var{q_i} + q_i^2 \right) -  \prod_{i=1}^L q_i^2, \label{eq:goodmanpath}
\end{align}
so we have $2^L-1$ terms in the sum that we need to calculate for a single path variance. For a tree that contains a number of paths $W$, calculation of the variance will take $W \times (2^L-1)$ additions and $L$ multiplications per summand. 

\textbf{Covariance.} The covariance of two subsequent random variables $q_j$ and $q_k$ of a path in an event tree is identical with zero since the latter variable $q_k$ is conditioned on the former $q_j$. In general, for four dependent random variables $X,Y,U,V$ that follow a normal distribution with expectation values $x,y,u,v$, it holds that
\begin{align}
    \Cov{XY,UV} &= xu \,\Cov{Y,V} + xv \,\Cov{Y,U} + yu \,\Cov{X,V} + yv \,\Cov{X,U} \notag \\ 
    &+ \Cov{X,U}\Cov{Y,V} + \Cov{X,V}\Cov{Y,U}. \label{eq:cov4}
\end{align}
It also follows directly from the definition of the covariance if $X,Y$ and $U,V$ are each independent that
\begin{align}
    \Cov{XY,UV} &= \Cov{X,U}\Cov{Y,V} + yv\Cov{X,U} + xu\Cov{Y,V}. \label{eq:covexp}
\end{align}
Also, we have the special case
\begin{align}
    \Cov{XY,XV} &= \Var{X}\,\Cov{Y,V} + yv\,\Var{X} + x^2\,\Cov{Y,V}. \label{eq:cov2same}
\end{align}
Let us remark that a covariance term can be upper-bounded using the Cauchy-Schwarz inequality as
\begin{align}
    \left| \Cov{X,Y} \right| \leq \sqrt{\Var{X}\,\Var{Y}}. \label{eq:cauchyschwarzcov}
\end{align}

\subsubsection{Examples for event tree covariance contributions}

\begin{figure}[!htbp]
    \centering

    \begin{tikzpicture}
	[
	mycircle/.style={circle, fill=blue!20,minimum size=1cm},
	mysquare/.style={regular polygon,regular polygon sides=4, fill=red!20,minimum size=1cm},
	level distance=1.5cm,
	sibling distance=1cm,
	edge from parent path={(\tikzparentnode) -- (\tikzchildnode)}
	]
	\Tree
	[.\node[style={draw,mycircle}]{$c_j$};
	\edge node[auto=right] {$A_w$};
	[.\node[style={draw,mysquare}]{$w_a$};
	\edge node[left] {$q_j$} node[right,yshift=-0.1cm] {$h_j$};
	[.\node[style={draw,mycircle}]{$c_k$};
	\edge node[left] {$B_w$};
	[.\node[style={draw,mysquare}]{$w_b$};
	\edge node[left] {$q_k$} node[right] {$h_k$};
	[.{\xmark} ]
	\edge node[right] {$n_k = 3$};
	[.{\cmark} ]
	]
	]
	\edge node[left,yshift=-0.2cm] {$n_j = 5$} node[right] {$1-q_j$};
	[.\node[style={draw,mycircle}]{$c_l$};
	\edge node[left] {$C_w$};
	[.\node[style={draw,mysquare}]{$w_c$};
	\edge node[left] {$q_l$} node[right] {$h_l$};
	[.{\xmark} ]
	\edge node[right] {$n_l = 2$};
	[.{\cmark} ]
	]
	]
	]
	]
    \end{tikzpicture}
    
    \caption{Example event tree with two stages and five samples (cf.~Fig.~\ref{fig:introDSS}(c)). Circuit nodes (blue circles) are labelled $c_j,c_k$ and $c_l$. Subset nodes (red squares) are labelled $w_a, w_b$ and $w_c$. The branching ratios $q_j$ and $q_k$ as well as $q_j$ and $q_l$ (and also $q_k$ and $q_l$) are uncorrelated.}
    \label{fig:covtree}
\end{figure}
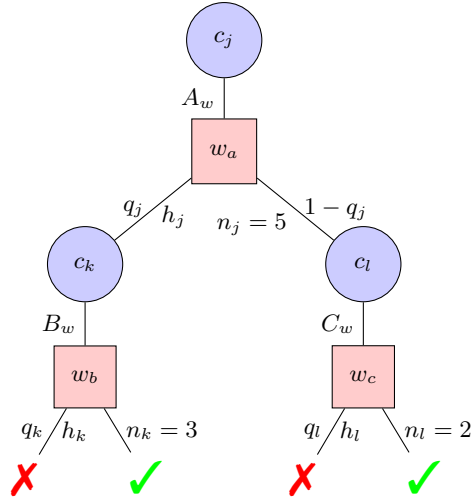

\textbf{Subsequent branching rates are uncorrelated.}  We now demonstrate, by an example, that subsequent branching rates $q_i$ in our event trees are indeed uncorrelated. Assume that the branching with $q_k$ in Fig.~\ref{fig:covtree} happens after, i.e.~further down in the tree and in the same path, as the branching with $q_j$. At the transition labelled with $q_j$ there is a list that keeps the measurement history $h_j$, e.g.~$h_j = [0,0,1,0,1]$ meaning for instance that, for the first shot, the measurement result for the transition rate $q_j$ was $0$ and, for the fifth shot, the measurement result for the transition rate $q_j$ was $1$. In this example, our current estimate of $q_j$ after 5 shots would then be $q_j = 3/5$. Let us assume that we branch off to $c_k$ whenever the measurement result was 0 and to another circuit node $c_l$ when the measurement result was 1. For $q_k$ then, there is another list $h_k$, e.g.~$h_k = [1,0,1]$. It must have the same length as the number of zeros in $h_j$ since only those measurement results lead us to circuit node $c_k$. At circuit node $c_l$ there will be a different history resulting from the measurement results 1 at $c_j$, e.g.~$h_l = [0, 1]$. We leave out the binomial factors for this discussion for simplicity. Now obviously, the sample covariance
\begin{align}
    \Cov{X,Y} = \frac{1}{N} \sum_{i=1}^N (X_i - \E(X))(Y_i - \E(Y))
\end{align}
is zero since we do not plug in $h_j$ and $h_k$ but only the sublist where all elements are zero $h_j^{(0)} = [0,0,0]$. Then
\begin{align}
    \E(h_j^{(0)}) &= 0,~\E(h_j^{(1)}) = 1,~\E(h_k) = \frac{2}{3},~\E(h_l) = \frac{1}{2} \\
    \Cov{h_j^{(0)}, h_k} &= \frac{1}{3} \left( (0 - 0)(1 - \frac{2}{3}) + (0 - 0)(0 - \frac{2}{3}) + (0 - 0)(1 - \frac{2}{3}) \right) = 0\\
    \Cov{h_j^{(1)}, h_l} &= \frac{1}{2} \left( (1 - 1)(0 - \frac{1}{2}) + (1 - 1)(1 - \frac{1}{2}) \right) = 0.
\end{align}

We now consider different types of minimal example trees to illustrate the relevance of non-zero covariance terms. We will see, as stated above, that covariance terms for branchings from subset nodes are negative. Covariance terms of branchings from circuit nodes (into fault-weight-subsets) can in general be non-negative.

\textbf{Subset node branching covariance is negative.} For a sum of products, we in fact calculate variances of correlated variables. This is because branching to the left, e.g.~in Fig.~\ref{fig:covtree}, is weighted with $q_j$ and branching to the right is weighted with $1-q_j$ (cf.~Eq.~\eqref{eq:cov_pm}). So for a generic branching we have terms like
\begin{align}
    \Var{Bq_jq_k + C(1-q_j)q_l} = B^2 \,\Var{q_jq_k} + C^2 \,\Var{(1-q_j)q_l} + 2BC\,\Cov{q_jq_k, (1-q_j)q_l}. \label{eq:var_3branch}
\end{align}

From Eq.~\eqref{eq:covexp} we can see that
\begin{align}
    \Cov{q_jq_k, (1-q_j)q_l} &= -q_kq_l \Var{q_j}. \label{eq:nice_cov}
\end{align}
with $X=q_j, Y=q_k, U=1-q_j, V=q_l$ since $\Cov{q_k, q_l} = 0$.

Denoting $\Var{q_j} = V_j$, we can continue simplifying Eq.~\eqref{eq:var_3branch} like
\begin{align}
    \Var{Bq_jq_k + C(1-q_j)q_l} &= B^2(q_j^2V_k + q_k^2V_j+ V_jV_k) + C^2(q_l^2V_j+(1-q_j)^2V_l + V_jV_l) - 2BCq_kq_lV_j \label{eq:var_3branch_withcov} \\
    &\leq B^2(q_j^2V_k + q_k^2V_j+ V_jV_k) + C^2(q_l^2V_j+(1-q_j)^2V_l + V_jV_l) \label{eq:var_3branch_withoutcov}
\end{align}
to obtain an expression that is symmetric under simultaneous exchange of $q_j \leftrightarrow 1-q_j$ and $q_k,V_k\leftrightarrow q_l,V_l$ if $B=C$ as expected. Since that last term of Eq.~\eqref{eq:var_3branch_withcov} contains only positive factors but a negative sign, we surely increase the value of the whole expression by neglecting it.

Note that we could also use the Cauchy-Schwarz inequality \eqref{eq:cauchyschwarzcov} to obtain a (worse) bound to Eq.~\eqref{eq:var_3branch} like
\begin{align}
    \Var{Bq_jq_k + C(1-q_j)q_l} &= B^2 \,\Var{q_jq_k} + C^2 \,\Var{(1-q_j)q_l} + 2BC\,\Cov{q_jq_k, (1-q_j)q_l} \notag \\
    &\leq B^2 \,\Var{q_jq_k} + C^2 \,\Var{(1-q_j)q_l} + 2BC \sqrt{\Var{q_jq_k}\Var{(1-q_j)q_l}},
\end{align}
which we just leave as a general comment.

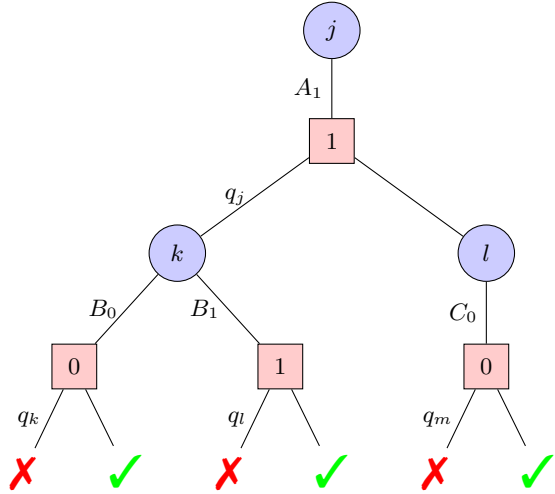
\begin{figure}[!htbp]
    \centering

    \begin{tikzpicture}
    [
        mycircle/.style={circle, fill=blue!20,minimum size=0.75cm},
    	mysquare/.style={regular polygon,regular polygon sides=4, fill=red!20,minimum size=0.5cm},
        level distance=1.5cm,
        sibling distance=0.75cm,
        edge from parent path={(\tikzparentnode) -- (\tikzchildnode)}
    ]
    \Tree
    [.\node[style={draw,mycircle}]{$j$};
        \edge node[auto=right] {$A_1$};
        [.\node[style={draw,mysquare}]{$1$};
            \edge node[left] {$q_j$} node[right] {};
            [.\node[style={draw,mycircle}]{$k$};
                \edge node[left] {$B_0$};
                [.\node[style={draw,mysquare}]{$0$}; 
                    \edge node[left] {$q_k$} node[right] {};
                    [.{\xmark} ]
                    \edge node[right] {};
                    [.{\cmark} ]
                ]
                \edge node[left] {$B_1$};
                [.\node[style={draw,mysquare}]{$1$};
                    \edge node[left] {$q_l$} node[right] {};
                    [.{\xmark} ]
                    \edge node[right] {};
                    [.{\cmark} ]
                ]
            ]
            \edge node[left] {} node[right] {};
            [.\node[style={draw,mycircle}]{$l$};
                \edge node[left] {$C_0$};
                [.\node[style={draw,mysquare}]{$0$}; 
                    \edge node[left] {$q_m$} node[right] {};
                    [.{\xmark} ]
                    \edge node[right] {};
                    [.{\cmark} ]
                ]
            ]
        ]
    ]
    \end{tikzpicture}
    
    \caption{Event tree example with non-zero covariance term between $q_k$ and $q_l$ because they both share the random variable $q_j$ in their path. We reduced the indexing of the nodes for better readability. Circuit nodes are labelled $j,k$ and $l$. Subset nodes are labelled $0$ and $1$ for this example.}
    \label{fig:covtree3}
\end{figure}

\textbf{Circuit node branching covariance can be non-negative.} In Fig.~\ref{fig:covtree}, we only looked at covariances at subset nodes. There is also a contribution from circuit nodes. Consider the tree in Fig.~\ref{fig:covtree3}, where we also included a weight-1 node, which will cause a non-negative Cov-term as we show in the following. The reason is that the branching does not involve two nodes that contain the branching ratios with opposite sign as above but both contributions are positive ($B_0$ and $B_1$ in the tree below). 

The failure rate from Fig.~\ref{fig:covtree3} is 
\begin{align}
    p_L &= A_1 (q_j(B_0q_k + B_1 q_l) + (1-q_j) C_0 q_m)
\end{align}
and the variance
\begin{align}
    \Var{p_L} &= A_1^2 \Var{q_j(B_0q_k+B_1q_l) + (1-q_j)C_0q_m} \notag \\
    &= A_1^2 \left[ \Var{q_j(B_0q_k+B_1q_l)} + \Var{(1-q_j)C_0q_m} + 2\,\Cov{q_j(B_0q_k+B_1q_l), (1-q_j)C_0q_m} \right] \notag  \\
    &= A_1^2 \left[ \Var{q_jB_0q_k}+\Var{q_jB_1q_l}+2\,\Cov{q_jB_0q_k, q_jB_1q_l} + C_0^2\Var{(1-q_j)q_m} \right. \notag \\
    &+ \left. 2\,\Cov{q_j(B_0q_k+B_1q_l), (1-q_j)C_0q_m} \right]\notag \\
    &= A_1^2 \left[ B_0^2\,\Var{q_jq_k}+B_1^2\,\Var{q_jq_l}+2B_0B_1\,\Cov{q_jq_k, q_jq_l} + C_0^2\Var{(1-q_j)q_m} \right. \notag \\ 
    &+ \left. 2\,\Cov{q_j(B_0q_k+B_1q_l), (1-q_j)C_0q_m} \right]
\end{align}
contains the negative Cov-term of the form shown before but there is also the term
\begin{align}
    \Cov{q_jq_k, q_jq_l} &= q_kq_l\,\Var{q_j}
\end{align}
for which we used Eq.~\eqref{eq:cov2same}. It clearly is a non-negative contribution that must be taken into account. We could use the Cauchy-Schwarz inequality \eqref{eq:cauchyschwarzcov} to estimate
\begin{align}
    B_0^2\,\Var{q_jq_k}+B_1^2\,\Var{q_jq_l}+2B_0B_1\,\Cov{q_jq_k, q_jq_l} \leq \left( B_0 \sqrt{\Var{q_jq_k}} + B_1\sqrt{\Var{q_jq_l}} \right)^2.
\end{align}
However, the appearance of such terms can be prevented by algebraic manipulation of the variance terms so that we never need to explicitly take them into account, e.g.~by recursively calculating subtree variances starting at the tree's leaves.

\subsection{Lower bound uncertainty}
It is straightforward to calculate the standard error $\sigma_L$ on the lower bound $p_L$ from Eq.~\eqref{eq:pL} using the general variance of linear combinations of random variables in Eq.~\eqref{eq:varlincomb} and Goodman's formula for the variance of products in Eq.~\eqref{eq:goodman}. An iterative algorithmic procedure to calculate $\Var{p_L}$ will be given in Sec.~\ref{sec:qsample}. 

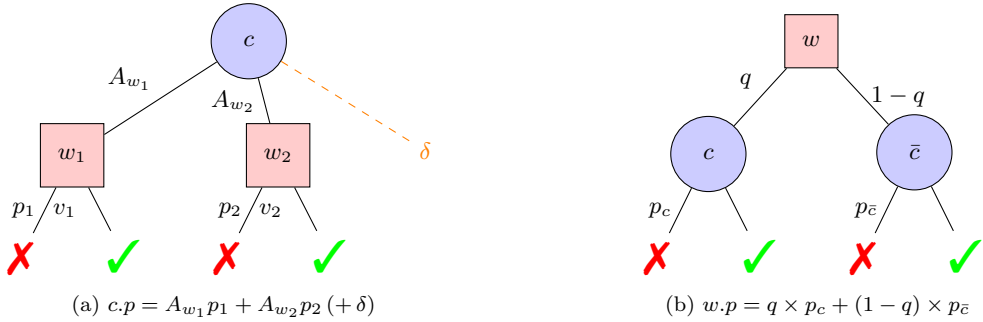
\begin{figure}[!htbp]

\subfloat[$c.p = A_{w_1}p_1 + A_{w_2} p_2\,(+\,\delta)$\label{fig:}]{\begin{tikzpicture}
    [
        mycircle/.style={circle, fill=blue!20,minimum size=1cm},
    	mysquare/.style={regular polygon,regular polygon sides=4, fill=red!20,minimum size=1cm},
        level distance=1.5cm,
        sibling distance=0.75cm,
        edge from parent path={(\tikzparentnode) -- (\tikzchildnode)}
    ]
    \Tree
    [.\node[style={draw,mycircle}]{$c$};
        \edge node[auto=right] {$A_{w_1}$};
        [.\node[style={draw,mysquare}]{$w_1$};
            \edge node[left] {$p_1$} node[right] {$v_1$};
            [.{\xmark} ] 
            \edge node[right] {};
            [.{\cmark} ]
        ]
        \edge node[auto=right,yshift=0.25cm] {$A_{w_2}$};
        [.\node[style={draw,mysquare}]{$w_2$};
            \edge node[left] {$p_2$} node[right] {$v_2$};
            [.{\xmark} ] 
            \edge node[right] {};
            [.{\cmark} ]
        ]
        \edge[draw=orange,dashed] node[auto=right, text=orange] {};
        [.\node[text=orange]{$\delta$}; ]
    ]
    \end{tikzpicture}
}
\hspace{2.5cm}\subfloat[$w.p = q \times p_c + (1-q) \times p_{\bar{c}}$\label{fig:}]{\begin{tikzpicture}
    [
        mycircle/.style={circle, fill=blue!20,minimum size=1cm},
    	mysquare/.style={regular polygon,regular polygon sides=4, fill=red!20,minimum size=1cm},
        level distance=1.5cm,
        sibling distance=0.75cm,
        edge from parent path={(\tikzparentnode) -- (\tikzchildnode)}
    ]
    \Tree
    [.\node[style={draw,mysquare}]{$w$};
        \edge node[auto=right] {$q$};
        [.\node[style={draw,mycircle}]{$c$};
            \edge node[left] {$p_{c}$} node[right] {};
            [.{\xmark} ] \edge node[left] {} node[right] {};
            [.{\cmark} ]
        ]
        \edge node[left] {} node[right] {$1-q$};
        [.\node[style={draw,mycircle}]{$\bar{c}$};
            \edge node[left] {$p_{\bar{c}}$} node[right] {};
            [.{\xmark} ] \edge node[left] {} node[right] {};
            [.{\cmark} ]
        ]
    ]
    \end{tikzpicture}
}
\caption{Any event tree can be constructed recursively by branchings from (a) circuit nodes $c$ to subset nodes $w_1, w_2, ...$ and (b) subset nodes $w$ to circuit nodes $c$ and $\bar{c}$. At higher levels, e.g.~if $c$ is the root node, the indicated failures correspond to coarse-grained subtree failure rates. If any $w$ does not have a circuit node successor but is followed by termination of the protocol, the base case of recursion is reached. Failure rates and variances of upper bounds can be calculated by including the cutoff error $\delta$ to a circuit node's failure rate. Its value must be determined according to the rules described at the end of Sec.~\ref{sec:cutoff_app}.}
\label{fig:recursive}

\end{figure}

Let us also remark that, alternatively, failure rates and variances may also be calculated neatly by exploiting the recursive structure of the tree, namely by \emph{recursively} ``propagating'' these quantities up from the leaf nodes of a tree until its root node. Any event tree can be built from the two components shown in Fig.~\ref{fig:recursive}. For any subset node $w$, we define its node rate $w.p$ (think of $p$ as a property or an attribute of $w$), which can be determined as 
\begin{align}
    w.p &= w.q \times c.p + (1-w.q) \times \bar{c}.p
\end{align}
where $w.q$ ($1-w.q$) is the transition rate $q$ from $w$ to its subsequent circuit node $c$\,($\bar{c}$) or to a ``fail'' (``success'') outcome if $w$ is a leaf node, i.e.~$w$ has no subsequent circuit node but the protocol terminates. Analogously, we define the node rate $c.p$ of a circuit node $c$ as the coarse-grained failure rate, or the total failure rate of the subtrees below $c$. It can be calculated as $c.p = \sum_{w \in c} A_w\,w.p$ where $w$ are all subset nodes directly subsequent to $c$ and $A_w$ is their respective binomial factor. The variance of $c.p$ could be calculated analogously via $c.v = \sum_{w \in c} A_w^2\,w.v$. The variance of any subset node $w$ is obtained by Eqs.~\eqref{eq:varlincomb} and \eqref{eq:goodman} as 
\begin{align}
    w.v &= \Var{w.p} = \Var{w.q \times c.p + (1-w.q) \times \bar{c}.p} \notag \\
    &= \Var{w.q \times c.p} + \Var{(1-w.q)\times \bar{c}.p} - 2\, c.p \times \bar{c}.p \, \Var{w.q} \notag \\
    &= w.q^2 \times c.v + \Var{w.q} \, c.p^2 + \Var{w.q}\,c.v + (1-w.q)^2\,\bar{c}.v + \Var{w.q} \,\bar{c}.p^2 + \Var{w.q}\,\bar{c}.v - 2\, c.p \times \bar{c}.p\,\Var{w.q} \notag \\
    &= \Var{w.q} \left[c.p^2 + c.v + \bar{c}.p^2 + \bar{c}.v - 2\, c.p \times \bar{c}.p \right] + w.q^2 \times c.v + (1-w.q)^2 \,\bar{c}.v \notag \\
    &= \Var{w.q} \left[(c.p - \bar{c}.p)^2 + c.v + \bar{c}.v \right] + w.q^2 \times c.v + (1-w.q)^2 \,\bar{c}.v
\end{align}
if $w$ is not a leaf node and $w.v = \Var{w.q}$ if $w$ is a leaf node.

This procedure can also be applied to calculate the upper bound $p_U$ and its variance by including the single-circuit cutoff error $c.\delta = 1-\sum_{w=0}^{w_\text{max}} A_w^c$ into $c.p$ such that $c.p = \sum_{w \in c} A_w\,w.p + c.\delta$. Special cases for the value of $\delta$ for FT protocols or small number of shots are discussed in Sec.~\ref{sec:cutoff_app}. 

\subsection{Upper bound uncertainty}

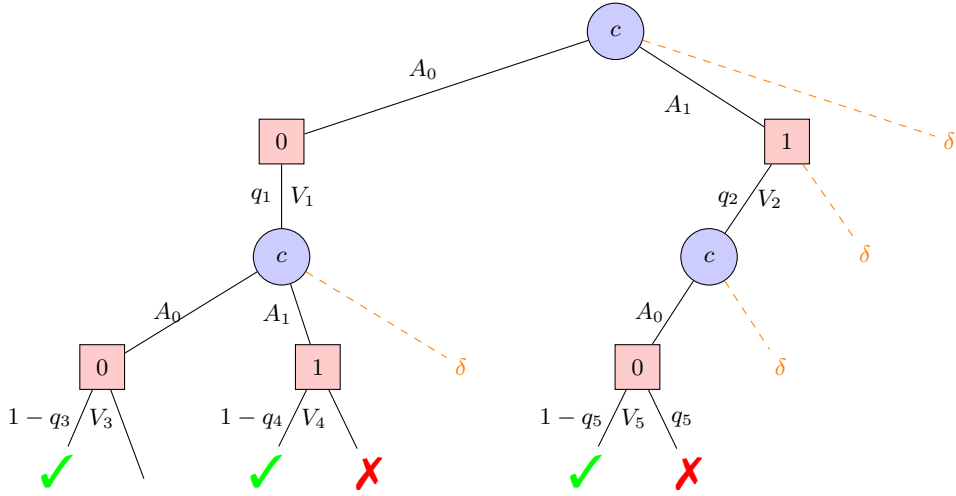
\begin{figure}[!htbp]
    \centering

    \begin{tikzpicture}
    [
        mycircle/.style={circle, fill=blue!20,minimum size=0.75cm},
    	mysquare/.style={regular polygon,regular polygon sides=4, fill=red!20,minimum size=0.5cm},
        level distance=1.5cm,
        sibling distance=0.75cm,
        edge from parent path={(\tikzparentnode) -- (\tikzchildnode)}
    ]
    \Tree
    [.\node[style={draw,mycircle}]{$c$};
        \edge node[auto=right] {$A_0$};
        [.\node[style={draw,mysquare}]{$0$}; 
            \edge node[left] {$q_1$} node[right] {$V_1$};
            [.\node[style={draw,mycircle}]{$c$};
                \edge node[left] {$A_0$};
                [.\node[style={draw,mysquare}]{$0$}; 
                    \edge node[left] {$1-q_3$} node[right] {$V_3$};
                    [.{\cmark} ]
                    \edge node[right] {};
                    [.{} ]
                ]
                \edge node[left] {$A_1$};
                [.\node[style={draw,mysquare}]{$1$}; 
                    \edge node[left] {$1-q_4$} node[right] {$V_4$};
                    [.{\cmark} ]
                    \edge node[right] {};
                    [.{\xmark} ]
                ]
                \edge[draw=orange,dashed] node[auto=right, text=orange] {};
                [.\node[text=orange]{$\delta$}; ]
            ]
        ]
        \edge node[auto=right] {$A_1$};
        [.\node[style={draw,mysquare}]{$1$}; 
            \edge node[left] {$q_2$} node[right] {$V_2$};
            [.\node[style={draw,mycircle}]{$c$};
                \edge node[left] {$A_0$};
                [.\node[style={draw,mysquare}]{$0$}; 
                    \edge node[left] {$1-q_5$} node[right] {$V_5$};
                    [.{\cmark} ]
                    \edge node[right] {$q_5$};
                    [.{\xmark} ]
                ]
                \edge[draw=orange,dashed] node[auto=right, text=orange] {};
                [.\node[text=orange]{$\delta$}; ]
            ]
            \edge[draw=orange,dashed] node[auto=right, text=orange] {};
            [.\node[text=orange]{$\delta$}; ]
        ]
        \edge[draw=orange,dashed] node[auto=right, text=orange] {};
        [.\node[text=orange]{$\delta$}; ]
    ]
    \end{tikzpicture}
    
    \caption{Example tree structure from running the circuit $c$ two consecutive times. No failure can happen at the end of the fault-free path, so $q_3 = V_3 = 0$. The paths indicated with dashed orange lines are yet unexplored but could potentially lead to failures and thus contribute to the cutoff error $\delta$.}
    \label{fig:incomplbranchtree}
\end{figure}

Calculating $\sigma_{U}$ iteratively requires careful consideration, namely all unexplored branchings in an event tree that could potentially lead to failures when running more samples must be incorporated. As an example, take the event tree in Fig.~\ref{fig:incomplbranchtree} that could result from a protocol where the circuit $c$ can be repeatedly run. Additionally, let us assume that after the first run of $c$, in principle, also another circuit $\bar{c}$ could be run. Not only do all unsampled subset nodes contribute to the cutoff error but also the incomplete branching with ratio $q_2$. Here we assume that, so far, only the branching to the left has been realized so that $q_2 = 1$ but the other branching possibility to the right, i.e.~running another circuit $\bar{c}$, has not been realized yet due to a finite number of samples.

The cutoff error must be calculated as 
\begin{align}
    \delta &= 1 - A_0 - A_1 + A_0q_1 (1-A_0-A_1) + A_1(q_2(1-A_0)+(1-q_2))
\end{align}
and the upper bound reads
\begin{align}
    p_L + \delta &= 1 - A_0-A_1 + A_0q_1(1-A_0-A_1(1-q_4)) + A_1(q_2(1-A_0(1-q_5)) + (1-q_2)),
\end{align}
where we explicitly have added the term $1-q_2 = 0$: Although it does not contribute to $\delta$, the finite variance $V_2 > 0$ leads to a contribution for the upper bound $\sigma_U$. We now calculate the variance of the upper bound
\begin{align}
    &\Var{1 - A_0-A_1 + A_0q_1(1-A_0-A_1(1-q_4)) + A_1(q_2(1-A_0(1-q_5)) + (1-q_2))} \equiv T_0 + T_1 \\
    T_0 &= A_0^2 \,\Var{q_1(1-A_0-A_1(1-q_4))} \notag \\
    &= A_0^2 ((1-A_0)^2V_1 + A_1^2 \,\Var{q_1(1-q_4)} - 2 (1-A_0)A_1(1-q_4)V_1) \sim p^2 \\
    T_1 &= A_1^2 \,\Var{q_2(1-A_0(1-q_5)) + (1-q_2)} \notag \\
    &= A_1^2 \left( \Var{q_2(1-A_0(1-q_5))} + V_2 - 2(1-A_0(1-q_5))V_2 \right) \sim p^2 
\end{align}
and observe that the standard error of the upper bound $\sigma_U = \sqrt{\Var{p_U}}$ scales linearly just as the failure rate estimator. Note that $V_1 = V_3 = 0$ since we assume that there can never be a branching off the fault-free path. If Fig.~\ref{fig:incomplbranchtree} represented an FT protocol, we would also have $q_5 = V_5 = 0$ since the weight-1 path can never lead to failure. However, $V_2 \neq 0$, which applies to QEC in the sense that the protocol can deal with different errors by realizing different circuit sequences. This means that we must properly deal with cutoff errors that arise from not having explored both possible circuit node successors from a subset node.

\subsection{Cutoff error for FT protocols}\label{sec:cutoff_app}
As mentioned in Secs.~\ref{sec:intro} and \ref{sec:examples}, protocols can be designed to have fault tolerance guarantees, in practice, e.g., FT-1 means that the occurrence of one single fault anywhere in the protocol cannot lead to failure. While the consequential improvement of the failure rate scaling $p^*$ from $\mathcal{O}(p)$ to $\mathcal{O}(p^2)$ is exactly what we want from an FT theory perspective, it poses the following challenge here: For any protocol, the cutoff error must upper bound the worst-case possible subset failure rates that could occur within so far unsampled parts of the protocol's event tree. When left as above, the cutoff error calculation will be ignorant to the FT guarantee, i.e.~in the calculation of $\delta$ we will assume that all unexplored paths lead to failure, which will give a scaling as $\delta = \mathcal{O}(p)$. But this would render the estimator overly pessimistic and the relative error would diverge for small $p$. However, we can exploit the fact that also the unexplored paths may contain cases where failure can be excluded when provided an FT guarantee. 

In the following, we show in two steps how to choose the upper bound such that its scaling is of the same order in $p$ as the failure rate, i.e.~the relative error stays constant when scaling $p$. We only consider the special case of fault tolerance towards a single fault in a QEC code of distance $d=3$ explicitly. The first step considers the branching ratios at a single stage within the tree. The second step then deals with circuit sequences, i.e.~the worst case tree depth.

\begin{figure}[!htbp]

\subfloat[What is the worst case contribution of the orange 1-subset to the failure rate? \label{fig:exft_a}]{\begin{tikzpicture}
    [
        mycircle/.style={circle, fill=blue!20,minimum size=1.25cm},
    	mysquare/.style={regular polygon,regular polygon sides=4, fill=red!20,minimum size=0.5cm},
        level distance=1.5cm,
        sibling distance=0.75cm,
        edge from parent path={(\tikzparentnode) -- (\tikzchildnode)}
    ]
    \Tree
    [.\node[style={draw,mycircle}]{ENC};
        \edge node[auto=right] {$A_0$};
        [.\node[style={draw,mysquare}]{$0$};
            \edge node[left] {$q_0 = 1/1$} node[right] {};
            [.\node[style={draw,mycircle}]{MEAS};
                \edge node[left] {$B_0$};
                [.\node[style={draw,mysquare}]{$0$}; 
                    \edge node[left] {$q_1 = 0/1$} node[right] {};
                    [.{} ] \edge node[right] {};
                    [.{\cmark} ]
                ]
            ]
        ]
        \edge[draw=orange,dashed] node[auto=right, text=orange] {$A_1$};
        [.\node[text=orange]{$1$};
            \edge[draw=orange,dashed] node[left] {} node[right] {};
            [.\node[text=orange]{?};
            ]
        ]
    ]
    \end{tikzpicture}
}
\hfill
\subfloat[Two possible circuit branchings within the unsampled tree have an FT path. \label{fig:exft_b}]{\begin{tikzpicture}
    [
        mycircle/.style={circle, fill=blue!20,minimum size=1.25cm},
    	mysquare/.style={regular polygon,regular polygon sides=4, fill=red!20,minimum size=0.5cm},
        level distance=1.5cm,
        sibling distance=0.75cm,
        edge from parent path={(\tikzparentnode) -- (\tikzchildnode)}
    ]
    \Tree
    [.\node[style={draw,mycircle}]{ENC};
        \edge node[auto=right] {$A_0$};
        [.\node[style={draw,mysquare}]{$0$};
            \edge node[left] {$q_0 = 1/1$} node[right] {};
            [.\node[style={draw,mycircle}]{MEAS};
                \edge node[left] {$B_0$};
                [.\node[style={draw,mysquare}]{$0$}; 
                    \edge node[left] {$q_1 = 0/1$} node[right] {};
                    [.{} ] \edge node[right] {};
                    [.{\cmark} ]
                ]
            ]
        ]
        \edge[draw=orange,dashed] node[auto=right, text=orange] {$A_1$};
        [.\node[text=orange]{$1$};
            \edge[draw=orange,dashed] node[left,text=orange] {$Q$} node[right] {};
            [.\node[text=orange]{MEAS};
                \edge[draw=orange,dashed] node[left,text=orange] {$B_0$};
                [.\node[text=orange]{$0$}; 
                    \edge[draw=orange,dashed] node[left] {} node[right] {};
                    [.{} ] \edge[draw=orange,dashed] node[right] {};
                    [.{\cmark} ]
                ]
            ]
            \edge[draw=orange,dashed] node[left] {} node[right,text=orange] {$1-Q$};
            [.\node[text=orange]{$\neg\text{MEAS}$};
                \edge[draw=orange,dashed] node[left,text=orange] {$C_0$};
                [.\node[text=orange]{$0$}; 
                    \edge[draw=orange,dashed] node[left] {} node[right] {};
                    [.{} ] \edge[draw=orange,dashed] node[right] {};
                    [.{\cmark} ]
                ]
            ]
        ]
    ]
    \end{tikzpicture}
}
\hfill
\subfloat[The FT paths can be upper bounded by a single node $\bullet$ with binomial factors $A_w^\bullet = M_w$. \label{fig:exft_c}]{\begin{tikzpicture}
    [
        mycircle/.style={circle, fill=blue!20,minimum size=1.25cm},
    	mysquare/.style={regular polygon,regular polygon sides=4, fill=red!20,minimum size=0.5cm},
        level distance=1.5cm,
        sibling distance=0.75cm,
        edge from parent path={(\tikzparentnode) -- (\tikzchildnode)}
    ]
    \Tree
    [.\node[style={draw,mycircle}]{ENC};
        \edge node[auto=right] {$A_0$};
        [.\node[style={draw,mysquare}]{$0$};
            \edge node[left] {$q_0 = 1/1$} node[right] {};
            [.\node[style={draw,mycircle}]{MEAS};
                \edge node[left] {$B_0$};
                [.\node[style={draw,mysquare}]{$0$}; 
                    \edge node[left] {$q_1 = 0/1$} node[right] {};
                    [.{} ] \edge node[right] {};
                    [.{\cmark} ]
                ]
            ]
        ]
        \edge[draw=orange,dashed] node[auto=right, text=orange] {$A_1$};
        [.\node[text=orange]{$1$};
            \edge[draw=orange,dashed] node[left] {} node[right] {};
            [.\node[text=orange]{$\bullet$};
                \edge[draw=orange,dashed] node[left,text=orange] {$M_0$};
                [.\node[text=orange]{$0$}; 
                    \edge[draw=orange,dashed] node[left] {} node[right] {};
                    [.{} ] \edge[draw=orange,dashed] node[right] {};
                    [.{\cmark} ]
                ]
            ]
        ]
    ]
    \end{tikzpicture}
}
\caption{Example trees illustrate the cutoff error (orange, no frames) after a single shot of a protocol containing the circuits ENC and MEAS has been run (black frames). It is assumed that the protocol consists of two stages which is not the case in general.}
\label{fig:exft}

\end{figure}
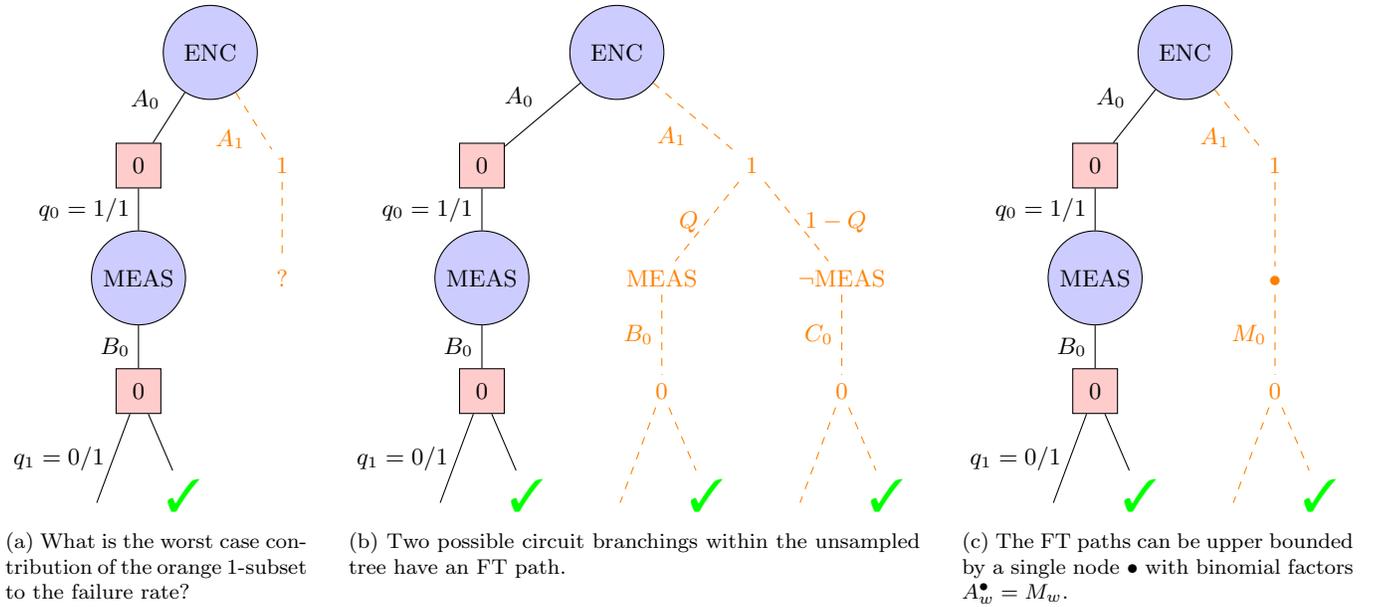

Figure \ref{fig:exft_a} shows the challenge, which is special to an FT protocol. While for a non-FT protocol the cutoff error contribution at a circuit node called ENC would easily be calculated as $1-A_0$, we now do not want to assume anymore that all paths branching off into the 1-subset of ENC lead to failure in the worst case. This is because we know beforehand that, since the protocol is FT, no 1-path can ever lead to failure. 

Two possible paths are shown in Fig.~\ref{fig:exft_b}. Their total weights are $A_1QB_0$ and $A_1(1-Q)C_0$. Reversely, if we do not exclude these paths from our cutoff error, we would estimate $\delta \sim 1-A_0 \sim p$. Although this would be a correct worst case bound to state, it unnecessarily overestimates the a priori known worst case and causes the relative error $\delta/p_L$ of $p_L \sim p^2$ to diverge at $p \rightarrow 0$ since then $\delta \sim p$. At the orange subset node ``1'', we cannot know a priori what the branching ratio $Q$ will be exactly without sampling (again, note that the point of $\delta$ is to make a statement about unsampled parts of the tree). It can also not be queried from the protocol since this would require testing all possible weight-1 faults at this stage. Also, here we are interested in analytically stating the worst case error. We do not even know what the two possible resulting circuits after the branching are since we cannot determine beforehand how the protocol deals with every individual fault. As an example, let us for the moment assume, the subsequent circuit could either be MEAS or $\neg$MEAS (a circuit other than MEAS). The cutoff error contribution of the orange subtree would then be $A_1(Q(1-B_0) + (1-Q)(1-C_0))$. One of the two binomial factors of the subsequent circuits will be smaller than the other. Here, let us assume for illustration (and without loss of generality) that $B_0 > C_0$. Then we can upper bound the cutoff error contribution as
\begin{align}
    A_1(Q(1-B_0) + (1-Q)(1-C_0)) \leq A_1(Q(1-C_0) + (1-Q)(1-C_0)) = A_1(1-C_0),
\end{align}
which has the desired effect to remove the unknown branching ratio $Q$ from our estimation. In practise, we know the smallest possible binomial factor independently from sampling. Thus, as shown in Fig.~\ref{fig:exft_c}, we can replace the branching with $Q$ by a single node with a binomial factor 
\begin{align}
    M_0 = \min_{c \in \C \in \Prot} (A_0^c).
\end{align}

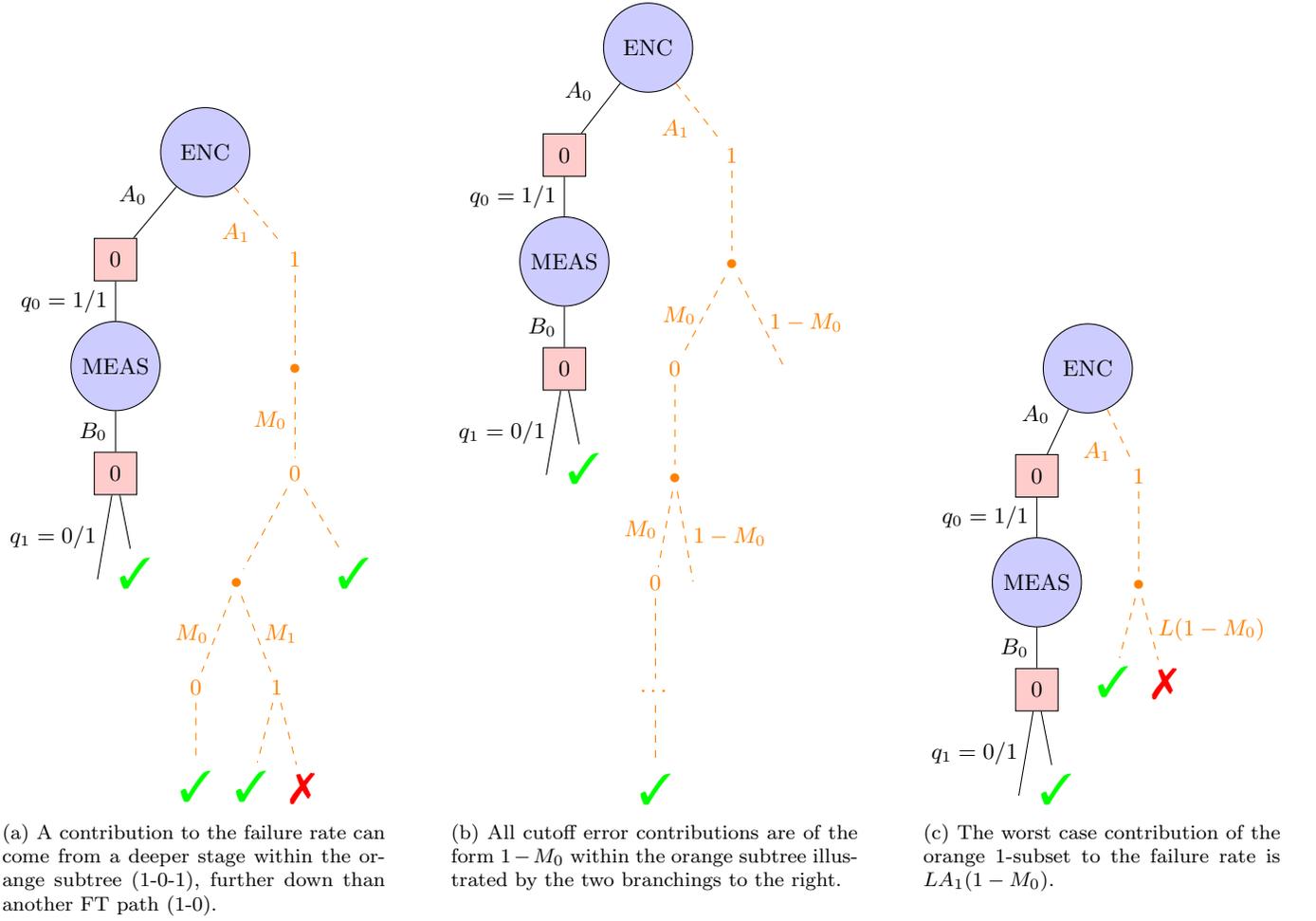
\begin{figure}[!htbp]

\subfloat[A contribution to the failure rate can come from a deeper stage within the orange subtree (1-0-1), further down than another FT path (1-0). \label{fig:exftdeep_a}]{\begin{tikzpicture}
    [
        mycircle/.style={circle, fill=blue!20,minimum size=1.25cm},
    	mysquare/.style={regular polygon,regular polygon sides=4, fill=red!20,minimum size=0.5cm},
        level distance=1.5cm,
        sibling distance=0.1cm,
        edge from parent path={(\tikzparentnode) -- (\tikzchildnode)}
    ]
    \Tree
    [.\node[style={draw,mycircle}]{ENC};
        \edge node[auto=right] {$A_0$};
        [.\node[style={draw,mysquare}]{$0$};
            \edge node[left] {$q_0 = 1/1$} node[right] {};
            [.\node[style={draw,mycircle}]{MEAS};
                \edge node[left] {$B_0$};
                [.\node[style={draw,mysquare}]{$0$}; 
                    \edge node[left] {$q_1 = 0/1$} node[right] {};
                    [.{} ] \edge node[right] {};
                    [.{\cmark} ]
                ]
            ]
        ]
        \edge[draw=orange,dashed] node[auto=right, text=orange] {$A_1$};
        [.\node[text=orange]{$1$};
            \edge[draw=orange,dashed] node[left] {} node[right] {};
            [.\node[text=orange]{$\bullet$};
                \edge[draw=orange,dashed] node[left,text=orange] {$M_0$};
                [.\node[text=orange]{$0$}; 
                    \edge[draw=orange,dashed] node[left] {} node[right] {};
                    [.\node[text=orange]{$\bullet$};
                        \edge[draw=orange,dashed] node[left,text=orange] {$M_0$};
                        [.\node[text=orange]{$0$}; 
                            \edge[draw=orange,dashed] node[right] {};
                            [.{\cmark} ]
                        ]
                        \edge[draw=orange,dashed] node[right,text=orange] {$M_1$};
                        [.\node[text=orange]{$1$}; 
                            \edge[draw=orange,dashed] node[right] {};
                            [.{\cmark} ]
                            \edge[draw=orange,dashed] node[right] {};
                            [.{\xmark} ]
                        ]
                    ]
                    \edge[draw=orange,dashed] node[right] {};
                    [.{\cmark} ]
                ]
            ]
        ]
    ]
    \end{tikzpicture}
}
\hfill
\subfloat[All cutoff error contributions are of the form $1-M_0$ within the orange subtree illustrated by the two branchings to the right. \label{fig:exftdeep_b}]{\begin{tikzpicture}
    [
        mycircle/.style={circle, fill=blue!20,minimum size=1.25cm},
    	mysquare/.style={regular polygon,regular polygon sides=4, fill=red!20,minimum size=0.5cm},
        level distance=1.5cm,
        sibling distance=0.1cm,
        edge from parent path={(\tikzparentnode) -- (\tikzchildnode)}
    ]
    \Tree
    [.\node[style={draw,mycircle}]{ENC};
        \edge node[auto=right] {$A_0$};
        [.\node[style={draw,mysquare}]{$0$};
            \edge node[left] {$q_0 = 1/1$} node[right] {};
            [.\node[style={draw,mycircle}]{MEAS};
                \edge node[left] {$B_0$};
                [.\node[style={draw,mysquare}]{$0$}; 
                    \edge node[left] {$q_1 = 0/1$} node[right] {};
                    [.{} ] \edge node[right] {};
                    [.{\cmark} ]
                ]
            ]
        ]
        \edge[draw=orange,dashed] node[auto=right, text=orange] {$A_1$};
        [.\node[text=orange]{$1$};
            \edge[draw=orange,dashed] node[left] {} node[right] {};
            [.\node[text=orange]{$\bullet$};
                \edge[draw=orange,dashed] node[left,text=orange] {$M_0$};
                [.\node[text=orange]{$0$}; 
                    \edge[draw=orange,dashed] node[left] {} node[right] {};
                    [.\node[text=orange]{$\bullet$};
                        \edge[draw=orange,dashed] node[left,text=orange] {$M_0$};
                        [.\node[text=orange]{$0$};
                            \edge[draw=orange,dashed] node[left] {} node[right] {};
                            [.\node[text=orange]{$\cdots$};
                                \edge[draw=orange,dashed] node[left] {} node[right] {};
                                [.\node[]{\cmark}; ]
                            ]
                        ]
                        \edge[draw=orange,dashed] node[right,text=orange] {$1-M_0$};
                            [.{} ] ]
                ]
                \edge[draw=orange,dashed] node[right,text=orange] {$1-M_0$};
                    [.{} ] ]
        ]
    ]
    \end{tikzpicture}
}
\hfill
\subfloat[The worst case contribution of the orange 1-subset to the failure rate is $LA_1(1-M_0)$. \label{fig:exftdeep_c}]{\begin{tikzpicture}
    [
        mycircle/.style={circle, fill=blue!20,minimum size=1.25cm},
    	mysquare/.style={regular polygon,regular polygon sides=4, fill=red!20,minimum size=0.5cm},
        level distance=1.5cm,
        sibling distance=0.1cm,
        edge from parent path={(\tikzparentnode) -- (\tikzchildnode)}
    ]
    \Tree
    [.\node[style={draw,mycircle}]{ENC};
        \edge node[auto=right] {$A_0$};
        [.\node[style={draw,mysquare}]{$0$};
            \edge node[left] {$q_0 = 1/1$} node[right] {};
            [.\node[style={draw,mycircle}]{MEAS};
                \edge node[left] {$B_0$};
                [.\node[style={draw,mysquare}]{$0$}; 
                    \edge node[left] {$q_1 = 0/1$} node[right] {};
                    [.{} ] \edge node[right] {};
                    [.{\cmark} ]
                ]
            ]
        ]
        \edge[draw=orange,dashed] node[auto=right, text=orange] {$A_1$};
        [.\node[text=orange]{$1$};
            \edge[draw=orange,dashed] node[left] {} node[right] {};
            [.\node[text=orange]{$\bullet$};
                \edge[draw=orange,dashed] node[left] {};
                [.{\cmark} ]
                \edge[draw=orange,dashed] node[right,text=orange] {$L(1-M_0)$};
                [.{\xmark} ]
            ]
        ]
    ]
    \end{tikzpicture}
}
\caption{Example trees illustrate the cutoff error contributions from longer circuit sequences. The actual circuit sequence is unimportant and can be bounded by $M_0$ as shown in Fig.~\ref{fig:exft}. Only the maximum length of all FT paths $L$ is needed.}
\label{fig:exft_deep}

\end{figure}

Now, as a second step, we need to consider longer circuit sequences, as for example in Fig.~\ref{fig:exft_c}. There could follow another circuit at the left leaf in the orange subtree, where, again, a second 1-subset could lead to failure. This scenario is depicted in Fig.~\ref{fig:exftdeep_a}. Although the weight sequence 1-0 can never lead to failure after termination of the protocol, the single fault could cause another branching so that a third circuit will be run. Then, the weight sequence 1-0-1 is not FT anymore and \emph{can} lead to failure. We must consider this case in our cutoff error. 

This event is covered by the tree in Fig.~\ref{fig:exftdeep_b}. There is the cutoff contribution $A_1M_0(1-M_0)$ that contains this particular example. Now, consider the fictitious case that the orange subtree could potentially be infinitely deep. Then, the cutoff contribution would be 
\begin{align}
    &A_1((1-M_0)+M_0(1-M_0)+M_0^2(1-M_0)+\cdots) \notag \\
    &= A_1(1-M_0)\sum_{k=0}^\infty M_0^k = A_1(1-M_0) \frac{1}{1-M_0} = A_1 \sim p
\end{align}
which destroys our initial goal to formulate a cutoff error that scales like $p_L$. However, we know beforehand that the tree can actually never be infinitely deep. All FT paths are finite because no real protocol would go on forever without dangerous faults and since the protocol was actively designed, its maximum length $L$ in case only correctable errors occur, is always known. This changes the cutoff contribution to read
\begin{align}
    &A_1((1-M_0)+M_0(1-M_0)+M_0^2(1-M_0)+\cdots+M_0^{L-1}(1-M_0)) \notag \\
    &= A_1(1-M_0)\sum_{k=0}^{L-1} M_0^k \leq L A_1 (1-M_0) \sim p^2
\end{align}
as we aimed for.

The maximum FT protocol lengths for the protocols considered here are given in Tab.~\ref{tab:protL} alongside with the circuits that belong to the protocol. 

\begin{table}\begin{center}
\begin{tabular}{c | c | c}
Protocol $\Prot$ & max FT length $L$ & circuits $c \in \C \in \Prot$ \\ \hline
GHZ state preparation & 1 & GHZ \\
Deterministic $\ket{0}_L$ & 4 & ENC, SZ, X7, MEAS \\ Stabilizer $\ket{0}_L$ & 8 & SX1a, SX2a, SX3a, SX1b, SX2b, SX3b, NFS, MEAS \\ \hline 
Repeat until success & 2+1 & ENC, MEAS
\end{tabular} 
\end{center}
\caption{Maximum FT protocol lengths for the protocols considered in Sec.~\ref{sec:examples}.}
\label{tab:protL}
\end{table}

There are different cases, illustrated in Fig.~\ref{fig:ft_cutoff_cases}, for how to consider the (FT) cutoff error and calculate the variance of the upper bound to the failure rate for a general tree:
\begin{enumerate}[(I)]
    \item If the total path weight up to a subset node is 0, we add no additional branching circuit because we assume there is only one fault-free path.
    \item Only for FT protocol: If the total path weight up to a subset node is 1, and the branching ratio $q$ of the subset node is equal to 1 and there is only one circuit node below the subset node but the other one is still unknown, then we have to add a complementary branching with factor $1-q$ and a $\delta$-node with value $L(1-M_0)$. While this node does not contribute directly to the cutoff error, it will contribute to the variance of the upper bound via $\Var{1-q} \neq 0$.
    \item If the total path weight up to a subset node is 2 (for an FT protocol) / 1 (for a non-FT protocol) or larger and the branching ratio $q$ of the subset node is equal to 1 and there is only one circuit node below the subset node but the other one is still unknown, then we have to add a complementary branching with factor $1-q$ and a $\delta$-node with value 1 below the complementary branching with $1-q$.
    \item Only for FT protocol: If the total path weight up to a circuit node is 0 and if the circuit node does not have a 1-subset, we have to add a virtual one. The virtual 1-subset has a $\delta$-node of value $L(1-M_0)$. The $\delta$-node of the circuit node then has the usual value.
\end{enumerate}

\begin{figure}[!htbp]
\begin{tikzpicture}
    [
        mycircle/.style={circle, fill=blue!20,minimum size=1.25cm},
    	mysquare/.style={regular polygon,regular polygon sides=4, fill=red!20,minimum size=0.5cm},
        level distance=1.5cm,
        sibling distance=0.25cm,
        edge from parent path={(\tikzparentnode) -- (\tikzchildnode)}
    ]
    \Tree
    [.\node[style={draw,mycircle}]{ENC};
        \edge node[auto=right] {$A_0$};
        [.\node[style={draw,mysquare}]{$0$};
            \edge node[left] {} node[right] {};
            [.\node[style={draw,mycircle}]{MEAS};
                \edge node[left,yshift=0.1cm] {$B_0$};
                [.\node[style={draw,mysquare}]{$0$}; 
                    \edge node[left] {} node[right] {};
                    [.{\cmark} ]
                    \edge[draw=red] node[left] {} node[right, text=magenta] {(I)};
                    [.\node[text=red]{};]
                ]
                \edge[draw=orange,dashed] node[auto=right, text=orange] {$B_1$};
                [.\node[text=orange]{$1$};
                    \edge[draw=orange,dashed] node[left] {} node[right] {};
                    [.\node[text=orange]{$\bullet$};
                        \edge[draw=orange,dashed] node[left] {} node[right, text=magenta] {(IV)};
                        [.\node[text=orange]{$L(1-M_0)$}; ]
                    ]
                ]
                \edge[draw=orange,dashed] node[auto=right, text=orange] {};
                [.\node[text=orange]{$1-B_0-B_1$};]
            ]
            \edge[draw=red] node[left] {} node[right, text=magenta] {(I)};
            [.\node[text=red]{};]
        ]
        \edge node[auto=right] {$A_1$};
        [.\node[style={draw,mysquare}]{$1$};
            \edge node[left,yshift=0.1cm] {$q_1$} node[right] {};
            [.\node[style={draw,mycircle}]{MEAS};
                \edge node[left,yshift=0.1cm] {$B_0$};
                [.\node[style={draw,mysquare}]{$0$}; 
                    \edge node[left] {$1-q_2$} node[right] {};
                    [.{\cmark} ]
                    \edge[draw=orange,dashed] node[auto=right, text=orange] {$q_2$};
                    [.\node[text=orange]{$\bullet$};
                        \edge[draw=orange,dashed] node[left] {} node[right, text=magenta] {(II)};
                        [.\node[text=orange]{$L(1-M_0)$}; ]
                    ]
                ]
                \edge node[left] {$B_1$};
                [.\node[style={draw,mysquare}]{$1$}; 
                    \edge node[left] {$1-q_3$} node[right] {};
                    [.{\cmark} ]
                    \edge[draw=orange,dashed] node[auto=right, text=orange] {$q_3$};
                    [.\node[text=orange]{$\bullet$};
                        \edge[draw=orange,dashed] node[left] {} node[right, text=magenta] {(III)};
                        [.\node[text=orange]{$\delta = 1$}; ]
                    ]
                ]
                \edge[draw=orange,dashed] node[auto=right, text=orange] {};
                [.\node[text=orange]{$1-B_0-B_1$};]
            ]
            \edge[draw=orange,dashed] node[auto=right, text=orange] {$1-q_1$};
            [.\node[text=orange]{$\bullet$};
                \edge[draw=orange,dashed] node[left] {} node[right, text=magenta] {(II)};
                [.\node[text=orange]{$L(1-M_0)$}; ]
            ]
        ]
        \edge[draw=orange,dashed] node[auto=right, text=orange] {};
        [.\node[text=orange]{$1-A_0-A_1$};]
    ]
\end{tikzpicture}
\caption{An FT protocol tree illustrating the four cases of the FT cutoff error. If the protocol were non-FT, the cases (II) and (IV) would be replaced with instances of case (III).}
\label{fig:ft_cutoff_cases}

\end{figure}

As a concluding remark, we note that to practically calculate the cutoff error within a given stage $i$, we can determine the factor $l = L - i$ to account for the longest possible ``rest of the sequence'' that could still follow from the current protocol stage, where $i$ circuits have been executed already. In Fig.~\ref{fig:exftdeep_c} we may replace with $L$ with $l$. However, \texttt{qsample} always uses the fixed value $L$ for the cutoff error calculation.

There is one extra peculiarity to a ``repeat until success'' protocol. Here we mean, for instance, a protocol that consists of repeating an encoding circuit ENC until a verification measurement signals success and then measuring all qubits as represented by the node MEAS (see Refs.~\cite{postler2021demonstration, heussen2023strategies} for examples). Since all qubits are always reset upon execution of the encoding circuit, we can only count its maximum length (of 2) from the last reset on. For example, a path ENC-1-ENC-0-MEAS-0 is possible to realize. However, it is not possible to realize an FT path longer than $2+t$, where $t = 1$. For example, a path ENC-1-ENC-1-ENC-0-MEAS-0 would be possible but it is of order $p^2$ and thus covered by the cutoff error contribution $A_1(1-M_0)$ in Fig.~\ref{fig:exftdeep_b}.
For a ``repeat until success'' protocol, one may achieve tighter bounds to the true logical failure rate by adjusting which leaf node variances are set to zero and which values are assigned to $\delta$-nodes based on the last qubit reset.

\section{Numerical implementation: \texttt{qsample}} \label{sec:qsample}

In this appendix, we describe the numerical implementation of dynamical subset sampling in the python package \texttt{qsample}, which is deployed together with this manuscript at \url{https://github.com/dpwinter/qsample}.

\subsection{Numerical calculation of estimators}
We model a quantum \emph{protocol} as a graph containing quantum circuits at its vertices and transition rules as edges, which define the order of circuit execution. At the beginning of sampling, the protocol graph starts at a predefined unique node, \textit{root}, and is subsequently traversed by choosing transitions dynamically during runtime conditioned on the measurement results of all circuits in the sequence of the current shot as visualized in Fig.~\ref{fig:introDSS}(d). Naturally, as our goal is to obtain failure rates of a protocol, we check at the end of the circuit sequence whether a failure has occurred. Protocol terminations that do not lead to failure are not explicitly represented as a node, i.e.~there is no \emph{success} node in a \texttt{qsample} protocol. 

Individual samples of a protocol are efficiently tracked in a tree data structure, analogous to Fig.~\ref{fig:introDSS}(c), the \textit{sample tree}. Repeatedly executing a noisy protocol thus non-deterministically yields circuit sequences, modelled as paths of the sample tree, which each start at a unique root node and end at a \textit{leaf} node. We restrict the maximum number of successors in the protocol graph to at most two, since we assume that any quantum protocol can be cast in such a form.
By this constraint, we can model the transition of a noisy circuit to the next by a Bernoulli random variable $X$ which can take on the values $0,1$ corresponding to which of two circuits is chosen. The transition rates to either circuit are then estimated by $\mathbb{E}[X]$ and $1-\mathbb{E}[X]$, respectively. Additionally, the binomial factors $A_w$ and cutoffs $\delta$, determined as the four cases in Sec.~\ref{sec:cutoff_app} prescribe, are included in the sample tree, branching off from any circuit node (see Fig.~\ref{fig:introDSS}(c)). Thus, we define a path $P\in\mathcal{P}$ as a unique sequence of circuits and binomial factors (one per circuit), represented by circuit nodes and subset nodes in the tree, ending on one of three terminal nodes, \textit{fail}, \textit{no-fail}, or $\delta$. As the fail and no-fail leaves represent the binary event whether or not a failure occurred, we denote a path ending on a fail leaf as \textit{fail path} $P_f\in\mathcal{P}_f$. Furthermore, we define a path ending on a $\delta$-node as $P_{\delta}\in\mathcal{P}_{\delta}$. The lower bound $p_L$ is then implemented as the sum over all fail paths $\mathcal{P}_f$ as
\begin{equation}
    p_L=\mathbb{E}\bigg[\sum_{P_f\in\mathcal{P}_f}\prod_{n\in P_f} A_w^n X_n\bigg]=\sum_{P_f\in\mathcal{P}_f}\prod_{n\in P_f} A_w^n \mathbb{E}[X_n],
    \label{eq:pL_2}
\end{equation}
where we denote the binomial factor for a subset of weight $w$ corresponding to the circuit node $n$ as $A_w^n$. The second equality of Eq.~\eqref{eq:pL_2} follows from independence of $X_n$ within the same path.
The upper bound $p_U$ additionally includes the sum over all path products $\mathcal{P}_{\delta}$, where the product is over all circuit transition rates $\mathbb{E}[X]$ and binomial factors $A_w$ in a path, i.e. 
\begin{equation}    
    p_L + \delta = p_L + \sum_{P_\delta\in\mathcal{P}_\delta}\prod_{n\in P_\delta} A_w^n f(n),
\end{equation}
where $f(n)=\mathbb{E}[X_n]$, except for $\delta$-nodes at which the value $f(n)=\delta_n$ is determined by the cutoff error of the particular circuit node (see a single yellow box in Fig.~\ref{fig:introDSS}(c) and Eq.~\eqref{eq:dssdelta}). The uncertainty on the failure rate bounds is calculated in a similar way as
\begin{equation}
    \text{Var}[p_L] = \text{Var}\bigg[\sum_{P_f\in\mathcal{P}_f}\prod_{n\in P_f} A_w^n X_n\bigg]
\end{equation}
\begin{equation}
    \text{Var}[p_L + \delta] = \text{Var}\bigg[\sum_{P_f\in\mathcal{P}_f}\prod_{n\in P_f} A_w^n X_n + \sum_{P_\delta\in\mathcal{P}_\delta}\prod_{n\in P_\delta} A_w^n X_n\bigg].
\end{equation}
If we express the result of a path product as a random variable $P_i=\prod_{n\in P}A_w^n X_n$, we have for the variance of the sum
\begin{equation}
    \text{Var}\bigg[\sum_i P_i\bigg] = \sum_{i,j} \text{Cov}[P_i,P_j] = \underbrace{\sum_i \text{Var}[P_i]}_{\text{path variance}} + \underbrace{2\sum_{i<j} \text{Cov}[P_i,P_j]}_{\text{overlap variance}},
    \label{eq:var}
\end{equation}
where we denote the first sum as the \textit{path variance} and the second as \textit{overlap variance}. It is worth mentioning that we have to consider the possibility that any path can potentially end in a failure. Even though a failure may not have been sampled for a path yet, its variance can be non-zero. These \textit{virtual} paths have contributions to both the path and overlap variance. Exempt from this are the fault-free path and any path with a total fault-weight of at most 1 for FT-1 protocols.

The path variances are calculated via Goodman's formula. We first consider the variance of a path $P\in\mathcal{P}$. Due to independence of all $X_i$ within a path $P$ we get

\begin{align}
    \text{Var}[P] &= \prod_i\big(A_w^i\big)^2\mathbb{E}[X_i^2]-\prod_i\big(A_w^i\big)^2\mathbb{E}[X_i]^2 = \bigg( \prod_i A_w^i\bigg)^2 \bigg(\prod_i\big(\Var{X_i} + \mathbb{E}[X_i]^2\big) - \prod_i\big(\mathbb{E}[X_i]^2\big)\bigg).
    \label{eq:pathvar}
\end{align}

It is straightforward to implement Eq.~\eqref{eq:pathvar} on our tree data structure as listed in Alg.~\ref{alg:pathvar}. Here, the value of a circuit node (subset node) is its associated transition rate (binomial factor). The values of $\delta$-nodes are determined according to Sec.~\ref{sec:cutoff_app}. The variance of a circuit node is determined by the Wilson score interval (see Eq.~\eqref{eq:wilsonint}) and the variance of a subset node is zero.

\begin{algorithm}[htp]
    \caption{Variance of a path}
    \begin{algorithmic}[1]
        \Statex \textbf{Input:} Path $P$
        \Statex \textbf{Output:} Variance of $P$, $\text{Var}[P]$
\For{node $n$ in path $P$}
                \If{$n$ is a subset node}
\State $A \gets A \times A^n_w$
                \ElsIf{$n$ is a circuit node}
\State $\mathbb{E}[P]^2 \gets \mathbb{E}[P]^2 \times \mathbb{E}[X_n]^2$
                    \State $\mathbb{E}[P^2] \gets \mathbb{E}[P^2] \times (\text{Var}[X_n]+\mathbb{E}[X_n]^2)$
                \EndIf
            \EndFor
            \State \Return $A^2\big( \mathbb{E}[P^2] - \mathbb{E}[P]^2 \big)$
    \end{algorithmic}
    \label{alg:pathvar}
\end{algorithm}

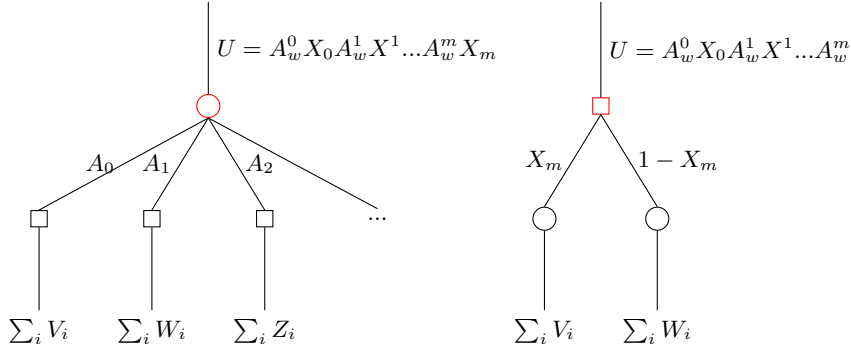
\begin{figure}[!htbp]
    \centering
    \begin{minipage}[b]{0.3\textwidth}
    
        \begin{tikzpicture}
        \node {}
		child {node [draw, shape=circle, color=red] {}
			child {node [draw, shape=rectangle] {}
				child {node {$\sum_i V_i$}}
				edge from parent node [left] {$A_0$}}
			child {node [draw, shape=rectangle] {}
				child {node {$\sum_i W_i$}}
				edge from parent node [left] {$A_1$}}
			child {node [draw, shape=rectangle] {}
				child {node {$\sum_i Z_i$}}
				edge from parent node [right] {$A_2$}}
            child {node [] {...}
edge from parent node [right] {}}
			edge from parent node [right] {$U=A_w^0 X_0 A_w^1 X^1 ... A_w^mX_m$}};
        \end{tikzpicture}

    \end{minipage}
    \begin{minipage}[b]{0.4\textwidth}

        \begin{tikzpicture}
            \node {}
    		child {node [draw, shape=rectangle, color=red] {}
    			child {node [draw, shape=circle] {}
    				child {node {$\sum_i V_i$}}
    				edge from parent node [left] {$X_m$}}
    			child {node [draw, shape=circle] {}
    				child {node {$\sum_i W_i$}}
    				edge from parent node [right] {$1-X_m$}}
    			edge from parent node [right] {$U=A_w^0 X_0 A_w^1 X^1 ... A_w^m$}};
       \end{tikzpicture}

    \end{minipage}
    
    \caption{Paths intersecting at a circuit node (circle) on the left and at subset node (rectangle) on the right marked in red. The common \textit{upstream} path contains all common random variables $X_i$ and constants $A_w^i$. Its path product can be interpreted as a new random variable $U$. The random variables $V_i, W_i, Z_i$ represent the rest of those paths \textit{downstream}.}
    \label{fig:divnodes}
\end{figure}

The covariance terms are harder to obtain. We note that only overlapping paths have a non-zero covariance and that overlapping paths share one unique node, below which the paths split and above which all nodes are common. Further, we can distinguish two situations in which the last common node is either a circuit node so that the subsequent branching is determined by the random variable $X$ or a subset node so that the subsequent branching is determined by binomial factor $A_w$. Both situations are depicted in Fig.~\ref{fig:divnodes}.\\

Consider first the case where paths split at a circuit node. If two paths $P_0=UA_0\sum_i V_i$ and $P_1=UA_1\sum_i W_i$ split at $X_m$ with common path $U$, the overlap variance reads
\begin{equation}
    \text{Cov}[P_0, P_1] = \underbrace{\mathbb{E}\bigg[A_0\sum_i V_i \bigg] \mathbb{E}\bigg[A_1\sum_i W_i\bigg]}_{\text{subtree sums }\mathcal{S}_w} \text{Var}[U].
\end{equation}
As the downstream paths $V_i$ and $W_i$ are all unique they are independent of each other, thus their covariance is fully determined by the variance of the upstream path scaled by the weights of its downstream \textit{subtrees}, i.e.~the sum of all paths starting at $X_m$ and ending on, e.g., failure nodes. By Eq.~\eqref{eq:var}, we have to consider all unique combinations of paths overlapping at this node which leads to the total contribution to the overlap variance from paths diverging at circuit nodes:
\begin{equation}
    2\sum_{i<j}\text{Cov}[P_i, P_j] = 2\sum_{i<j} \mathbb{E}\bigg[A_w^i\sum_k V_k^i\bigg] \mathbb{E}\bigg[A_w^j\sum_l W_l^j\bigg] \text{Var}[U_{i}]
    \label{eq:tot_X}
\end{equation}

In case paths split at subset nodes, there are always only two downstream paths. However, as the last common random variable in both paths are complements of each other, we obtain for the two paths $P_0=UX_m\sum_i V_i$ and $P_1=U(1-X_m)\sum_i W_i$ the covariance
\begin{equation}
    \text{Cov}[P_0, P_1] = \underbrace{\mathbb{E}\bigg[\sum_i V_i\bigg] \mathbb{E}\bigg[\sum_i W_i\bigg]}_{\text{subtree sums }\mathcal{S}_k} \Cov{U X_m, U(1-X_m)} = \mathbb{E}\bigg[\sum_i V_i\bigg] \mathbb{E}\bigg[\sum_i W_i\bigg]\big(\text{Var}[U]\mathbb{E}[X_m] - \text{Var}(UX_m)\big).
\end{equation}
Thus, the total contribution to the overlap variance for paths overlapping at a subset node is 
\begin{equation}
    2\sum_{i<j}\text{Cov}[P_i,P_j] = 2\sum_{i<j} \mathbb{E}\bigg[\sum_k V_k^i\bigg] \mathbb{E}\bigg[\sum_l W_l^j\bigg]\big(\text{Var}[U_i]\mathbb{E}[X_m^i] - \text{Var}(U_iX_m^i)\big).
    \label{eq:tot_A}
\end{equation}
In the python package \texttt{qsample} the calculation of variances of $p_L$ and $p_L + \delta$ is performed as described in pseudocode in Alg.~\ref{alg:var}. 

\begin{algorithm}[htp]
    \caption{Variance on lower bound $p_L$ (upper bound $p_L + \delta$) to failure rate}
    \begin{algorithmic}[1]
        \Statex \textbf{Input:} Sample tree $\mathcal{T}$
        \Statex \textbf{Output:} Variance $\text{Var}[p_L]$ ($\text{Var}[p_L+\delta]$)
            \State Get $\mathcal{P}=\mathcal{P}_f$ ($\mathcal{P}_f \cup \mathcal{P}_\delta$) from $\mathcal{T}$
            \State Calculate sum of path variances $V_1\gets\sum_{P\in\mathcal{P}} \text{Var}[P]$
            \State Get overlaps $O\gets\{O_i=\max_d{(P\cap P^\prime)} \forall P,P^\prime\in \mathcal{P}\}$ with node depth $d$
            \For{$O_i \in O$}
                \If{$O_i$ is a circuit node}
                    \State Calculate subtree sums $\mathcal{S}_w=A_w^m\sum_j \mathbb{E}[X_j^w]$ for $j>m$
                    \State Calculate upstream path variance $\text{Var}[O_i]$
\State $V_2^i\gets 2\text{Var}[O_i] \sum_{w<w^\prime} \mathcal{S}_w\mathcal{S}_{w^\prime}$ 

                \ElsIf{$O_i$ is a subset node}
                    \State Calculate subtree sums $\mathcal{S}_k=\sum_j \mathbb{E}\big[X^k_j\big]$ for $j>m$
                    \State Calculate upstream path variance $\text{Var}[O_i]$
                    \State Calculate path variance $\text{Var}[O_iX_m]$
                    \State Get expectation value $\mathbb{E}[X_m]$
\State $V_2^i\gets 2 (\text{Var}[O_i]\mathbb{E}[X_m]-\text{Var}[O_iX_m])  \prod_k \mathcal{S}_k$
                \EndIf
                \State $V_2\gets V_2 + V_2^i$
            \EndFor
        \State \Return $V_1 + V_2$
    \end{algorithmic}
    \label{alg:var}
\end{algorithm}

\subsection{Uncertainty intervals}\label{sec:wilson}
The sampling error for MC sampling can be estimated by the Wald interval
\begin{align}
    \varepsilon_\text{MC} = \sqrt{\frac{\hat{p}\left(1-\hat{p}\right)}{N}} \label{eq:stddev_MC}
\end{align}
so that for a large number of samples $N \rightarrow \infty$ the true failure rate $p^*$ is likely to be found in the 68\% confidence interval $\left[ \hat{p} - \varepsilon_\text{MC},\, \hat{p} + \varepsilon_\text{MC}\right]$. It is known that for $\hat{p}$ estimations that are close to or equal to zero or one after a finite but potentially small number of samples the Wald interval suffers from irregularities. These can be prevented using the Wilson score interval \cite{wilson1927probable} instead, which is bounded by
\begin{align}
    p_\pm &= \frac{1}{1+\frac{z_{\alpha/2}^2}{N}} \left( \hat{p} + \frac{z_{\alpha/2}^2}{2N} \pm z_{\alpha/2} \sqrt{ \frac{\hat{p}(1-\hat{p})}{N} + \frac{z_{\alpha/2}^2}{4N^2}} \right) \label{eq:wilsonint}
\end{align}
at confidence level $1-\alpha$ where $z$ is the quantile function of the normal distribution. The sampling error for all numerical simulations in \texttt{qsample} is given as the Wilson score interval \eqref{eq:wilsonint} at a confidence level of 68\% ($z_{\alpha/2} = 1$) in a symmetric form $[\hat{p} - \frac{p_+-p_-}{2}, \hat{p} + \frac{p_+-p_-}{2}]$.

\subsection{Selection criterion}\label{sec:choices}

\subsubsection{Binomial-factor-based choice}
The default procedure in \texttt{qsample} simply selects the subsets randomly based on the binomial distribution $A_w(p_\text{max})$ of subsets for each circuit at run-time. For each circuit, we draw a random number $r \in [0, 1]$ and choose the weight $w$ for which
\begin{align}
    \sum_{i=0}^{w-1} A_i \leq r \leq \sum_{i=0}^{w} A_i
    \label{eq:ss_sel_prob}
\end{align}
where the left sum is understood to be equal to zero if $w = 0$.

Furthermore, sampling of the weight-0 subset of a circuit can be prohibited if the circuit under fault-free conditions always yields the same deterministic outcome, for instance for a ``repeat until success'' protocol. In this case, the sums in Eq.~\eqref{eq:ss_sel_prob} should start from $i=1$ while each term is rescaled such that the overall subset selection probability remains unity.

\subsubsection{ERU choice}\label{sec:eruapp}
We implement the ERU criterion whose defining feature is that it always chooses the subset so that the next shot maximizes the expected reduction of the total uncertainty of $\hat{p}$. It balances the trade-off between continuing to sample in a subset that is already known -- thereby decreasing the sampling error --  or going to a new subset where, initially, we would end up with a relatively large sampling uncertainty but we get to reduce the cutoff error $\delta$. 

It requires calculation of the shifted transition rates $q_i^\pm$ as given by Eq.~\eqref{eq:shiftq}, where it is assumed that the next sample yields a $\pm$ measurement outcome. Let us assume that any $q_i$ is determined by 
\begin{align}
    q_i = \frac{m_{\C_i}}{N_{\C_i}}
\end{align}
where $m_{\C_i}$ counts the numbers of positive measurement outcomes and $N_{\C_i}$ is the total number of samples of the particular node $q_i$ belongs to. Then
\begin{align}
    q_i^+ &\equiv \frac{m_{\C_i}+1}{N_{\C_i}+1} = \frac{m_{\C_i} + 1}{N_{\C_i}} \frac{N_{\C_i}}{N_{\C_i} + 1} = \left( \frac{m_{\C_i}}{N_{\C_i}} + \frac{1}{N_{\C_i}} \right) \frac{N_{\C_i}}{N_{\C_i} + 1} = \left( q_i + \frac{1}{N_{\C_i}} \right) \frac{N_{\C_i}}{N_{\C_i} + 1} ~~~\text{and}\\
    q_i^- &\equiv \frac{m_{\C_i}}{N_{\C_i}+1} = q_i \frac{N_{\C_i}}{N_{\C_i} + 1}
\end{align}
as stated in Eq.~\eqref{eq:shiftq}.

\begin{figure}\centering
    \includegraphics[width=0.6\linewidth]{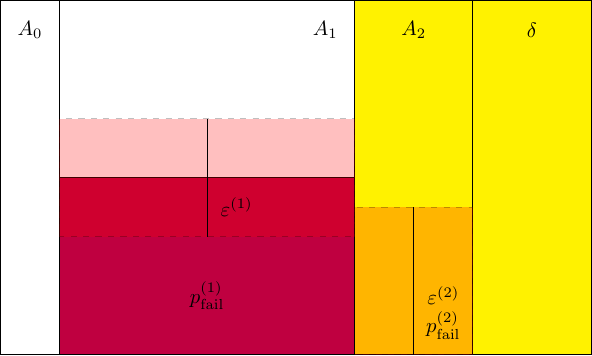}
    \caption{Box representation of different fault-weight-subsets for a protocol that consists of a single circuit with binomial factors $A_i$. The subset failure rate $p_\text{fail}^{(1)}$ has a sampling uncertainty $\varepsilon^{(1)}$. Opening the 2-fault-subset amounts to reducing the cutoff error $\delta$ by the binomial factor $A_2$ but in turn increasing the overall sampling uncertainty. This is because an initial estimate from few shots of $p_\text{fail}^{(2)}$ comes with a large uncertainty $\varepsilon^{(2)}$. (The 0-fault-subset is assumed to be the largest subset, but we draw it on the side for better visibility of the higher-weight subsets.)}
    \label{fig:fxd_boxes}
\end{figure}

The expected reduction of uncertainty (ERU) at a node $i$ is defined by Eq.~\eqref{eq:erv}. An illustration is given in Fig.~\ref{fig:fxd_boxes}. It depicts a known subset failure rate $p_\text{fail}^{(1)}$ in the 1-subset with the yellow marked cutoff error $\delta$, which includes $A_2$. $p_\text{fail}^{(1)}$ has a sampling uncertainty of $\varepsilon^{(1)}$ so that the total length of the uncertainty interval $\eta$ about $\hat{p}$, i.e.~the difference between upper and lower bound according to Eq.~\eqref{eq:eta}, is $2\varepsilon^{(1)} + \delta$ (Note that the estimator $\hat{p}$ is centered between upper and lower bound according to Eq.~\eqref{eq:dssestim} and that, since we only consider a single circuit, the sampling uncertainties of the upper and lower bound are identical so that $2\varepsilon^{(1)} = \sigma_L + \sigma_U$ in this example). If we were to spend the next shot in the 2-subset, the total uncertainty would change to read $2\varepsilon^{(1)} + 2\varepsilon^{(2)} + \delta - A_2$. Another shot in the 1-subset would just decrease $\varepsilon^{(1)}$ and keep $\delta$ unchanged. The choice that reduces the \emph{expected} total uncertainty the most, is the one that is taken according to the ERU criterion. In this example we have 
\begin{align}
    \Delta(2) &= \left(2\varepsilon^{(1)} + \delta \right) - \left[ p_\text{fail}^{(2)} \cdot \left(2(\varepsilon^{(1)} + \varepsilon^{(2)}|_+) + \delta - A_2 \right) + (1-p_\text{fail}^{(2)}) \cdot \left(2(\varepsilon^{(1)} + \varepsilon^{(2)}|_-) + \delta - A_2 \right) \right] = A_2 - 2\varepsilon^{(2)}|_+ \label{eq:delta2}\\
    \Delta(1) &= \left(2\varepsilon^{(1)} + \delta \right) - \left[ p_\text{fail}^{(1)} \cdot \left(2\varepsilon^{(1)}|_+ + \delta \right) + (1-p_\text{fail}^{(1)}) \cdot \left(2\varepsilon^{(1)}|_- +\delta \right) \right] = 2\varepsilon^{(1)} - \left[ p_\text{fail}^{(1)} \cdot 2\varepsilon^{(1)}|_+ + (1-p_\text{fail}^{(1)}) \cdot 2\varepsilon^{(1)}|_- \right].
\end{align}
Note that, although there is no current value for $p_\text{fail}^{(2)}$ yet, this is not a problem for evaluating Eq.~\eqref{eq:delta2} since we have $\varepsilon^{(2)}|_+ = \varepsilon^{(2)}|_-$, which is always the case when we open a new subset.

\begin{figure}[!htbp]
    \centering
    \includegraphics[width=0.6\linewidth]{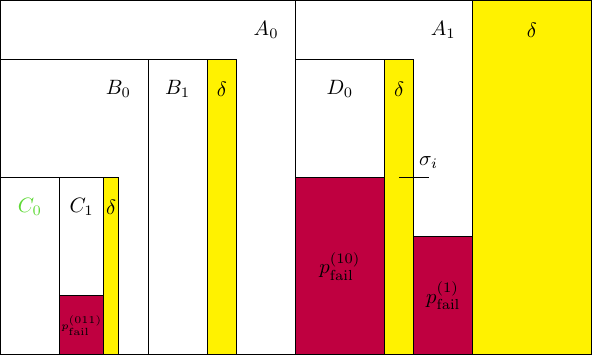}
    \caption{Circuit sequences can be represented by nested boxes as in Fig.~\ref{fig:introDSS}(c). A subbox can take up at most all the space provided by the enclosing box. The fault-free path has the total binomial factor $A_0B_0C_0$ here. Within the subbox labelled by the green $C_0$, no failure can occur (compare to the green path in Fig.~\ref{fig:introDSS}(a)).}
    \label{fig:dyn_boxes}
\end{figure}
\begin{figure}[!htbp]
    \centering
    \includegraphics[width=0.6\linewidth]{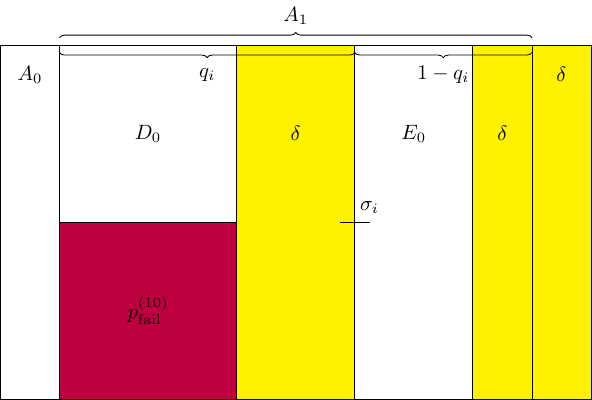}
    \caption{Subboxes of the $A_1$-box, labelled $D$ and $E$, are weighted by the branching ratio $q_i$ or $1-q_i$ respectively (cf.~Fig.~\ref{fig:introDSS}(c)). Since the branching ratio estimate $q_i$ has an uncertainty $\sigma_i$, the size of any subbox is not determined with total confidence.}
    \label{fig:dyn_boxes_qsplit}
\end{figure}

The more complicated situation of circuit sequences is illustrated in Fig.~\ref{fig:dyn_boxes}. When choosing a weight-subset at a given stage, we must consider the total uncertainty of all the contained boxes (and the cutoff error of the given stage). Choosing 0 in the outermost box, for example, will be done under consideration of $p_\text{fail}^{(011)}$ and contributions to $\delta$ from all stages. For choosing the 1-subset, we must consider the total uncertainty of failure in all subboxes of $A_1$. This especially now includes the uncertainty about the size of the subboxes, as indicated by $\sigma_i$, the uncertainty of the branching ratio $q_i$ that leads from $A_1$ to the circuits with binomial factors $D_w$. Note that this uncertainty does not exist for the 0-subsets since the fault-free path is deterministic. The splitting is further illustrated in Fig.~\ref{fig:dyn_boxes_qsplit}.

When probing a new subset, i.e.~reducing the total cutoff error, for the calculation of the ERU, it is reasonable to assume that a newly opened subset will lead to failure with probability $1/2$. Since nothing is known about the subtree below this subset node and we do not get a measurement outcome that would reveal some new information about the tree, we can only assume failure with a finite probability, e.g.~$1/2$. Note that while assuming a failure probability of $0$ is unrealistic for a non-FT path, assuming a failure probability of $1$ would cause the cutoff error to not change at all compared to the current tree value. This is different from the single-circuit case discussed above since there we can immediately evaluate the effect of the failure or no-failure outcome after choosing a subset, which cannot be done for a circuit sequence. It is not strictly necessary to assume an initial failure rate of $1/2$ but any finite value is fine. It could, for example, be set as the current average value of all subset failure rates $\frac{1}{N_w} \sum_w \pfail^{(w)}$. This is not yet implemented in \texttt{qsample}.

\section{Noise model}\label{sec:noise}

For the examples that we show in Sec.~\ref{sec:examples}, we employ circuit-level depolarizing noise. We apply the fault-operators
\begin{align}
	F_1 &\in \{\sigma_i, \forall i \in [1,2,3]\} \\
	F_2 &\in \{\sigma_i \otimes \sigma_j, \forall i,j \in [0,1,2,3]\}~\backslash~I \otimes I \label{eq:error2}
\end{align}
where $\sigma_i$ are the Pauli matrices and $\sigma_0 = I$ is the identity operation. Single-qubit faults $F_1$ are applied uniformly with probability $p_1/3$ and are not applied with probability $1-p_1$. The two-qubit faults $F_2$ are applied uniformly with probability $p_2/15$ and are not applied with probability $1-p_2$. This means that 
\begin{enumerate}
    \item a single-qubit gate is followed by a Pauli fault $F_1$ drawn uniformly and independently from $\{X,Y,Z\}$ with probability $p_1/3$,
    \item a two-qubit gate is followed by a two-Pauli fault $F_2$ drawn uniformly and independently from $\{I,X,Y,Z\}^{\otimes 2} \backslash I \otimes I$ with probability $p_2/15$,
    \item qubit initialization is flipped (e.g.~$\ket{0} \rightarrow \ket{1},\,\ket{+} \rightarrow \ket{-}$) with probability $2p_1/3$ and
    \item qubit measurements yield a flipped result ($\pm 1 \rightarrow \mp 1$) with probability $2p_1/3$.\end{enumerate}

\section{Tighter bounds for flag-FT stabilizer measurements}\label{sec:app_flagstab2}

\begin{figure}\centering\includegraphics[width=0.7\linewidth]{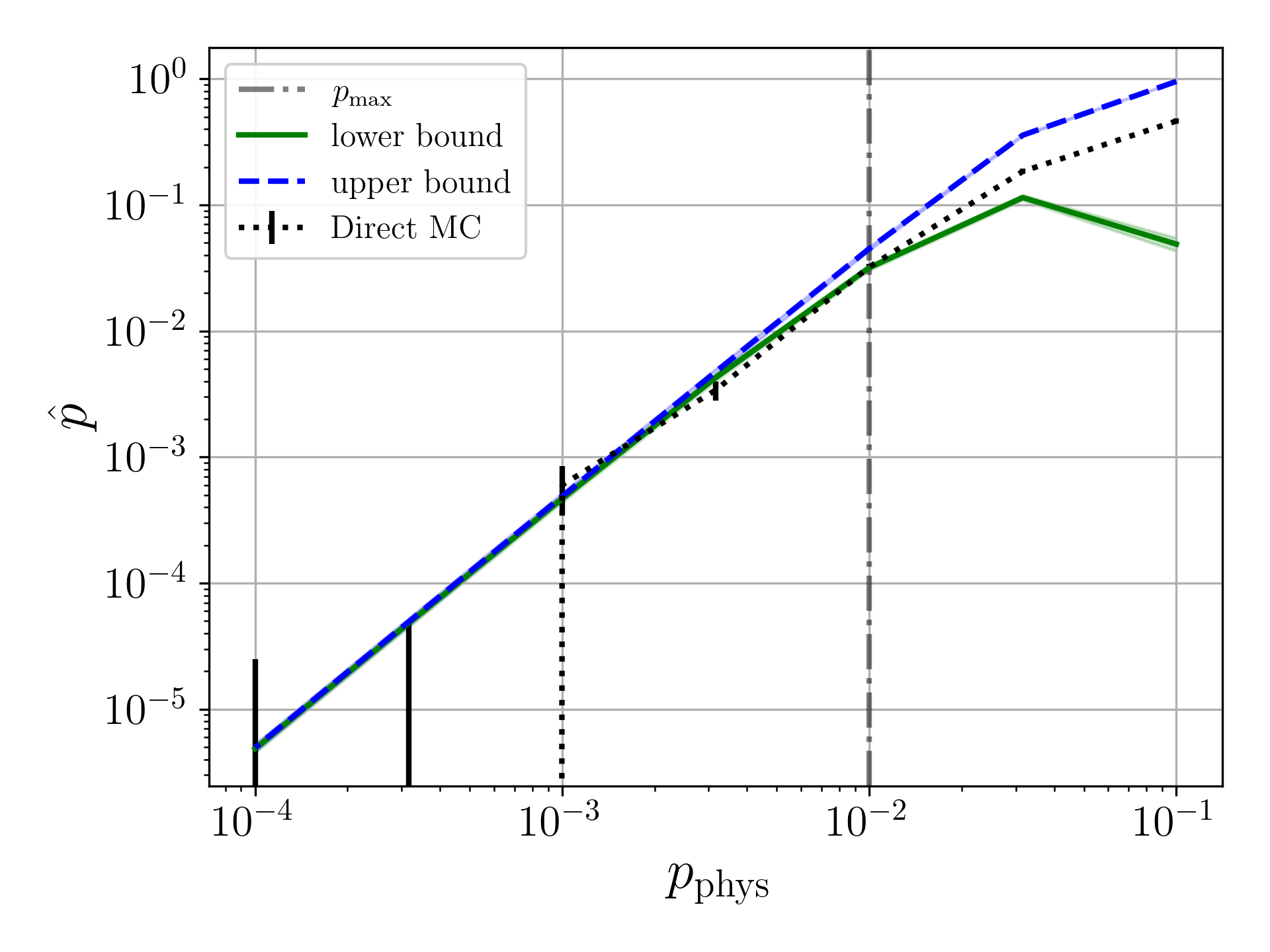}
	\caption{Improved bounds for the logical failure rate estimation for the state preparation protocol via flag-FT stabilizer measurements. The only difference to Fig.~\ref{fig:flag_scaling} is that we sampled at an increased $p_\text{max} = 10^{-2}$ (gray vertical dash-dotted line). Here, the cutoff error vanishes in the limit $\pphy\rightarrow 0$ since $10^4$ shots with DSS and binomial factor choice suffice to sample all relevant subsets.}
	\label{fig:flag_scaling_2}
\end{figure}

The relatively large cutoff error that we found for the flag-FT stabilizer measurement example protocol in Sec.~\ref{sec:examples_flag} can be reduced in practice without changing the dynamical subset sampling algorithm. In Fig.~\ref{fig:flag_scaling_2} we show a much smaller cutoff error that we obtain by increasing the maximal physical fault rate to $p_\text{max} = 10^{-2}$ instead of sampling at $p_\text{max} = 10^{-3}$. While the cutoff error remains large at $p_\text{max} = 10^{-2}$, this increase has the effect that more different fault-weight-subsets are chosen by the binomial factor choice criterion. As a consequence, the space of possible paths is more thoroughly explored and the bounds to the true protocol failure rate tighten as we lower $\pphy$, in agreement with the direct MC data points, which have a growing uncertainty interval as in Fig.~\ref{fig:flag_scaling}.

\twocolumngrid 
\clearpage 
\bibliographystyle{bibstyle}
\bibliography{references}

\begin{thebibliography}{70}%
\makeatletter
\providecommand \@ifxundefined [1]{%
 \@ifx{#1\undefined}
}%
\providecommand \@ifnum [1]{%
 \ifnum #1\expandafter \@firstoftwo
 \else \expandafter \@secondoftwo
 \fi
}%
\providecommand \@ifx [1]{%
 \ifx #1\expandafter \@firstoftwo
 \else \expandafter \@secondoftwo
 \fi
}%
\providecommand \natexlab [1]{#1}%
\providecommand \enquote  [1]{``#1''}%
\providecommand \bibnamefont  [1]{#1}%
\providecommand \bibfnamefont [1]{#1}%
\providecommand \citenamefont [1]{#1}%
\providecommand \href@noop [0]{\@secondoftwo}%
\providecommand \href [0]{\begingroup \@sanitize@url \@href}%
\providecommand \@href[1]{\@@startlink{#1}\@@href}%
\providecommand \@@href[1]{\endgroup#1\@@endlink}%
\providecommand \@sanitize@url [0]{\catcode `\\12\catcode `\$12\catcode
  `\&12\catcode `\#12\catcode `\^12\catcode `\_12\catcode `\%12\relax}%
\providecommand \@@startlink[1]{}%
\providecommand \@@endlink[0]{}%
\providecommand \url  [0]{\begingroup\@sanitize@url \@url }%
\providecommand \@url [1]{\endgroup\@href {#1}{\urlprefix }}%
\providecommand \urlprefix  [0]{URL }%
\providecommand \Eprint [0]{\href }%
\providecommand \doibase [0]{https://doi.org/}%
\providecommand \selectlanguage [0]{\@gobble}%
\providecommand \bibinfo  [0]{\@secondoftwo}%
\providecommand \bibfield  [0]{\@secondoftwo}%
\providecommand \translation [1]{[#1]}%
\providecommand \BibitemOpen [0]{}%
\providecommand \bibitemStop [0]{}%
\providecommand \bibitemNoStop [0]{.\EOS\space}%
\providecommand \EOS [0]{\spacefactor3000\relax}%
\providecommand \BibitemShut  [1]{\csname bibitem#1\endcsname}%
\let\auto@bib@innerbib\@empty
\bibitem [{\citenamefont {Preskill}(2018)}]{preskill2018quantum}%
  \BibitemOpen
  \bibfield  {author} {\bibinfo {author} {J.~Preskill},\ }\emph {Quantum
  computing in the {NISQ} era and beyond},\ \href
  {https://doi.org/https://doi.org/10.22331/q-2018-08-06-79} {\bibfield
  {journal} {\bibinfo  {journal} {Quantum}\ }\textbf {\bibinfo {volume} {2}},\
  \bibinfo {pages} {79} (\bibinfo {year} {2018})}\BibitemShut {NoStop}%
\bibitem [{\citenamefont {Briegel}\ \emph {et~al.}(2009)\citenamefont
  {Briegel}, \citenamefont {Browne}, \citenamefont {D{\"u}r}, \citenamefont
  {Raussendorf},\ and\ \citenamefont {Van~den Nest}}]{briegel2009measurement}%
  \BibitemOpen
  \bibfield  {author} {\bibinfo {author} {H.~J. Briegel}, \bibinfo {author}
  {D.~E. Browne}, \bibinfo {author} {W.~D{\"u}r}, \bibinfo {author}
  {R.~Raussendorf},\ and\ \bibinfo {author} {M.~Van~den Nest},\ }\emph
  {Measurement-based quantum computation},\ \href
  {https://doi.org/https://doi.org/10.1038/nphys1157} {\bibfield  {journal}
  {\bibinfo  {journal} {Nature Physics}\ }\textbf {\bibinfo {volume} {5}},\
  \bibinfo {pages} {19} (\bibinfo {year} {2009})}\BibitemShut {NoStop}%
\bibitem [{\citenamefont {Jozsa}(2006)}]{jozsa2006introduction}%
  \BibitemOpen
  \bibfield  {author} {\bibinfo {author} {R.~Jozsa},\ }\emph {An introduction
  to measurement based quantum computation},\ \href
  {https://doi.org/https://doi.org/10.48550/arXiv.quant-ph/0508124} {\bibfield
  {journal} {\bibinfo  {journal} {NATO Science Series, III: Computer and
  Systems Sciences. Quantum Information Processing-From Theory to Experiment}\
  }\textbf {\bibinfo {volume} {199}},\ \bibinfo {pages} {137} (\bibinfo {year}
  {2006})}\BibitemShut {NoStop}%
\bibitem [{\citenamefont {Duan}\ and\ \citenamefont
  {Monroe}(2010)}]{duan2010colloquium}%
  \BibitemOpen
  \bibfield  {author} {\bibinfo {author} {L.-M. Duan}\ and\ \bibinfo {author}
  {C.~Monroe},\ }\emph {Colloquium: Quantum networks with trapped ions},\ \href
  {https://doi.org/10.1103/RevModPhys.82.1209} {\bibfield  {journal} {\bibinfo
  {journal} {Reviews of Modern Physics}\ }\textbf {\bibinfo {volume} {82}},\
  \bibinfo {pages} {1209} (\bibinfo {year} {2010})}\BibitemShut {NoStop}%
\bibitem [{\citenamefont {Kozlowski}\ and\ \citenamefont
  {Wehner}(2019)}]{kozlowski2019towards}%
  \BibitemOpen
  \bibfield  {author} {\bibinfo {author} {W.~Kozlowski}\ and\ \bibinfo {author}
  {S.~Wehner},\ }in\ \href
  {https://doi.org/https://doi.org/10.1145/3345312.3345497} {\emph {\bibinfo
  {booktitle} {Proceedings of the Sixth Annual ACM International Conference on
  Nanoscale Computing and Communication}}}\ (\bibinfo {year} {2019})\
  p.~\bibinfo {pages} {1}\BibitemShut {NoStop}%
\bibitem [{\citenamefont {Bravyi}\ and\ \citenamefont
  {Vargo}(2013)}]{bravyi2013simulation}%
  \BibitemOpen
  \bibfield  {author} {\bibinfo {author} {S.~Bravyi}\ and\ \bibinfo {author}
  {A.~Vargo},\ }\emph {Simulation of rare events in quantum error correction},\
  \href {https://doi.org/10.1103/PhysRevA.88.062308} {\bibfield  {journal}
  {\bibinfo  {journal} {Physical Review A}\ }\textbf {\bibinfo {volume} {88}},\
  \bibinfo {pages} {062308} (\bibinfo {year} {2013})}\BibitemShut {NoStop}%
\bibitem [{\citenamefont {Steane}(2003)}]{steane2003overhead}%
  \BibitemOpen
  \bibfield  {author} {\bibinfo {author} {A.~M. Steane},\ }\emph {Overhead and
  noise threshold of fault-tolerant quantum error correction},\ \href
  {https://doi.org/10.1103/PhysRevA.68.042322} {\bibfield  {journal} {\bibinfo
  {journal} {Physical Review A}\ }\textbf {\bibinfo {volume} {68}},\ \bibinfo
  {pages} {042322} (\bibinfo {year} {2003})}\BibitemShut {NoStop}%
\bibitem [{\citenamefont {Raussendorf}\ \emph {et~al.}(2007)\citenamefont
  {Raussendorf}, \citenamefont {Harrington},\ and\ \citenamefont
  {Goyal}}]{raussendorf2007topological}%
  \BibitemOpen
  \bibfield  {author} {\bibinfo {author} {R.~Raussendorf}, \bibinfo {author}
  {J.~Harrington},\ and\ \bibinfo {author} {K.~Goyal},\ }\emph {Topological
  fault-tolerance in cluster state quantum computation},\ \href
  {https://doi.org/https://doi.org/10.1088/1367-2630/9/6/199} {\bibfield
  {journal} {\bibinfo  {journal} {New Journal of Physics}\ }\textbf {\bibinfo
  {volume} {9}},\ \bibinfo {pages} {199} (\bibinfo {year} {2007})}\BibitemShut
  {NoStop}%
\bibitem [{\citenamefont {Rynbach}\ \emph {et~al.}(2012)\citenamefont
  {Rynbach}, \citenamefont {Muhammad}, \citenamefont {Mehta}, \citenamefont
  {Hussmann},\ and\ \citenamefont {Kim}}]{van2012quantum}%
  \BibitemOpen
  \bibfield  {author} {\bibinfo {author} {A.~V. Rynbach}, \bibinfo {author}
  {A.~Muhammad}, \bibinfo {author} {A.~C. Mehta}, \bibinfo {author}
  {J.~Hussmann},\ and\ \bibinfo {author} {J.~Kim},\ }\href@noop {} {\emph {A
  Quantum Performance Simulator based on fidelity and fault-path counting}}
  (\bibinfo {year} {2012}),\ \Eprint {https://arxiv.org/abs/1212.0845}
  {arXiv:1212.0845 [quant-ph]} \BibitemShut {NoStop}%
\bibitem [{\citenamefont {Fowler}(2013)}]{fowler2013analytic}%
  \BibitemOpen
  \bibfield  {author} {\bibinfo {author} {A.~G. Fowler},\ }\emph {Analytic
  asymptotic performance of topological codes},\ \href
  {https://doi.org/10.1103/PhysRevA.87.040301} {\bibfield  {journal} {\bibinfo
  {journal} {Physical Review A}\ }\textbf {\bibinfo {volume} {87}},\ \bibinfo
  {pages} {040301} (\bibinfo {year} {2013})}\BibitemShut {NoStop}%
\bibitem [{\citenamefont {Li}\ \emph {et~al.}(2017)\citenamefont {Li},
  \citenamefont {Guti{\'e}rrez}, \citenamefont {David}, \citenamefont
  {Hernandez},\ and\ \citenamefont {Brown}}]{li2017fault}%
  \BibitemOpen
  \bibfield  {author} {\bibinfo {author} {M.~Li}, \bibinfo {author}
  {M.~Guti{\'e}rrez}, \bibinfo {author} {S.~E. David}, \bibinfo {author}
  {A.~Hernandez},\ and\ \bibinfo {author} {K.~R. Brown},\ }\emph {Fault
  tolerance with bare ancillary qubits for a [[7, 1, 3]] code},\ \href
  {https://doi.org/10.1103/PhysRevA.96.032341} {\bibfield  {journal} {\bibinfo
  {journal} {Physical Review A}\ }\textbf {\bibinfo {volume} {96}},\ \bibinfo
  {pages} {032341} (\bibinfo {year} {2017})}\BibitemShut {NoStop}%
\bibitem [{\citenamefont {Trout}\ \emph {et~al.}(2018)\citenamefont {Trout},
  \citenamefont {Li}, \citenamefont {Guti{\'e}rrez}, \citenamefont {Wu},
  \citenamefont {Wang}, \citenamefont {Duan},\ and\ \citenamefont
  {Brown}}]{trout2018simulating}%
  \BibitemOpen
  \bibfield  {author} {\bibinfo {author} {C.~J. Trout}, et~al.,\ }\emph
  {Simulating the performance of a distance-3 surface code in a linear ion
  trap},\ \href {https://doi.org/10.1088/1367-2630/aab341} {\bibfield
  {journal} {\bibinfo  {journal} {New Journal of Physics}\ }\textbf {\bibinfo
  {volume} {20}},\ \bibinfo {pages} {043038} (\bibinfo {year}
  {2018})}\BibitemShut {NoStop}%
\bibitem [{\citenamefont {Guti{\'e}rrez}\ \emph {et~al.}(2019)\citenamefont
  {Guti{\'e}rrez}, \citenamefont {M{\"u}ller},\ and\ \citenamefont
  {Berm{\'u}dez}}]{gutierrez2019transversality}%
  \BibitemOpen
  \bibfield  {author} {\bibinfo {author} {M.~Guti{\'e}rrez}, \bibinfo {author}
  {M.~M{\"u}ller},\ and\ \bibinfo {author} {A.~Berm{\'u}dez},\ }\emph
  {Transversality and lattice surgery: Exploring realistic routes toward
  coupled logical qubits with trapped-ion quantum processors},\ \href
  {https://doi.org/10.1103/PhysRevA.99.022330} {\bibfield  {journal} {\bibinfo
  {journal} {Physical Review A}\ }\textbf {\bibinfo {volume} {99}},\ \bibinfo
  {pages} {022330} (\bibinfo {year} {2019})}\BibitemShut {NoStop}%
\bibitem [{\citenamefont {Li}(2020)}]{li2020fault}%
  \BibitemOpen
  \bibfield  {author} {\bibinfo {author} {M.~Li},\ }\emph {\bibinfo {title}
  {Fault-tolerance on near-term quantum computers and subsystem quantum error
  correcting codes}},\ \href {http://hdl.handle.net/1853/62750} {Ph.D.
  thesis},\ \bibinfo  {school} {Georgia Institute of Technology} (\bibinfo
  {year} {2020})\BibitemShut {NoStop}%
\bibitem [{\citenamefont {Paetznick}\ \emph {et~al.}(2023)\citenamefont
  {Paetznick}, \citenamefont {Knapp}, \citenamefont {Delfosse}, \citenamefont
  {Bauer}, \citenamefont {Haah}, \citenamefont {Hastings},\ and\ \citenamefont
  {da~Silva}}]{paetznick2022performance}%
  \BibitemOpen
  \bibfield  {author} {\bibinfo {author} {A.~Paetznick}, et~al.,\ }\emph
  {Performance of planar Floquet codes with Majorana-based qubits},\ \href
  {https://doi.org/10.1103/PRXQuantum.4.010310} {\bibfield  {journal} {\bibinfo
   {journal} {PRX Quantum}\ }\textbf {\bibinfo {volume} {4}},\ \bibinfo {pages}
  {010310} (\bibinfo {year} {2023})}\BibitemShut {NoStop}%
\bibitem [{\citenamefont {Heu{\ss}en}\ \emph {et~al.}(2023)\citenamefont
  {Heu{\ss}en}, \citenamefont {Postler}, \citenamefont {Rispler}, \citenamefont
  {Pogorelov}, \citenamefont {Marciniak}, \citenamefont {Monz}, \citenamefont
  {Schindler},\ and\ \citenamefont {M{\"u}ller}}]{heussen2023strategies}%
  \BibitemOpen
  \bibfield  {author} {\bibinfo {author} {S.~Heu{\ss}en}, et~al.,\ }\emph
  {Strategies for a practical advantage of fault-tolerant circuit design in
  noisy trapped-ion quantum computers},\ \href
  {https://doi.org/10.1103/PhysRevA.107.042422} {\bibfield  {journal} {\bibinfo
   {journal} {Physical Review A}\ }\textbf {\bibinfo {volume} {107}},\ \bibinfo
  {pages} {042422} (\bibinfo {year} {2023})}\BibitemShut {NoStop}%
\bibitem [{\citenamefont {Terhal}(2015)}]{terhal2015quantum}%
  \BibitemOpen
  \bibfield  {author} {\bibinfo {author} {B.~M. Terhal},\ }\emph {Quantum error
  correction for quantum memories},\ \href
  {https://doi.org/10.1103/RevModPhys.87.307} {\bibfield  {journal} {\bibinfo
  {journal} {Reviews of Modern Physics}\ }\textbf {\bibinfo {volume} {87}},\
  \bibinfo {pages} {307} (\bibinfo {year} {2015})}\BibitemShut {NoStop}%
\bibitem [{\citenamefont {Winter}\ and\ \citenamefont
  {Heu{\ss}en}(2023)}]{qsample}%
  \BibitemOpen
  \bibfield  {author} {\bibinfo {author} {D.~Winter}\ and\ \bibinfo {author}
  {S.~Heu{\ss}en},\ }\href@noop {} {\emph {qsample}},\ \bibinfo {howpublished}
  {\url{https://github.com/dpwinter/qsample}} (\bibinfo {year}
  {2023})\BibitemShut {NoStop}%
\bibitem [{\citenamefont {Greenberger}\ \emph {et~al.}(1989)\citenamefont
  {Greenberger}, \citenamefont {Horne},\ and\ \citenamefont
  {Zeilinger}}]{greenberger1989going}%
  \BibitemOpen
  \bibfield  {author} {\bibinfo {author} {D.~M. Greenberger}, \bibinfo {author}
  {M.~A. Horne},\ and\ \bibinfo {author} {A.~Zeilinger},\ }in\ \href
  {https://doi.org/https://doi.org/10.1007/978-94-017-0849-4} {\emph {\bibinfo
  {booktitle} {Bell’s theorem, quantum theory and conceptions of the
  universe}}}\ (\bibinfo  {publisher} {Springer},\ \bibinfo {year} {1989})\
  p.~\bibinfo {pages} {69}\BibitemShut {NoStop}%
\bibitem [{\citenamefont {Shor}(1996)}]{shor1996fault}%
  \BibitemOpen
  \bibfield  {author} {\bibinfo {author} {P.~W. Shor},\ }in\ \href
  {https://doi.org/10.1109/SFCS.1996.548464} {\emph {\bibinfo {booktitle}
  {Proceedings of 37th Conference on Foundations of Computer Science}}}\
  (\bibinfo  {publisher} {IEEE},\ \bibinfo {year} {1996})\ p.~\bibinfo {pages}
  {56}\BibitemShut {NoStop}%
\bibitem [{\citenamefont {Steane}(1996{\natexlab{a}})}]{steane1996multiple}%
  \BibitemOpen
  \bibfield  {author} {\bibinfo {author} {A.~Steane},\ }\emph
  {{Multiple-particle interference and quantum error correction}},\ \href
  {https://doi.org/10.1098/rspa.1996.0136} {\bibfield  {journal} {\bibinfo
  {journal} {Proceedings of the Royal Society of London. Series A:
  Mathematical, Physical and Engineering Sciences}\ }\textbf {\bibinfo {volume}
  {452}},\ \bibinfo {pages} {2551} (\bibinfo {year}
  {1996}{\natexlab{a}})}\BibitemShut {NoStop}%
\bibitem [{\citenamefont {Steane}(1996{\natexlab{b}})}]{steane1996error}%
  \BibitemOpen
  \bibfield  {author} {\bibinfo {author} {A.~M. Steane},\ }\emph {Error
  correcting codes in quantum theory},\ \href
  {https://doi.org/10.1103/PhysRevLett.77.793} {\bibfield  {journal} {\bibinfo
  {journal} {Physical Review Letters}\ }\textbf {\bibinfo {volume} {77}},\
  \bibinfo {pages} {793} (\bibinfo {year} {1996}{\natexlab{b}})}\BibitemShut
  {NoStop}%
\bibitem [{\citenamefont {Goto}(2016)}]{goto2016minimizing}%
  \BibitemOpen
  \bibfield  {author} {\bibinfo {author} {H.~Goto},\ }\emph {Minimizing
  resource overheads for fault-tolerant preparation of encoded states of the
  Steane code},\ \href {https://doi.org/10.1038/srep19578} {\bibfield
  {journal} {\bibinfo  {journal} {Scientific Reports}\ }\textbf {\bibinfo
  {volume} {6}},\ \bibinfo {pages} {1} (\bibinfo {year} {2016})}\BibitemShut
  {NoStop}%
\bibitem [{\citenamefont {Chao}\ and\ \citenamefont
  {Reichardt}(2018{\natexlab{a}})}]{chao2018quantum}%
  \BibitemOpen
  \bibfield  {author} {\bibinfo {author} {R.~Chao}\ and\ \bibinfo {author}
  {B.~W. Reichardt},\ }\emph {{Quantum error correction with only two extra
  qubits}},\ \href {https://doi.org/10.1103/PhysRevLett.121.050502} {\bibfield
  {journal} {\bibinfo  {journal} {Physical Review Letters}\ }\textbf {\bibinfo
  {volume} {121}},\ \bibinfo {pages} {050502} (\bibinfo {year}
  {2018}{\natexlab{a}})}\BibitemShut {NoStop}%
\bibitem [{\citenamefont {Chamberland}\ and\ \citenamefont
  {Beverland}(2018)}]{chamberland2018flag}%
  \BibitemOpen
  \bibfield  {author} {\bibinfo {author} {C.~Chamberland}\ and\ \bibinfo
  {author} {M.~E. Beverland},\ }\emph {{Flag fault-tolerant error correction
  with arbitrary distance codes}},\ \href
  {https://doi.org/10.22331/q-2018-02-08-53} {\bibfield  {journal} {\bibinfo
  {journal} {Quantum}\ }\textbf {\bibinfo {volume} {2}},\ \bibinfo {pages} {53}
  (\bibinfo {year} {2018})}\BibitemShut {NoStop}%
\bibitem [{\citenamefont {Ryan-Anderson}\ \emph {et~al.}(2021)\citenamefont
  {Ryan-Anderson}, \citenamefont {Bohnet}, \citenamefont {Lee}, \citenamefont
  {Gresh}, \citenamefont {Hankin}, \citenamefont {Gaebler}, \citenamefont
  {Francois}, \citenamefont {Chernoguzov}, \citenamefont {Lucchetti},
  \citenamefont {Brown}, \citenamefont {Gatterman}, \citenamefont {Halit},
  \citenamefont {Gilmore}, \citenamefont {Gerber}, \citenamefont {Neyenhuis},
  \citenamefont {Hayes},\ and\ \citenamefont {Stutz}}]{ryan2021realization}%
  \BibitemOpen
  \bibfield  {author} {\bibinfo {author} {C.~Ryan-Anderson}, et~al.,\ }\emph
  {{Realization of real-time fault-tolerant quantum error correction}},\ \href
  {https://doi.org/10.1103/PhysRevX.11.041058} {\bibfield  {journal} {\bibinfo
  {journal} {Physical Review X}\ }\textbf {\bibinfo {volume} {11}},\ \bibinfo
  {pages} {041058} (\bibinfo {year} {2021})}\BibitemShut {NoStop}%
\bibitem [{\citenamefont {Postler}\ \emph {et~al.}(2022)\citenamefont
  {Postler}, \citenamefont {Heu{\ss}en}, \citenamefont {Pogorelov},
  \citenamefont {Rispler}, \citenamefont {Feldker}, \citenamefont {Meth},
  \citenamefont {Marciniak}, \citenamefont {Stricker}, \citenamefont
  {Ringbauer}, \citenamefont {Blatt}, \citenamefont {Schindler}, \citenamefont
  {Müller},\ and\ \citenamefont {Monz}}]{postler2021demonstration}%
  \BibitemOpen
  \bibfield  {author} {\bibinfo {author} {L.~Postler}, et~al.,\ }\emph
  {Demonstration of fault-tolerant universal quantum gate operations},\ \href
  {https://doi.org/10.1038/s41586-022-04721-1} {\bibfield  {journal} {\bibinfo
  {journal} {Nature}\ }\textbf {\bibinfo {volume} {605}},\ \bibinfo {pages}
  {675} (\bibinfo {year} {2022})}\BibitemShut {NoStop}%
\bibitem [{\citenamefont {Brown}\ \emph {et~al.}(2001)\citenamefont {Brown},
  \citenamefont {Cai},\ and\ \citenamefont {DasGupta}}]{brown2001interval}%
  \BibitemOpen
  \bibfield  {author} {\bibinfo {author} {L.~D. Brown}, \bibinfo {author}
  {T.~T. Cai},\ and\ \bibinfo {author} {A.~DasGupta},\ }\emph {Interval
  estimation for a binomial proportion},\ \href
  {https://www.jstor.org/stable/2676784} {\bibfield  {journal} {\bibinfo
  {journal} {Statistical Science}\ }\textbf {\bibinfo {volume} {16}},\ \bibinfo
  {pages} {101} (\bibinfo {year} {2001})}\BibitemShut {NoStop}%
\bibitem [{\citenamefont {Dauphin}\ \emph {et~al.}(2014)\citenamefont
  {Dauphin}, \citenamefont {M{\"u}ller},\ and\ \citenamefont
  {Martin-Delgado}}]{dauphin2014efficient}%
  \BibitemOpen
  \bibfield  {author} {\bibinfo {author} {A.~Dauphin}, \bibinfo {author}
  {M.~M{\"u}ller},\ and\ \bibinfo {author} {M.~A. Martin-Delgado},\ }\emph
  {Efficient algorithm to compute the Berry conductivity},\ \href
  {https://doi.org/10.1088/1367-2630/16/7/073016} {\bibfield  {journal}
  {\bibinfo  {journal} {New Journal of Physics}\ }\textbf {\bibinfo {volume}
  {16}},\ \bibinfo {pages} {073016} (\bibinfo {year} {2014})}\BibitemShut
  {NoStop}%
\bibitem [{\citenamefont {Wilson}(1927)}]{wilson1927probable}%
  \BibitemOpen
  \bibfield  {author} {\bibinfo {author} {E.~B. Wilson},\ }\emph {Probable
  inference, the law of succession, and statistical inference},\ \href
  {https://doi.org/10.2307/2276774} {\bibfield  {journal} {\bibinfo  {journal}
  {Journal of the American Statistical Association}\ }\textbf {\bibinfo
  {volume} {22}},\ \bibinfo {pages} {209} (\bibinfo {year} {1927})}\BibitemShut
  {NoStop}%
\bibitem [{\citenamefont {DiVincenzo}\ and\ \citenamefont
  {Shor}(1996)}]{divincenzo1996fault}%
  \BibitemOpen
  \bibfield  {author} {\bibinfo {author} {D.~P. DiVincenzo}\ and\ \bibinfo
  {author} {P.~W. Shor},\ }\emph {Fault-tolerant error correction with
  efficient quantum codes},\ \href
  {https://doi.org/10.1103/PhysRevLett.77.3260} {\bibfield  {journal} {\bibinfo
   {journal} {Physical Review Letters}\ }\textbf {\bibinfo {volume} {77}},\
  \bibinfo {pages} {3260} (\bibinfo {year} {1996})}\BibitemShut {NoStop}%
\bibitem [{\citenamefont {Plenio}\ \emph {et~al.}(1997)\citenamefont {Plenio},
  \citenamefont {Vedral},\ and\ \citenamefont
  {Knight}}]{plenio1997conditional}%
  \BibitemOpen
  \bibfield  {author} {\bibinfo {author} {M.~Plenio}, \bibinfo {author}
  {V.~Vedral},\ and\ \bibinfo {author} {P.~Knight},\ }\emph {Conditional
  generation of error syndromes in fault-tolerant error correction},\ \href
  {https://doi.org/10.1103/PhysRevA.55.4593} {\bibfield  {journal} {\bibinfo
  {journal} {Physical Review A}\ }\textbf {\bibinfo {volume} {55}},\ \bibinfo
  {pages} {4593} (\bibinfo {year} {1997})}\BibitemShut {NoStop}%
\bibitem [{\citenamefont {Devitt}\ \emph {et~al.}(2013)\citenamefont {Devitt},
  \citenamefont {Munro},\ and\ \citenamefont {Nemoto}}]{devitt2013quantum}%
  \BibitemOpen
  \bibfield  {author} {\bibinfo {author} {S.~J. Devitt}, \bibinfo {author}
  {W.~J. Munro},\ and\ \bibinfo {author} {K.~Nemoto},\ }\emph {Quantum error
  correction for beginners},\ \href
  {https://doi.org/10.1088/0034-4885/76/7/076001} {\bibfield  {journal}
  {\bibinfo  {journal} {Reports on Progress in Physics}\ }\textbf {\bibinfo
  {volume} {76}},\ \bibinfo {pages} {076001} (\bibinfo {year}
  {2013})}\BibitemShut {NoStop}%
\bibitem [{\citenamefont {Aharonov}\ and\ \citenamefont
  {Ben-Or}(2008)}]{Aharonov2008}%
  \BibitemOpen
  \bibfield  {author} {\bibinfo {author} {D.~Aharonov}\ and\ \bibinfo {author}
  {M.~Ben-Or},\ }\emph {Fault-tolerant quantum computation with constant error
  rate},\ \href {https://doi.org/10.1137/S0097539799359385} {\bibfield
  {journal} {\bibinfo  {journal} {SIAM Journal on Computing}\ }\textbf
  {\bibinfo {volume} {38}},\ \bibinfo {pages} {1207--1282} (\bibinfo {year}
  {2008})}\BibitemShut {NoStop}%
\bibitem [{\citenamefont {Aaronson}\ and\ \citenamefont
  {Gottesman}(2004)}]{aaronson2004improved}%
  \BibitemOpen
  \bibfield  {author} {\bibinfo {author} {S.~Aaronson}\ and\ \bibinfo {author}
  {D.~Gottesman},\ }\emph {Improved simulation of stabilizer circuits},\ \href
  {https://doi.org/10.1103/PhysRevA.70.052328} {\bibfield  {journal} {\bibinfo
  {journal} {Physical Review A}\ }\textbf {\bibinfo {volume} {70}},\ \bibinfo
  {pages} {052328} (\bibinfo {year} {2004})}\BibitemShut {NoStop}%
\bibitem [{\citenamefont {Gidney}(2021)}]{gidney2021stim}%
  \BibitemOpen
  \bibfield  {author} {\bibinfo {author} {C.~Gidney},\ }\emph {Stim: a fast
  stabilizer circuit simulator},\ \href
  {https://doi.org/10.22331/q-2021-07-06-497} {\bibfield  {journal} {\bibinfo
  {journal} {Quantum}\ }\textbf {\bibinfo {volume} {5}},\ \bibinfo {pages}
  {497} (\bibinfo {year} {2021})}\BibitemShut {NoStop}%
\bibitem [{\citenamefont {Gottesman}(2016)}]{Gottesman2016}%
  \BibitemOpen
  \bibfield  {author} {\bibinfo {author} {D.~Gottesman},\ }\href@noop {} {\emph
  {Quantum fault tolerance in small experiments}} (\bibinfo {year} {2016}),\
  \Eprint {https://arxiv.org/abs/1610.03507} {arXiv:1610.03507 [quant-ph]}
  \BibitemShut {NoStop}%
\bibitem [{\citenamefont {Kitaev}(2003)}]{kitaev2003fault}%
  \BibitemOpen
  \bibfield  {author} {\bibinfo {author} {A.~Y. Kitaev},\ }\emph
  {Fault-tolerant quantum computation by anyons},\ \href
  {https://doi.org/https://doi.org/10.1016/S0003-4916(02)00018-0} {\bibfield
  {journal} {\bibinfo  {journal} {Annals of Physics}\ }\textbf {\bibinfo
  {volume} {303}},\ \bibinfo {pages} {2} (\bibinfo {year} {2003})}\BibitemShut
  {NoStop}%
\bibitem [{\citenamefont {Fowler}\ \emph {et~al.}(2012)\citenamefont {Fowler},
  \citenamefont {Mariantoni}, \citenamefont {Martinis},\ and\ \citenamefont
  {Cleland}}]{fowler2012surface}%
  \BibitemOpen
  \bibfield  {author} {\bibinfo {author} {A.~G. Fowler}, \bibinfo {author}
  {M.~Mariantoni}, \bibinfo {author} {J.~M. Martinis},\ and\ \bibinfo {author}
  {A.~N. Cleland},\ }\emph {Surface codes: Towards practical large-scale
  quantum computation},\ \href {https://doi.org/10.1103/PhysRevA.86.032324}
  {\bibfield  {journal} {\bibinfo  {journal} {Physical Review A}\ }\textbf
  {\bibinfo {volume} {86}},\ \bibinfo {pages} {032324} (\bibinfo {year}
  {2012})}\BibitemShut {NoStop}%
\bibitem [{\citenamefont {Bombin}\ and\ \citenamefont
  {Martin-Delgado}(2006)}]{bombin2006topological}%
  \BibitemOpen
  \bibfield  {author} {\bibinfo {author} {H.~Bombin}\ and\ \bibinfo {author}
  {M.~A. Martin-Delgado},\ }\emph {Topological quantum distillation},\ \href
  {https://doi.org/10.1103/PhysRevLett.97.180501} {\bibfield  {journal}
  {\bibinfo  {journal} {Physical Review Letters}\ }\textbf {\bibinfo {volume}
  {97}},\ \bibinfo {pages} {180501} (\bibinfo {year} {2006})}\BibitemShut
  {NoStop}%
\bibitem [{\citenamefont {Fowler}(2011)}]{fowler2011two}%
  \BibitemOpen
  \bibfield  {author} {\bibinfo {author} {A.~G. Fowler},\ }\emph
  {Two-dimensional color-code quantum computation},\ \href
  {https://doi.org/10.1103/PhysRevA.83.042310} {\bibfield  {journal} {\bibinfo
  {journal} {Physical Review A}\ }\textbf {\bibinfo {volume} {83}},\ \bibinfo
  {pages} {042310} (\bibinfo {year} {2011})}\BibitemShut {NoStop}%
\bibitem [{\citenamefont {Gottesman}(1998)}]{gottesman1998heisenberg}%
  \BibitemOpen
  \bibfield  {author} {\bibinfo {author} {D.~Gottesman},\ }\href@noop {} {\emph
  {The Heisenberg Representation of Quantum Computers}} (\bibinfo {year}
  {1998}),\ \Eprint {https://arxiv.org/abs/quant-ph/9807006}
  {arXiv:quant-ph/9807006 [quant-ph]} \BibitemShut {NoStop}%
\bibitem [{\citenamefont {Dennis}\ \emph {et~al.}(2002)\citenamefont {Dennis},
  \citenamefont {Kitaev}, \citenamefont {Landahl},\ and\ \citenamefont
  {Preskill}}]{dennis2002topological}%
  \BibitemOpen
  \bibfield  {author} {\bibinfo {author} {E.~Dennis}, \bibinfo {author}
  {A.~Kitaev}, \bibinfo {author} {A.~Landahl},\ and\ \bibinfo {author}
  {J.~Preskill},\ }\emph {Topological quantum memory},\ \href
  {https://doi.org/10.1063/1.1499754} {\bibfield  {journal} {\bibinfo
  {journal} {Journal of Mathematical Physics}\ }\textbf {\bibinfo {volume}
  {43}},\ \bibinfo {pages} {4452} (\bibinfo {year} {2002})}\BibitemShut
  {NoStop}%
\bibitem [{\citenamefont {Ohno}\ \emph {et~al.}(2004)\citenamefont {Ohno},
  \citenamefont {Arakawa}, \citenamefont {Ichinose},\ and\ \citenamefont
  {Matsui}}]{ohno2004phase}%
  \BibitemOpen
  \bibfield  {author} {\bibinfo {author} {T.~Ohno}, \bibinfo {author}
  {G.~Arakawa}, \bibinfo {author} {I.~Ichinose},\ and\ \bibinfo {author}
  {T.~Matsui},\ }\emph {Phase structure of the random-plaquette Z2 gauge model:
  accuracy threshold for a toric quantum memory},\ \href
  {https://doi.org/https://doi.org/10.1016/j.nuclphysb.2004.07.003} {\bibfield
  {journal} {\bibinfo  {journal} {Nuclear physics B}\ }\textbf {\bibinfo
  {volume} {697}},\ \bibinfo {pages} {462} (\bibinfo {year}
  {2004})}\BibitemShut {NoStop}%
\bibitem [{\citenamefont {Katzgraber}\ \emph {et~al.}(2009)\citenamefont
  {Katzgraber}, \citenamefont {Bombin},\ and\ \citenamefont
  {Martin-Delgado}}]{katzgraber2009error}%
  \BibitemOpen
  \bibfield  {author} {\bibinfo {author} {H.~G. Katzgraber}, \bibinfo {author}
  {H.~Bombin},\ and\ \bibinfo {author} {M.~Martin-Delgado},\ }\emph {Error
  threshold for color codes and random three-body Ising models},\ \href
  {https://doi.org/10.1103/PhysRevLett.103.090501} {\bibfield  {journal}
  {\bibinfo  {journal} {Physical Review Letters}\ }\textbf {\bibinfo {volume}
  {103}},\ \bibinfo {pages} {090501} (\bibinfo {year} {2009})}\BibitemShut
  {NoStop}%
\bibitem [{\citenamefont {Raussendorf}\ and\ \citenamefont
  {Harrington}(2007)}]{raussendorf2007fault}%
  \BibitemOpen
  \bibfield  {author} {\bibinfo {author} {R.~Raussendorf}\ and\ \bibinfo
  {author} {J.~Harrington},\ }\emph {Fault-tolerant quantum computation with
  high threshold in two dimensions},\ \href
  {https://doi.org/10.1103/PhysRevLett.98.190504} {\bibfield  {journal}
  {\bibinfo  {journal} {Physical Review Letters}\ }\textbf {\bibinfo {volume}
  {98}},\ \bibinfo {pages} {190504} (\bibinfo {year} {2007})}\BibitemShut
  {NoStop}%
\bibitem [{\citenamefont {Landahl}\ \emph {et~al.}(2011)\citenamefont
  {Landahl}, \citenamefont {Anderson},\ and\ \citenamefont
  {Rice}}]{landahl2011fault}%
  \BibitemOpen
  \bibfield  {author} {\bibinfo {author} {A.~J. Landahl}, \bibinfo {author}
  {J.~T. Anderson},\ and\ \bibinfo {author} {P.~R. Rice},\ }\href@noop {}
  {\emph {Fault-tolerant quantum computing with color codes}} (\bibinfo {year}
  {2011}),\ \Eprint {https://arxiv.org/abs/1108.5738} {arXiv:1108.5738
  [quant-ph]} \BibitemShut {NoStop}%
\bibitem [{\citenamefont {Tomita}\ and\ \citenamefont
  {Svore}(2014)}]{tomita2014low}%
  \BibitemOpen
  \bibfield  {author} {\bibinfo {author} {Y.~Tomita}\ and\ \bibinfo {author}
  {K.~M. Svore},\ }\emph {Low-distance surface codes under realistic quantum
  noise},\ \href {https://doi.org/10.1103/PhysRevA.90.062320} {\bibfield
  {journal} {\bibinfo  {journal} {Physical Review A}\ }\textbf {\bibinfo
  {volume} {90}},\ \bibinfo {pages} {062320} (\bibinfo {year}
  {2014})}\BibitemShut {NoStop}%
\bibitem [{\citenamefont {Bermudez}\ \emph {et~al.}(2017)\citenamefont
  {Bermudez}, \citenamefont {Xu}, \citenamefont {Nigmatullin}, \citenamefont
  {O’Gorman}, \citenamefont {Negnevitsky}, \citenamefont {Schindler},
  \citenamefont {Monz}, \citenamefont {Poschinger}, \citenamefont {Hempel},
  \citenamefont {Home} \emph {et~al.}}]{bermudez2017assessing}%
  \BibitemOpen
  \bibfield  {author} {\bibinfo {author} {A.~Bermudez}, et~al.,\ }\emph
  {Assessing the progress of trapped-ion processors towards fault-tolerant
  quantum computation},\ \href {https://doi.org/10.1103/PhysRevX.7.041061}
  {\bibfield  {journal} {\bibinfo  {journal} {Physical Review X}\ }\textbf
  {\bibinfo {volume} {7}},\ \bibinfo {pages} {041061} (\bibinfo {year}
  {2017})}\BibitemShut {NoStop}%
\bibitem [{\citenamefont {Debroy}\ \emph {et~al.}(2020)\citenamefont {Debroy},
  \citenamefont {Li}, \citenamefont {Huang},\ and\ \citenamefont
  {Brown}}]{debroy2020logical}%
  \BibitemOpen
  \bibfield  {author} {\bibinfo {author} {D.~M. Debroy}, \bibinfo {author}
  {M.~Li}, \bibinfo {author} {S.~Huang},\ and\ \bibinfo {author} {K.~R.
  Brown},\ }\emph {Logical performance of 9 qubit compass codes in ion traps
  with crosstalk errors},\ \href {https://doi.org/10.1088/2058-9565/ab7e80}
  {\bibfield  {journal} {\bibinfo  {journal} {Quantum Science and Technology}\
  }\textbf {\bibinfo {volume} {5}},\ \bibinfo {pages} {034002} (\bibinfo {year}
  {2020})}\BibitemShut {NoStop}%
\bibitem [{\citenamefont {Parrado-Rodr{\'\i}guez}\ \emph
  {et~al.}(2021)\citenamefont {Parrado-Rodr{\'\i}guez}, \citenamefont
  {Ryan-Anderson}, \citenamefont {Bermudez},\ and\ \citenamefont
  {M{\"u}ller}}]{parrado2021crosstalk}%
  \BibitemOpen
  \bibfield  {author} {\bibinfo {author} {P.~Parrado-Rodr{\'\i}guez}, \bibinfo
  {author} {C.~Ryan-Anderson}, \bibinfo {author} {A.~Bermudez},\ and\ \bibinfo
  {author} {M.~M{\"u}ller},\ }\emph {Crosstalk suppression for fault-tolerant
  quantum error correction with trapped ions},\ \href
  {https://doi.org/10.22331/q-2021-06-29-487} {\bibfield  {journal} {\bibinfo
  {journal} {Quantum}\ }\textbf {\bibinfo {volume} {5}},\ \bibinfo {pages}
  {487} (\bibinfo {year} {2021})}\BibitemShut {NoStop}%
\bibitem [{\citenamefont {Guti{\'e}rrez}\ \emph {et~al.}(2013)\citenamefont
  {Guti{\'e}rrez}, \citenamefont {Svec}, \citenamefont {Vargo},\ and\
  \citenamefont {Brown}}]{gutierrez2013approximation}%
  \BibitemOpen
  \bibfield  {author} {\bibinfo {author} {M.~Guti{\'e}rrez}, \bibinfo {author}
  {L.~Svec}, \bibinfo {author} {A.~Vargo},\ and\ \bibinfo {author} {K.~R.
  Brown},\ }\emph {Approximation of realistic errors by Clifford channels and
  Pauli measurements},\ \href {https://doi.org/10.1103/PhysRevA.87.030302}
  {\bibfield  {journal} {\bibinfo  {journal} {Physical Review A}\ }\textbf
  {\bibinfo {volume} {87}},\ \bibinfo {pages} {030302} (\bibinfo {year}
  {2013})}\BibitemShut {NoStop}%
\bibitem [{\citenamefont {Plenio}\ and\ \citenamefont
  {Knight}(1998)}]{plenio1998quantum}%
  \BibitemOpen
  \bibfield  {author} {\bibinfo {author} {M.~B. Plenio}\ and\ \bibinfo {author}
  {P.~L. Knight},\ }\emph {The quantum-jump approach to dissipative dynamics in
  quantum optics},\ \href {https://doi.org/10.1103/RevModPhys.70.101}
  {\bibfield  {journal} {\bibinfo  {journal} {Reviews of Modern Physics}\
  }\textbf {\bibinfo {volume} {70}},\ \bibinfo {pages} {101} (\bibinfo {year}
  {1998})}\BibitemShut {NoStop}%
\bibitem [{\citenamefont {Jones}\ \emph {et~al.}(2019)\citenamefont {Jones},
  \citenamefont {Brown}, \citenamefont {Bush},\ and\ \citenamefont
  {Benjamin}}]{jones2019quest}%
  \BibitemOpen
  \bibfield  {author} {\bibinfo {author} {T.~Jones}, \bibinfo {author}
  {A.~Brown}, \bibinfo {author} {I.~Bush},\ and\ \bibinfo {author} {S.~C.
  Benjamin},\ }\emph {QuEST and high performance simulation of quantum
  computers},\ \href {https://doi.org/10.1038/s41598-019-47174-9} {\bibfield
  {journal} {\bibinfo  {journal} {Scientific Reports}\ }\textbf {\bibinfo
  {volume} {9}},\ \bibinfo {pages} {1} (\bibinfo {year} {2019})}\BibitemShut
  {NoStop}%
\bibitem [{\citenamefont {Chao}\ and\ \citenamefont
  {Reichardt}(2018{\natexlab{b}})}]{Chao2018}%
  \BibitemOpen
  \bibfield  {author} {\bibinfo {author} {R.~Chao}\ and\ \bibinfo {author}
  {B.~W. Reichardt},\ }\emph {{Quantum error correction with only two extra
  qubits}},\ \href {https://doi.org/10.1103/PhysRevLett.121.050502} {\bibfield
  {journal} {\bibinfo  {journal} {Physical Review Letters}\ }\textbf {\bibinfo
  {volume} {121}},\ \bibinfo {pages} {050502} (\bibinfo {year}
  {2018}{\natexlab{b}})}\BibitemShut {NoStop}%
\bibitem [{\citenamefont {Foss-Feig}\ \emph {et~al.}(2023)\citenamefont
  {Foss-Feig}, \citenamefont {Tikku}, \citenamefont {Lu}, \citenamefont
  {Mayer}, \citenamefont {Iqbal}, \citenamefont {Gatterman}, \citenamefont
  {Gerber}, \citenamefont {Gilmore}, \citenamefont {Gresh}, \citenamefont
  {Hankin}, \citenamefont {Hewitt}, \citenamefont {Horst}, \citenamefont
  {Matheny}, \citenamefont {Mengle}, \citenamefont {Neyenhuis}, \citenamefont
  {Dreyer}, \citenamefont {Hayes}, \citenamefont {Hsieh},\ and\ \citenamefont
  {Kim}}]{foss2023experimental}%
  \BibitemOpen
  \bibfield  {author} {\bibinfo {author} {M.~Foss-Feig}, et~al.,\ }\href@noop
  {} {\emph {Experimental demonstration of the advantage of adaptive quantum
  circuits}} (\bibinfo {year} {2023}),\ \Eprint
  {https://arxiv.org/abs/2302.03029} {arXiv:2302.03029 [quant-ph]} \BibitemShut
  {NoStop}%
\bibitem [{\citenamefont {Chamberland}\ and\ \citenamefont
  {Cross}(2019)}]{chamberland2019fault}%
  \BibitemOpen
  \bibfield  {author} {\bibinfo {author} {C.~Chamberland}\ and\ \bibinfo
  {author} {A.~W. Cross},\ }\emph {Fault-tolerant magic state preparation with
  flag qubits},\ \href
  {https://doi.org/https://doi.org/10.22331/q-2019-05-20-143} {\bibfield
  {journal} {\bibinfo  {journal} {Quantum}\ }\textbf {\bibinfo {volume} {3}},\
  \bibinfo {pages} {143} (\bibinfo {year} {2019})}\BibitemShut {NoStop}%
\bibitem [{\citenamefont {Tuckett}\ \emph {et~al.}(2018)\citenamefont
  {Tuckett}, \citenamefont {Bartlett},\ and\ \citenamefont
  {Flammia}}]{tuckett2018ultrahigh}%
  \BibitemOpen
  \bibfield  {author} {\bibinfo {author} {D.~K. Tuckett}, \bibinfo {author}
  {S.~D. Bartlett},\ and\ \bibinfo {author} {S.~T. Flammia},\ }\emph {Ultrahigh
  error threshold for surface codes with biased noise},\ \href
  {https://doi.org/10.1103/PhysRevLett.120.050505} {\bibfield  {journal}
  {\bibinfo  {journal} {Physical Review Letters}\ }\textbf {\bibinfo {volume}
  {120}},\ \bibinfo {pages} {050505} (\bibinfo {year} {2018})}\BibitemShut
  {NoStop}%
\bibitem [{\citenamefont {Tuckett}\ \emph {et~al.}(2019)\citenamefont
  {Tuckett}, \citenamefont {Darmawan}, \citenamefont {Chubb}, \citenamefont
  {Bravyi}, \citenamefont {Bartlett},\ and\ \citenamefont
  {Flammia}}]{tuckett2019tailoring}%
  \BibitemOpen
  \bibfield  {author} {\bibinfo {author} {D.~K. Tuckett}, et~al.,\ }\emph
  {Tailoring surface codes for highly biased noise},\ \href
  {https://doi.org/10.1103/PhysRevX.9.041031} {\bibfield  {journal} {\bibinfo
  {journal} {Physical Review X}\ }\textbf {\bibinfo {volume} {9}},\ \bibinfo
  {pages} {041031} (\bibinfo {year} {2019})}\BibitemShut {NoStop}%
\bibitem [{\citenamefont {Tuckett}\ \emph {et~al.}(2020)\citenamefont
  {Tuckett}, \citenamefont {Bartlett}, \citenamefont {Flammia},\ and\
  \citenamefont {Brown}}]{tuckett2020fault}%
  \BibitemOpen
  \bibfield  {author} {\bibinfo {author} {D.~K. Tuckett}, \bibinfo {author}
  {S.~D. Bartlett}, \bibinfo {author} {S.~T. Flammia},\ and\ \bibinfo {author}
  {B.~J. Brown},\ }\emph {Fault-tolerant thresholds for the surface code in
  excess of 5\% under biased noise},\ \href
  {https://doi.org/10.1103/PhysRevLett.124.130501} {\bibfield  {journal}
  {\bibinfo  {journal} {Physical Review Letters}\ }\textbf {\bibinfo {volume}
  {124}},\ \bibinfo {pages} {130501} (\bibinfo {year} {2020})}\BibitemShut
  {NoStop}%
\bibitem [{\citenamefont {Bonilla~Ataides}\ \emph {et~al.}(2021)\citenamefont
  {Bonilla~Ataides}, \citenamefont {Tuckett}, \citenamefont {Bartlett},
  \citenamefont {Flammia},\ and\ \citenamefont {Brown}}]{bonilla2021xzzx}%
  \BibitemOpen
  \bibfield  {author} {\bibinfo {author} {J.~P. Bonilla~Ataides}, \bibinfo
  {author} {D.~K. Tuckett}, \bibinfo {author} {S.~D. Bartlett}, \bibinfo
  {author} {S.~T. Flammia},\ and\ \bibinfo {author} {B.~J. Brown},\ }\emph {The
  XZZX surface code},\ \href {https://doi.org/10.1038/s41467-021-22274-1}
  {\bibfield  {journal} {\bibinfo  {journal} {Nature Communications}\ }\textbf
  {\bibinfo {volume} {12}},\ \bibinfo {pages} {1} (\bibinfo {year}
  {2021})}\BibitemShut {NoStop}%
\bibitem [{\citenamefont {Darmawan}\ \emph {et~al.}(2021)\citenamefont
  {Darmawan}, \citenamefont {Brown}, \citenamefont {Grimsmo}, \citenamefont
  {Tuckett},\ and\ \citenamefont {Puri}}]{darmawan2021practical}%
  \BibitemOpen
  \bibfield  {author} {\bibinfo {author} {A.~S. Darmawan}, \bibinfo {author}
  {B.~J. Brown}, \bibinfo {author} {A.~L. Grimsmo}, \bibinfo {author} {D.~K.
  Tuckett},\ and\ \bibinfo {author} {S.~Puri},\ }\emph {Practical quantum error
  correction with the XZZX code and Kerr-cat qubits},\ \href
  {https://doi.org/10.1103/PRXQuantum.2.030345} {\bibfield  {journal} {\bibinfo
   {journal} {PRX Quantum}\ }\textbf {\bibinfo {volume} {2}},\ \bibinfo {pages}
  {030345} (\bibinfo {year} {2021})}\BibitemShut {NoStop}%
\bibitem [{\citenamefont {Xu}\ \emph {et~al.}(2023)\citenamefont {Xu},
  \citenamefont {Mannucci}, \citenamefont {Seif}, \citenamefont {Kubica},
  \citenamefont {Flammia},\ and\ \citenamefont {Jiang}}]{xu2022tailored}%
  \BibitemOpen
  \bibfield  {author} {\bibinfo {author} {Q.~Xu}, et~al.,\ }\emph {Tailored
  XZZX codes for biased noise},\ \href
  {https://doi.org/10.1103/PhysRevResearch.5.013035} {\bibfield  {journal}
  {\bibinfo  {journal} {Physical Review Research}\ }\textbf {\bibinfo {volume}
  {5}},\ \bibinfo {pages} {013035} (\bibinfo {year} {2023})}\BibitemShut
  {NoStop}%
\bibitem [{\citenamefont {Gottesman}\ and\ \citenamefont
  {Chuang}(1999)}]{gottesman1999demonstrating}%
  \BibitemOpen
  \bibfield  {author} {\bibinfo {author} {D.~Gottesman}\ and\ \bibinfo {author}
  {I.~L. Chuang},\ }\emph {Demonstrating the viability of universal quantum
  computation using teleportation and single-qubit operations},\ \href
  {https://doi.org/https://doi.org/10.1038/46503} {\bibfield  {journal}
  {\bibinfo  {journal} {Nature}\ }\textbf {\bibinfo {volume} {402}},\ \bibinfo
  {pages} {390} (\bibinfo {year} {1999})}\BibitemShut {NoStop}%
\bibitem [{\citenamefont {Hilder}\ \emph {et~al.}(2022)\citenamefont {Hilder},
  \citenamefont {Pijn}, \citenamefont {Onishchenko}, \citenamefont {Stahl},
  \citenamefont {Orth}, \citenamefont {Lekitsch}, \citenamefont
  {Rodriguez-Blanco}, \citenamefont {M{\"u}ller}, \citenamefont
  {Schmidt-Kaler},\ and\ \citenamefont {Poschinger}}]{hilder2022fault}%
  \BibitemOpen
  \bibfield  {author} {\bibinfo {author} {J.~Hilder}, et~al.,\ }\emph
  {Fault-tolerant parity readout on a shuttling-based trapped-ion quantum
  computer},\ \href {https://doi.org/10.1103/PhysRevX.12.011032} {\bibfield
  {journal} {\bibinfo  {journal} {Physical Review X}\ }\textbf {\bibinfo
  {volume} {12}},\ \bibinfo {pages} {011032} (\bibinfo {year}
  {2022})}\BibitemShut {NoStop}%
\bibitem [{\citenamefont {Kirkpatrick}\ \emph {et~al.}(1983)\citenamefont
  {Kirkpatrick}, \citenamefont {Gelatt~Jr},\ and\ \citenamefont
  {Vecchi}}]{kirkpatrick1983optimization}%
  \BibitemOpen
  \bibfield  {author} {\bibinfo {author} {S.~Kirkpatrick}, \bibinfo {author}
  {C.~D. Gelatt~Jr},\ and\ \bibinfo {author} {M.~P. Vecchi},\ }\emph
  {Optimization by simulated annealing},\ \href
  {https://doi.org/https://doi.org/10.1126/science.220.4598.671} {\bibfield
  {journal} {\bibinfo  {journal} {Science}\ }\textbf {\bibinfo {volume}
  {220}},\ \bibinfo {pages} {671} (\bibinfo {year} {1983})}\BibitemShut
  {NoStop}%
\bibitem [{\citenamefont {Krinner}\ \emph {et~al.}(2022)\citenamefont
  {Krinner}, \citenamefont {Lacroix}, \citenamefont {Remm}, \citenamefont
  {Di~Paolo}, \citenamefont {Genois}, \citenamefont {Leroux}, \citenamefont
  {Hellings}, \citenamefont {Lazar}, \citenamefont {Swiadek}, \citenamefont
  {Herrmann} \emph {et~al.}}]{krinner2021realizing}%
  \BibitemOpen
  \bibfield  {author} {\bibinfo {author} {S.~Krinner}, et~al.,\ }\emph
  {Realizing repeated quantum error correction in a distance-three surface
  code},\ \href {https://doi.org/https://doi.org/10.1038/s41586-022-04566-8}
  {\bibfield  {journal} {\bibinfo  {journal} {Nature}\ }\textbf {\bibinfo
  {volume} {605}},\ \bibinfo {pages} {669} (\bibinfo {year}
  {2022})}\BibitemShut {NoStop}%
\bibitem [{\citenamefont {Soch}(2023)}]{statproofbook}%
  \BibitemOpen
  \bibfield  {author} {\bibinfo {author} {J.~Soch},\ }\href@noop {} {\emph {The
  book of statistical proofs}},\ \bibinfo {howpublished}
  {\url{https://statproofbook.github.io}} (\bibinfo {year} {2023})\BibitemShut
  {NoStop}%
\bibitem [{\citenamefont {Goodman}(1960)}]{goodman1960exact}%
  \BibitemOpen
  \bibfield  {author} {\bibinfo {author} {L.~A. Goodman},\ }\emph {On the exact
  variance of products},\ \href
  {https://doi.org/10.1080/01621459.1960.10483369} {\bibfield  {journal}
  {\bibinfo  {journal} {Journal of the American Statistical Association}\
  }\textbf {\bibinfo {volume} {55}},\ \bibinfo {pages} {708} (\bibinfo {year}
  {1960})}\BibitemShut {NoStop}%
\bibitem [{\citenamefont {Goodman}(1962)}]{goodman1962variance}%
  \BibitemOpen
  \bibfield  {author} {\bibinfo {author} {L.~A. Goodman},\ }\emph {The variance
  of the product of $K$ random variables},\ \href
  {https://doi.org/10.2307/2282440} {\bibfield  {journal} {\bibinfo  {journal}
  {Journal of the American Statistical Association}\ }\textbf {\bibinfo
  {volume} {57}},\ \bibinfo {pages} {54} (\bibinfo {year} {1962})}\BibitemShut
  {NoStop}%
\end{thebibliography}%

\end{document}